\documentclass[12pt,preprint]{aastex}
\usepackage{epsf}
\newcommand{\beq}{\begin{equation}} 
\newcommand{\Halpha}{H${\alpha}$} 
\newcommand{\Hbeta}{H${\beta}$} 
\newcommand{\Hgamma}{H${\gamma}$} 
\newcommand{\enq}{\end{equation}}   

\newcommand{\Msun}{\mathrm{M}_{\odot}}
\newcommand{\chandra}{\textit{Chandra}}

\newcommand{\ergsec}{\ensuremath{\mathrm{erg}~\mathrm{s}^{-1}}}
\newcommand{\cmsq}{\ensuremath{\mathrm{cm}^2}}
\newcommand{\pcmsq}{\ensuremath{\mathrm{cm}^{-2}}}

\begin{document}

\shortauthors{Prieto et~al.}

\title{A Study of the Type~Ia/IIn Supernova 2005gj from X-ray to the
Infrared: Paper I \altaffilmark{1,2,3,4}}

\author{
J.~L.~Prieto\altaffilmark{5},
P.~M.~Garnavich\altaffilmark{6}, 
M.~M.~Phillips\altaffilmark{7},
D.~L.~DePoy\altaffilmark{5},
J.~Parrent\altaffilmark{8},
D.~Pooley\altaffilmark{9,10}, 
V.~V.~Dwarkadas\altaffilmark{11},
E.~Baron\altaffilmark{8,12},
B.~Bassett\altaffilmark{13,14},
A.~Becker\altaffilmark{15},
D.~Cinabro\altaffilmark{16},
F.~DeJongh\altaffilmark{17},
B.~Dilday\altaffilmark{18,19},
M.~Doi\altaffilmark{20},
J.~A.~Frieman\altaffilmark{11,17,18},
C.~J.~Hogan\altaffilmark{15},
J.~Holtzman\altaffilmark{21},
S.~Jha\altaffilmark{22},
R.~Kessler\altaffilmark{19,23},
K.~Konishi\altaffilmark{24},
H.~Lampeitl\altaffilmark{25},
J.~Marriner\altaffilmark{17},
J.~L.~Marshall\altaffilmark{5},
G.~Miknaitis\altaffilmark{17},
R.~C.~Nichol\altaffilmark{26},
A.~G.~Riess\altaffilmark{25,27},
M.~W.~Richmond\altaffilmark{28},
R.~Romani\altaffilmark{22},
M.~Sako\altaffilmark{29},
D.~P.~Schneider\altaffilmark{30},
M.~Smith\altaffilmark{26},
N.~Takanashi\altaffilmark{20},
K.~Tokita\altaffilmark{20},
K.~van~der~Heyden\altaffilmark{13},
N.~Yasuda\altaffilmark{24},
C.~Zheng\altaffilmark{22},
J.~C.~Wheeler, \altaffilmark{31},
J.~Barentine\altaffilmark{31,32},
J.~Dembicky\altaffilmark{32},
J.~Eastman\altaffilmark{5},
S.~Frank\altaffilmark{5},
W.~Ketzeback\altaffilmark{32},
R.~J.~McMillan\altaffilmark{32},
N.~Morrell\altaffilmark{7},
G.~Folatelli\altaffilmark{7},
C.~Contreras\altaffilmark{7},
C.~R.~Burns\altaffilmark{33},
W.~L.~Freedman\altaffilmark{33},
S.~Gonz\'alez\altaffilmark{7},
M.~Hamuy\altaffilmark{34},
W.~Krzeminski\altaffilmark{7},
B.~F.~Madore\altaffilmark{33},
D.~Murphy\altaffilmark{13},
S.~E.~Persson\altaffilmark{33},
M.~Roth\altaffilmark{7},
N.~B.~Suntzeff\altaffilmark{35}
}

\altaffiltext{1}{Based in part on observations obtained with the Apache
Point Observatory 3.5-meter telescope, which is owned and operated
by the Astrophysical Research Consortium.}
\altaffiltext{2}{Based in part on observations taken at the Cerro Tololo
Inter-American Observatory, National Optical Astronomy Observatory, 
which is operated by the Association of Universities for Research in 
Astronomy, Inc. (AURA) under cooperative agreement with the National 
Science Foundation.}
\altaffiltext{3}{This paper includes data gathered with the 6.5 meter 
Magellan Telescopes located at Las Campanas Observatory, Chile.}
\altaffiltext{4}{Partly based on observations collected at the European 
Southern Observatory, Chile, in the course of programme 076.A-0156.}
\altaffiltext{5}{
  Department of Astronomy,
   Ohio State University, 140 West 18th Avenue, Columbus, OH 43210-1173.
}
\altaffiltext{6}{
  University of Notre Dame, 225 Nieuwland Science, Notre Dame, IN 46556-5670.
}
\altaffiltext{7}{
  Las Campanas Observatory, Carnegie Observatories, La Serena, Chile
}
\altaffiltext{8}{
  Holmer L. Dodge Department of Physics and Astronomy, University of Oklahoma, 440 West Brooks, Room 100, Norman, OK 73019-2061
}
\altaffiltext{9}{
  Astronomy Department, University of California at Berkeley, 601 Campbell Hall, Berkeley, CA 94720
}
\altaffiltext{10}{
  Chandra Fellow
}
\altaffiltext{11}{
  Department of Astronomy and Astrophysics,
   The University of Chicago, 5640 South Ellis Avenue, Chicago, IL 60637.
}
\altaffiltext{12}{
  Computational Research Division, Lawrence Berkeley
  National Laboratory, MS 50F-1650, 1 Cyclotron Rd, Berkeley, CA
  94720 USA
}
\altaffiltext{13}{
Department of Mathematics and Applied Mathematics,
University of Cape Town, Rondebosch 7701, South Africa.
}
\altaffiltext{14}{
  South African Astronomical Observatory,
   P.O. Box 9, Observatory 7935, South Africa.
}
\altaffiltext{15}{
  Department of Astronomy,
   University of Washington, Box 351580, Seattle, WA 98195.
}
\altaffiltext{16}{
Department of Physics, 
Wayne State University, Detroit, MI 48202.
}
\altaffiltext{17}{
Center for Particle Astrophysics, 
  Fermi National Accelerator Laboratory, P.O. Box 500, Batavia, IL 60510.
}
\altaffiltext{18}{
  Kavli Institute for Cosmological Physics, 
   The University of Chicago, 5640 South Ellis Avenue Chicago, IL 60637.
}
\altaffiltext{19}{
Department of Physics, 
University of Chicago, Chicago, IL 60637.
}
\altaffiltext{20}{
  Institute of Astronomy, Graduate School of Science,
   University of Tokyo 2-21-1, Osawa, Mitaka, Tokyo 181-0015, Japan.
}
\altaffiltext{21}{
  Department of Astronomy,
   MSC 4500,
   New Mexico State University, P.O. Box 30001, Las Cruces, NM 88003.
}
\altaffiltext{22}{
 Kavli Institute for Particle Astrophysics \& Cosmology, 
  Stanford University, Stanford, CA 94305-4060.
}
\altaffiltext{23}{
Enrico Fermi Institute,
University of Chicago, 5640 South Ellis Avenue, Chicago, IL 60637.
}
\altaffiltext{24}{
Institute for Cosmic Ray Research,
University of Tokyo, 5-1-5, Kashiwanoha, Kashiwa, Chiba, 277-8582, Japan.
}
\altaffiltext{25}{
  Space Telescope Science Institute,
   3700 San Martin Drive, Baltimore, MD 21218.
}
\altaffiltext{26}{
  Institute of Cosmology and Gravitation,
   Mercantile House,
   Hampshire Terrace, University of Portsmouth, Portsmouth PO1 2EG, UK.
}
\altaffiltext{27}{
Department of Physics and Astronomy,
Johns Hopkins University, 3400 North Charles Street, Baltimore, MD 21218.
}
\altaffiltext{28}{
  Physics Department,
   Rochester Institute of Technology,
   85 Lomb Memorial Drive, Rochester, NY 14623-5603.
}
\altaffiltext{29}{
Department of Physics and Astronomy,
University of Pennsylvania, 203 South 33rd Street, Philadelphia, PA  19104.
}
\altaffiltext{30}{
  Department of Astronomy and Astrophysics,
   The Pennsylvania State University,
   525 Davey Laboratory, University Park, PA 16802.
}
\altaffiltext{31}{
  Department of Astronomy,
   McDonald Observatory, University of Texas, Austin, TX 78712
}
\altaffiltext{32}{
  Apache Point Observatory, P.O. Box 59, Sunspot, NM 88349.
}
\altaffiltext{33}{
  Carnegie Institution of Washington, 813 Santa Barbara St., Pasadena, CA 91101
}
\altaffiltext{34}{
  Universidad de Chile, Departamento de Astronom\'ia, Santiago, Chile
}
\altaffiltext{35}{
  Texas A\&M University Physics Department, College Station, TX
}
\email{prieto@astronomy.ohio-state.edu}

\begin{abstract}
 We present extensive $ugriz\,YHJK_{s}$ photometry and optical
 spectroscopy of SN~2005gj obtained by the SDSS-II and CSP Supernova
 Projects, which give excellent coverage during the first 150~days after
 the time of explosion. These data show that SN~2005gj is the second
 clear case, after SN~2002ic, of a thermonuclear explosion in a dense
 circumstellar environment. Both the presence of singly and doubly
 ionized iron-peak elements (Fe~III and weak S~II, Si~II) near maximum
 light as well as the spectral evolution show that SN~2002ic-like events
 are Type~Ia explosions. Independent evidence comes from the exponential
 decay in luminosity of SN~2005gj, pointing to an exponential density
 distribution of the ejecta. The interaction of the supernova ejecta
 with the dense circumstellar medium is stronger than in SN~2002ic: (1)
 the supernova lines are weaker; (2) the Balmer emission lines are more
 luminous; and (3) the bolometric luminosity is higher close to maximum
 light. The velocity evolution of the \Halpha\ components suggest that
 the CSM around SN~2005gj is clumpy and it has a flatter density
 distribution compared with the steady wind solution, in agreement with
 SN~2002ic. An early X-ray observation with {\em Chandra} gives an
 upper-limit on the mass loss rate from the companion of $\dot{M}\la
 2\times 10^{-4}\,\,\Msun\,{\rm yr^{-1}}$.
 
\end{abstract}


\section{Introduction}
\label{sec:intro}

Thermonuclear supernova explosions (Type~Ia supernovae, SN~Ia hereafter)
are believed to be the detonation or deflagration of a white
dwarf accreting matter from a companion star \citep{arnett82}. The mass
of the white dwarf slowly increases until it approaches the
Chandrasekhar limit where the star becomes thermally unstable.  At this
point fusion of Carbon and Oxygen begins near the center and quickly
moves through most of the star before degeneracy is lifted. The result
is a spectacular and powerful explosion that is visible across much of
the Universe. Since SN~Ia arise from a narrow range of white dwarf
masses, their peak luminosities are very consistent and they make
excellent distance indicators \citep[e.g.,][]{phillips93}.  SN~Ia are
powerful probes of cosmology and have been instrumental in narrowing the
uncertainty in the Hubble parameter, discovery of the accelerating
universe, and constraining dark energy models \citep{hamuy95,hamuy96,riess05,
riess98,perlmutter99,riess04,astier06,riess07,wood-vasey07}.

But the use of SN~Ia as reliable distance indicators will always be
questioned until the progenitor and explosion physics are
well-understood. What types of binaries create SN~Ia? How is matter
transferred to the white dwarf without causing thermonuclear runaways on
the surface? Are there several types of progenitors?. These big
questions remain to be answered and detailed observations of hundreds of
events have yielded few clues. 

In 2002, \citet{hamuy03} identified a new kind of supernova. The early
spectrum of SN~2002ic was a cross between a Type~Ia event and a Type~IIn
\citep[][have called this type of supernova a ``IIa'']{deng04}, showing
P-Cygni features similar to SN~Ia and resolved Balmer lines in
emission. Type~IIn supernovae are core-collapse explosions going off in
dense circumstellar environments \citep{schlegel90,chevalier94}. They
are relatively common since the massive stars that create core-collapse
supernovae often have thick winds. If the interpretations of the
pre-explosion observations of SN~2005gl are correct, SN~IIn could be
associated in some cases with luminous blue variables \citep{gal-yam06}. 

In the case of SN~2002ic, the presence of Balmer lines with profiles
characteristic of SN~IIn, and the high luminosities and slow decline
after maximum lead to the conclusion that most of the energy came
from the interaction of the ejecta with a dense circumstellar medium
(CSM). Other Type~IIn events (SN~1997cy, \citealt{germany00}; SN~1999E,
\citealt{rigon03}) have been re-classified as SN~2002ic-like that were
caught late in their evolution \citep{hamuy03,wood-vasey04}.

SN~2002ic provided the first direct evidence that thermonuclear
explosions can also occur in a dense medium, but in this case the
circumstellar medium is probably generated by an Asymptotic Giant
companion (\citealt{hamuy03}; \citealt{wang04};
\citealt{han06}). However, there is still debate in the literature about
the origin of SN~2002ic. \citet{livio_riess03} proposed the merger of
two white dwarfs as a possible progenitor, with the explosion occurring
in the common envelope phase. \citet{chugai04} concluded that the
properties of the circumstellar interaction in 2002ic-like events can be
broadly explained by the SN~1.5 scenario \citep{iben83}: the
thermonuclear explosion of the degenerate core of a massive AGB
star. Recently, \citet{benetti06} questioned the earlier interpretation
of the observations and proposed that SN~2002ic can be equally well
explained by the core collapse of a stripped-envelope massive star in a
dense medium.
 
SN~2005gj was discovered on 2005 September 28.6 (UT) by the SDSS-II
Supernova Collaboration \citep{frieman07} in $gri$ images obtained with
the SDSS-2.5m telescope at Apache Point Observatory (APO). The new
supernova \citep{barentine05} was $\sim$1$\arcsec$ from the center of
its host galaxy at the position $\alpha$~=~$03^{\rm h}01^{\rm m}12\fs0$,
$\delta$~=~$+00\arcdeg33\arcmin13\farcs9$ (J2000.0). It had SDSS
magnitudes $(g, r, i)=(18.6, 18.6, 18.7)$~mag, obtained from PSF
photometry after kernel-matching and subtraction of a template image in
each band. The SN was independently discovered by the Nearby Supernova
Factory on 2005 September 29 \citep{aldering06}.

SN~2005gj was classified as a Type~Ia candidate from the first three
epochs of the $gri$ light curve using a light curve fitting program, and
was sent to the queue of the MDM-2.4m telescope for spectroscopic
confirmation. The optical spectrum obtained on 2005 October 1 (UT)
showed a blue continuum with resolved Hydrogen Balmer lines in emission,
very similar to the spectrum of a young Type~IIn supernova, but also
with an unusual continuum showing broad and weak absorption
features. Further spectroscopic follow-up showed a dramatic
evolution. The continuum became substantially redder and developed
broad, P-Cygni features probably associated with blended lines of
Fe-peak mass elements, similar to a Type~Ia SN a few weeks after
maximum. The spectrum obtained on 2005 Nov. 12 (UT) was remarkably
similar to that of the unusual Type~Ia supernova SN~2002ic obtained on
2002 Dec. 27 (UT) \citep{prieto05}.

\citet{aldering06} presented optical photometry and spectroscopy of this
SN. Through detailed analysis they confirmed its photometric and
spectroscopic resemblance to SN~2002ic, confirming it was a new case of
a Type~Ia explosion interacting with a dense circumstellar
environment. From a spectrum obtained with the slit oriented to overlap
with its host galaxy, they calculated a redshift for the host of
$z=0.0616 \pm 0.0002$. SN~2005gj was not detected in the radio with the
Very Large Array \citep{soderberg05} or in the X-ray with {\it Swift}
\citep{immler05}.

Here we present extensive follow-up photometry and spectroscopy of the
Type~Ia/Type~IIn SN~2005gj during the first $\sim$150~days after
discovery. These are the most detailed observations ever obtained of a
SN~2002ic-like event, and provide insight into the early evolution,
progenitor, and variety of these events. We also present a sensitive,
early X-ray observation with \chandra\ that gives an upper limit on the
X-ray luminosity of this peculiar object. We describe the optical and NIR
photometry of SN~2005gj in \S~\ref{sec:phot} and the optical
spectroscopy in \S~\ref{sec:spec}. We describe the X-ray observation
with \chandra\ in \S~\ref{sec:xray}. An analysis of the photometric and
spectroscopic data are presented in \S~\ref{sec:results}. Finally, we
discuss the significance of the results in \S~\ref{sec:discus}. We will
present later photometric follow-up and Spitzer/IRAC observations in
Paper~II. We adopt a cosmology with $H_{0}=72\pm 8\,\, {\rm km\ s^{-1}\
Mpc^{-1}}$, $\Omega_{M}=0.3$, and $\Omega_{\Lambda} = 0.7$ throughout
the paper (see \citealt{spergel06}), which yields a distance modulus of
$\mu=37.15$~mag to the host of SN~2005gj.

\section{Photometry}
\label{sec:phot}

\subsection{SDSS and MDM}
\label{subsec:phot_sdss}

The Sloan Digital Sky Survey (SDSS; \citealt{york00}) uses a wide-field,
2.5-meter telescope \citep{gunn06} and mosaic CCD camera \citep{gunn98}
at APO to survey the sky. The SDSS-II Supernova Survey, part of a 3-year
extension of the original SDSS, uses the APO-2.5m telescope to detect
and measure light-curves for a large number of supernovae through repeat
scans of the SDSS Southern equatorial stripe (about 2.5~deg wide by
$\sim$120~deg long) over the course of three 3-month campaigns
(Sept-Nov. 2005-2007). SN~2005gj was discovered in the second month of
the first campaign (October 2005). Twenty epochs of $ugriz$ photometry
were obtained between 2005 Sep 26-Nov 30 (U.T.). Details of the
photometric system, magnitude system, astrometry and calibration are
given in \citet{fukugita96}, \citet{lupton99}, \citet{hogg01},
\citet{smith02}, \citet{pier03}, \citet{ivezic04}, and
\citet{tucker06}. Additional $griz$ imaging of SN~2005gj was obtained
with the MDM Observatory 2.4m telescope using a facility CCD imager
(RETROCAM; see \citealt{Morgan05}, for a complete description of the
imager).

Photometry of SN~2005gj on the SDSS images was carried out using the
scene modeling code developed for SDSS-II as described in
\citet{holtzman07}. A sequence of stars around the supernova were taken
from the list of \citet{ivezic07}, who derived standard SDSS magnitudes
from multiple observations taken during the main SDSS survey under
photometric conditions. Using these stars, frame scalings and
astrometric solutions were derived for each of the supernova frames, as
well as for the twenty five pre-supernova frames taken as part of either
the main SDSS survey or the SN survey. Finally, the entire stack of
frames was simultaneously fit for a single supernova position, a fixed
galaxy background in each filter (characterized by a grid of galaxy
intensities), and the supernova brightness in each frame.

The supernova photometry in the MDM frames was also determined using the
scene modeling code. Since the MDM observations had different response
functions from the standard SDSS bandpasses, the photometric frame
solutions included color terms from the SDSS standard magnitudes.  To
prevent uncertainties in the frame parameters and color terms from
possibly corrupting the galaxy model (here affecting the SDSS
photometry), the MDM data were not included in the galaxy determination,
but the galaxy model as determined from the SDSS was used (with color
terms) to subtract the galaxy in the MDM frames. The resulting SN
photometry from the MDM frames is reported on the native MDM system,
since the color terms derived from stars are likely not to apply to the
spectrum of the supernova.

\begin{figure*}[t]
\epsscale{1.0}
\plotone{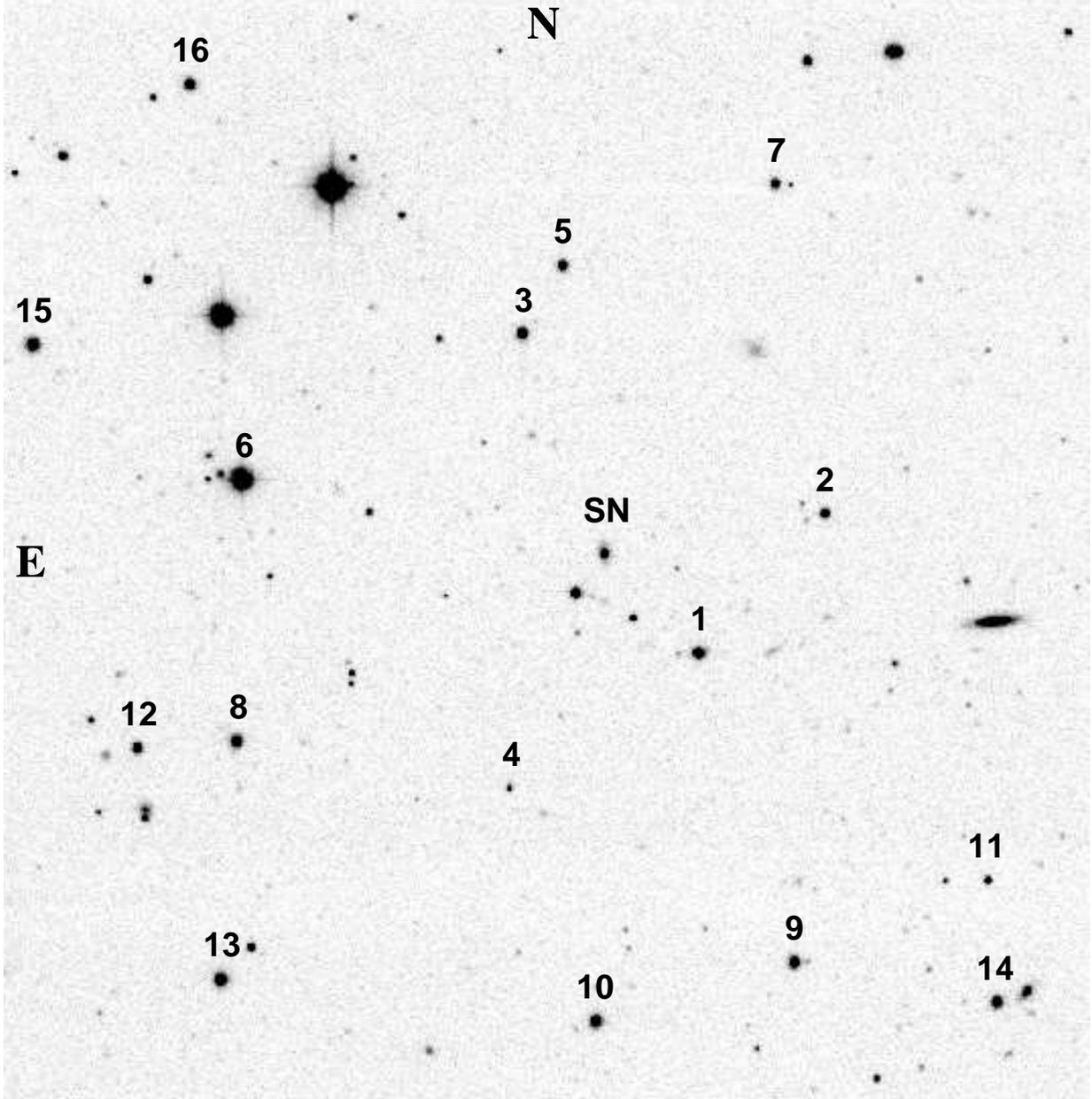}
\caption{$r'$-band image ($3.5'\times 3.5'$) of the field around
SN~2005gj obtained with the Swope-1m telescope at LCO. North is up and
east is to the left. Sixteen comparison stars in common between SDSS and
CSP used to derive differential photometry of the SN are labeled as in
Table~\ref{tab:stds1}. The SN is close to the center of the field.}
\label{fig:fc}
\end{figure*}

Figure~\ref{fig:fc} shows a $3.5\arcmin \times 3.5\arcmin$ field around
SN~2005gj and 16 comparison stars used for calibration of the SN
magnitudes by SDSS and the Carnegie Supernova Group (CSP; see
\S~\ref{subsec:phot_csp} below for details). In Table~\ref{tab:stds1} we
present the SDSS $ugriz$ and CSP $u'g'r'i'$ photometry of the comparison
stars in common. The final SDSS and MDM $griz$ photometry is given in
Table~\ref{tab:phot1}.

\subsection{CSP}
\label{subsec:phot_csp}

Optical photometry from the CSP was obtained with the Swope-1m telescope
at LCO, using a SITe CCD and a set of $u'g'r'i'$ filters. A subraster of
$1200 \times 1200$~pixels was read from the center of the CCD, which, at
a scale of 0.435~$\arcsec\,{\rm pixel^{-1}}$, yielded a field of view of
$8.7\arcmin \times 8.7\arcmin$. Typical image quality ranged between
1$\arcsec$ and 2$\arcsec$ (FWHM). A photometric sequence of comparison
stars in the SN field was calibrated with the Swope telescope from
observations of SDSS standard stars from \citet{smith02} during four
photometric nights. SN magnitudes in the $u'g'r'i'$ system were obtained
differentially relative to the comparison stars using PSF photometry.
In order to minimize the contamination from the host galaxy light in the
SN magnitudes, PSF-matched $ugri$ template images from SDSS were
subtracted from the SN images. On every galaxy-subtracted image, a PSF
was fitted to the SN and comparison stars within a radius of
3$\arcsec$. See \citet{hamuy06} for further details about the
measurement procedures.

The NIR photometry of SN~2005gj was obtained by the CSP using three
different instruments/telescopes. A total of 15 epochs in $Y$, $J$ and
$H$ filters were obtained using RetroCam, mounted on the Swope-1m
telescope at LCO. A few additional epochs in $YJHK_{s}$ were obtained with 
the Wide Infrared Camera (WIRC; \citealt{persson02}) mounted on the
duPont-2.5m telescope, and the PANIC camera \citep{martini04} mounted on
the Magellan-6.5m Baade telescope, both at LCO. We refer to
\citet{hamuy06} and \citet{phillips06} for details of the imagers and
procedures to extract the SN photometry. The host galaxy was not
subtracted from the NIR frames; therefore the SN photometry contains an
unknown galaxy contamination component.

The final CSP $u'g'r'i'$ photometry of SN~2005gj is given in
Table~\ref{tab:phot2} and $YJHK_{s}$ photometry in
Table~\ref{tab:phot3}.  A minimum uncertainty of 0.015~mag in the
optical bandpasses and 0.02~mag in the NIR is assumed for a single
measurement based on the typical scatter in the transformation from
instrumental to standard magnitudes of bright stars \citep{hamuy06}.

\section{Spectroscopy}
\label{sec:spec}

The spectroscopic observations of SN~2005gj are summarized in
Table~\ref{tab:spec}. They were obtained using five different
telescopes/instruments at four observatories.

A total of twelve spectra were obtained between early-October 2005 and
late-January 2006 with the Boller \& Chivens CCD Spectrograph (CCDS)
mounted at the MDM-2.4m telescope. This instrument uses a Loral
1200$\times$800 pixel CCD with 15~$\micron\,{\rm pixel^{-1}}$ and a
150~l/mm grating (blazed at 4700~\AA). We used a 2$\arcsec$ slit which
gives a dispersion of 3.1~${\rm \AA\,pixel^{-1}}$ in the wavelength
range $\sim$3800--7300~\AA.

Eight spectra were obtained at LCO with the Wide Field Reimaging Camera
(WFCCD) and the Modular Spectrograph (ModSpec) at the duPont-2.5m
telescope, and LDSS-3 at the Clay-6.5m telescope. In the case of WFCCD a
400~l/mm grism and a 1.6$\arcsec$ slit were used, reaching a dispersion
of 3~${\rm \AA\,pixel^{-1}}$ in the wavelength range
$\sim$3800--9200~\AA. For ModSpec we used a 300~l/mm grating and a
1$\arcsec$ slit that gave a dispersion of 2.45~${\rm \AA\,pixel^{-1}}$
in the range $\sim$3800--7300~\AA. Further details of WFCCD and ModSpec
can be found in \citet{hamuy06}. For LDSS-3, which employs a STA0500A
$4064\times 4064$~pixel CCD, we used the VPH blue and red grisms, the
latter with an OG590 filter to block second order contamination, with a
0.75$\arcsec$ slit reaching a dispersion of 0.70~${\rm \AA\,pixel^{-1}}$
and 1.12~${\rm \AA\,pixel^{-1}}$ in the blue and red sides of the
spectrum, respectively.

Additionally we obtained two early spectra with the Double Imaging
Spectrograph (DIS) mounted on the ARC-3.5m telescope at APO, and one
spectrum with the Intermediate dispersion Spectrograph and Imaging
System (ISIS) at the WHT-4.2m telescope at the Roque de Los Muchachos
Observatory in La Palma, Spain. The DIS spectrograph has blue and red
detectors, each uses a Marconi $2048\times 1024$~pixel CCD with
13.5~$\micron\,{\rm pixel^{-1}}$. We used a 300~l/mm grating and a
1.5$\arcsec$ slit which gives a dispersion of 2.4~${\rm
\AA\,pixel^{-1}}$. The ISIS spectrograph has a blue and red arm, the
blue arm using a EEV12 CCD and the red using a MARCONI2 detector.  We
used a 300~l/mm grating in both the blue and red arm and a 1$\arcsec$
slit, which gives a dispersion of 0.86~${\rm \AA\,pixel^{-1}}$ in the
blue and 1.47~${\rm \AA\,pixel^{-1}}$ in the red.
 
Most of the spectra were obtained close to the parallactic angle to
minimize relative changes in the calibration of the blue and red parts
of the spectrum due to differential refraction through the
atmosphere. The spectroscopic reductions were performed using standard
IRAF tasks and included: bias and overscan subtraction, flat-fielding,
combination of 2-4 individual 2D spectra to reach the best
signal-to-noise ratio in the final image, tracing and extraction of a 1D
spectrum from the combined 2D image, subtracting the background sky
around the selected aperture, wavelength calibration using an arc-lamp,
and flux calibration. In order to flux calibrate the spectra we observed
1-2 spectrophotometric standard stars per night. The spectra from LCO,
WHT and MDM were corrected by atmospheric telluric lines using the
spectrum of a hot spectrophotometric standard star and the spectra from
APO were corrected using a model atmosphere. This correction is not
optimal for some of the spectra and there are evident residuals left in
the corrected spectra.

\begin{figure*}[t]
\epsscale{1.0}
\plotone{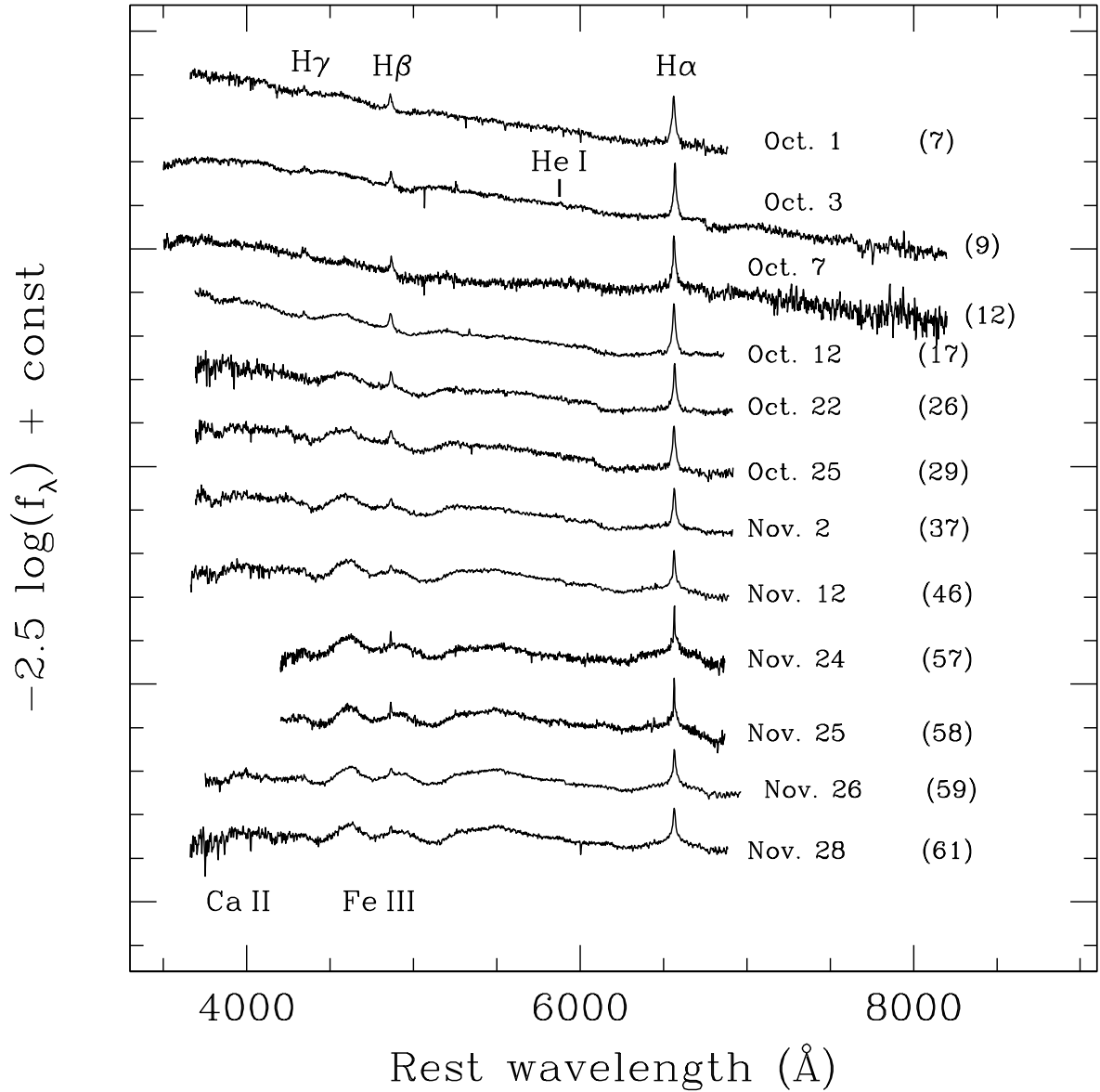}
\caption{Spectra of SN~2005gj obtained from Oct. 1 ($\sim$7~days after
explosion) to Nov. 28 ($\sim$61~days after explosion) of 2005. The
sequence show the dramatic spectral evolution of the SN from a very blue
continuum with strong Hydrogen-Balmer lines in emission in the early
phases, resembling the spectrum of a Type~IIn SN, to a Type~Ia
supernova-dominated continuum with broad absorption and emission
features (P-cygni profiles) of blended \ion{Fe}{2} and \ion{Fe}{3}
profiles. The spectra are shown in logarithmic flux scale and a constant
shift has been applied for clarity. The wavelength is in the rest-frame
corrected using $z=0.0616$ for the host galaxy. We show the UT date when
the spectra were obtained and the epoch (rest-frame days after
explosion) in parenthesis.}
\label{fig:spec1}
\end{figure*}

\begin{figure*}[t]
\epsscale{1.0}
\plotone{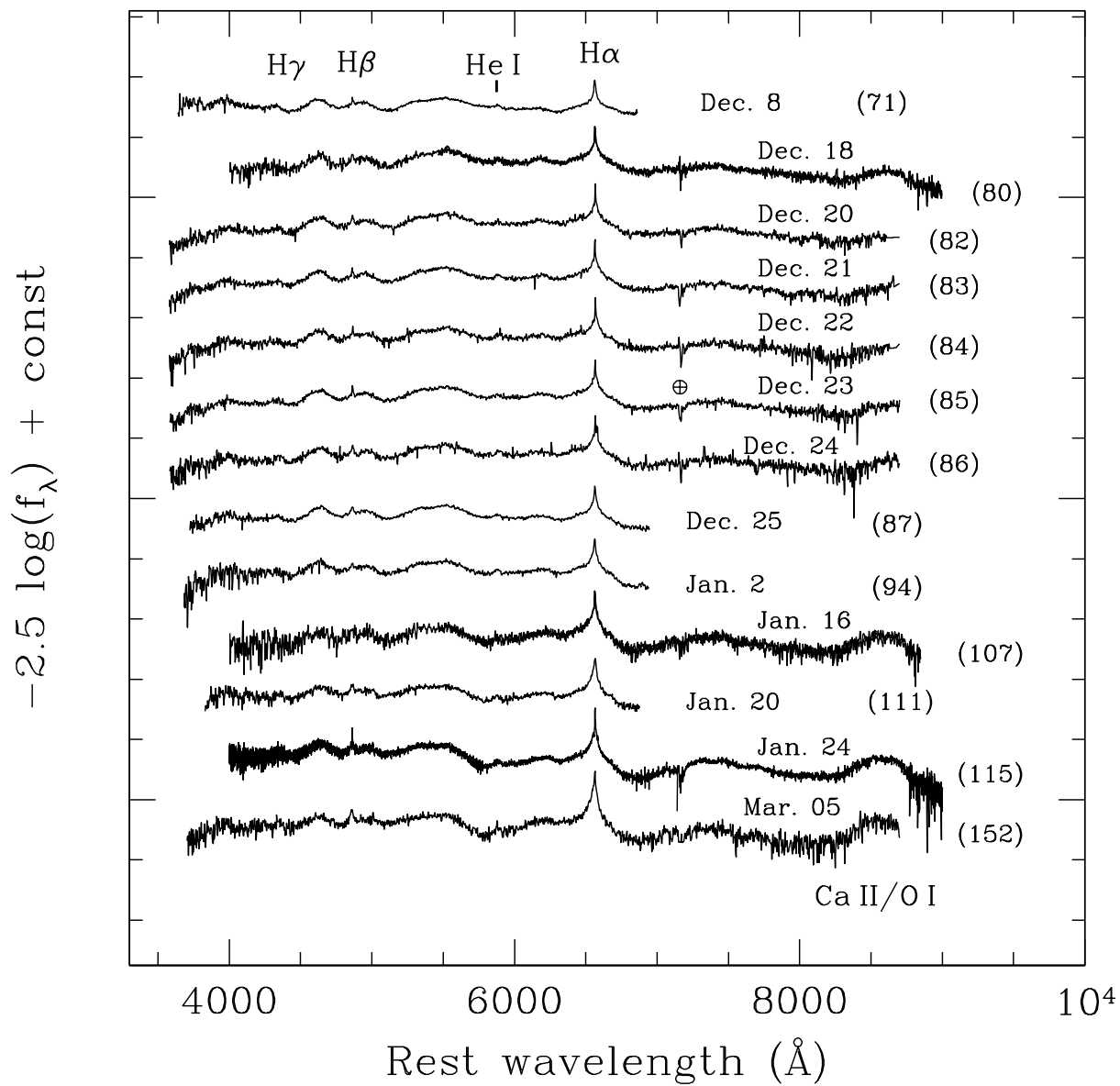}
\caption{Late time spectra of SN~2005gj obtained between Dec. 8, 2005
($\sim$71 days after explosion) and Mar. 6, 2006 ($\sim$152 days after
explosion). The labels, axis and symbols are the same as in
Figure~\ref{fig:spec1}. The {\it earth} symbol shows the position of
a telluric feature present in some of the spectra.}
\label{fig:spec2}
\end{figure*}

Figure~\ref{fig:spec1} and \ref{fig:spec2} show a montage of the optical
spectra of SN~2005gj obtained from October~2005 to March~2006. We have
split them in two figures to avoid crowding. The position of the most
conspicuous spectral features have been indicated in this figure.

\section{X-ray Observation}
\label{sec:xray}

SN~2005gj was observed under Director's Discretionary Time for 49.5 ks
on 2005 Dec 11/12 (ObsID 7241) with the \textit{Chandra X-ray
Observatory's} Advanced CCD Imaging Spectrometer (ACIS).  The data were
taken in timed-exposure mode with an integration time of 3.2 s per
frame, and the telescope aimpoint was on the back-side illuminated S3
chip.  The data were telemetered to the ground in ``very faint'' mode.

Data reduction was performed using the CIAO\,3.3 software provided by
the \chandra\ X-ray Center\footnote{\url{http://asc.harvard.edu}}.  The
data were reprocessed using the CALDB\,3.2.2 set of calibration files
(gain maps, quantum efficiency, quantum efficiency uniformity, effective
area) including a new bad pixel list made with the acis\_run\_hotpix
tool.  The reprocessing was done without including the pixel
randomization that is added during standard processing.  This omission
slightly improves the point spread function.  The data were filtered
using the standard {\it ASCA} grades (0, 2, 3, 4, and 6) and excluding
both bad pixels and software-flagged cosmic ray events.  Intervals of
strong background flaring were searched for, but none were found.

Absolute \chandra\ astrometry is typically good to $\sim$0\farcs5, and
we sought to improve it by registering the \chandra\ image with an SDSS
image.  \chandra\ point sources were found using the wavdetect tool, and
their positions were refined using ACIS Extract version 3.101.  Fourteen
X-ray sources had SDSS counterparts, which we used to shift the
\chandra\ frame by a small amount (0\farcs15 in RA and 0\farcs07 in
DEC).  After the shift, the residual differences between the \chandra\
and SDSS sources had rms values of 0\farcs19 in RA and 0\farcs12 in DEC.

We extracted counts in the 0.5--8 keV bandpass from the position of the
supernova using a standard extraction region ($\sim$1\arcsec\ radius),
and we constructed response files with the CAIO tools and ACIS Extract.
The background region is a source-free annulus centered on the position
of the supernova with inner and outer radii of 6\arcsec\ and 32\arcsec.
Based on the 300 photons detected in this region, we expect 0.3
background counts in our source extraction region.

We detect only two counts from the location of the supernova, but
neither may be associated with the supernova itself.  The counts had
energies of 4.0 keV and 6.5 keV, but one would expect some emission in
the 0.5--2.5 keV range since this is where \chandra\ has the most
collecting area.  For example, the average effective area in each of the
0.5--2.5 keV, 4.0--6.0 keV, and 6.0--8.0 keV bands is 470 \cmsq, 270
\cmsq, and 90 \cmsq, respectively.

We calculate our upper limits using the Bayesian method of
\citet{Kraft91}.  For the 0.5--4 keV band, the 68\% (95.5\%)
upper limit to the source counts is 1.14 (3.05).  For the 0.5--8 keV
band, the 68\% (95.5\%) upper limit is 3.52 (6.14).

Since no Type~Ia supernova has been conclusively detected in X-ray, we
have no {\it a priori} expectation of the spectral shape.  We therefore
adopt a simple absorbed power law with a photon index of 2 and an
absorbing column of $n_H = 7.08\times10^{20}$ \pcmsq.  For this choice
of spectrum, the count rate to flux conversion is $5.2\times10^{-12}$
erg/count in the 0.5--4 keV band and $6.9\times10^{-12}$ erg/count in
the 0.5--8 keV band.

We therefore arrive at 68\% (95.5\%) upper limits on the X-ray
luminosity of $1.1\, (2.9) \times 10^{39}$ \ergsec\ in the 0.5--4 keV
band and $4.3\, (7.6) \times 10^{39}$ \ergsec\ in the 0.5--8 keV band.
Based on the above statements concerning the \chandra\ effective area,
we feel the 0.5--4 keV limit is a more appropriate limit.

\section{Results}
\label{sec:results}

\subsection{Optical light curves and colors}
\label{sec:lc_res}

\begin{figure*}[t]
\epsscale{1.0}
\plotone{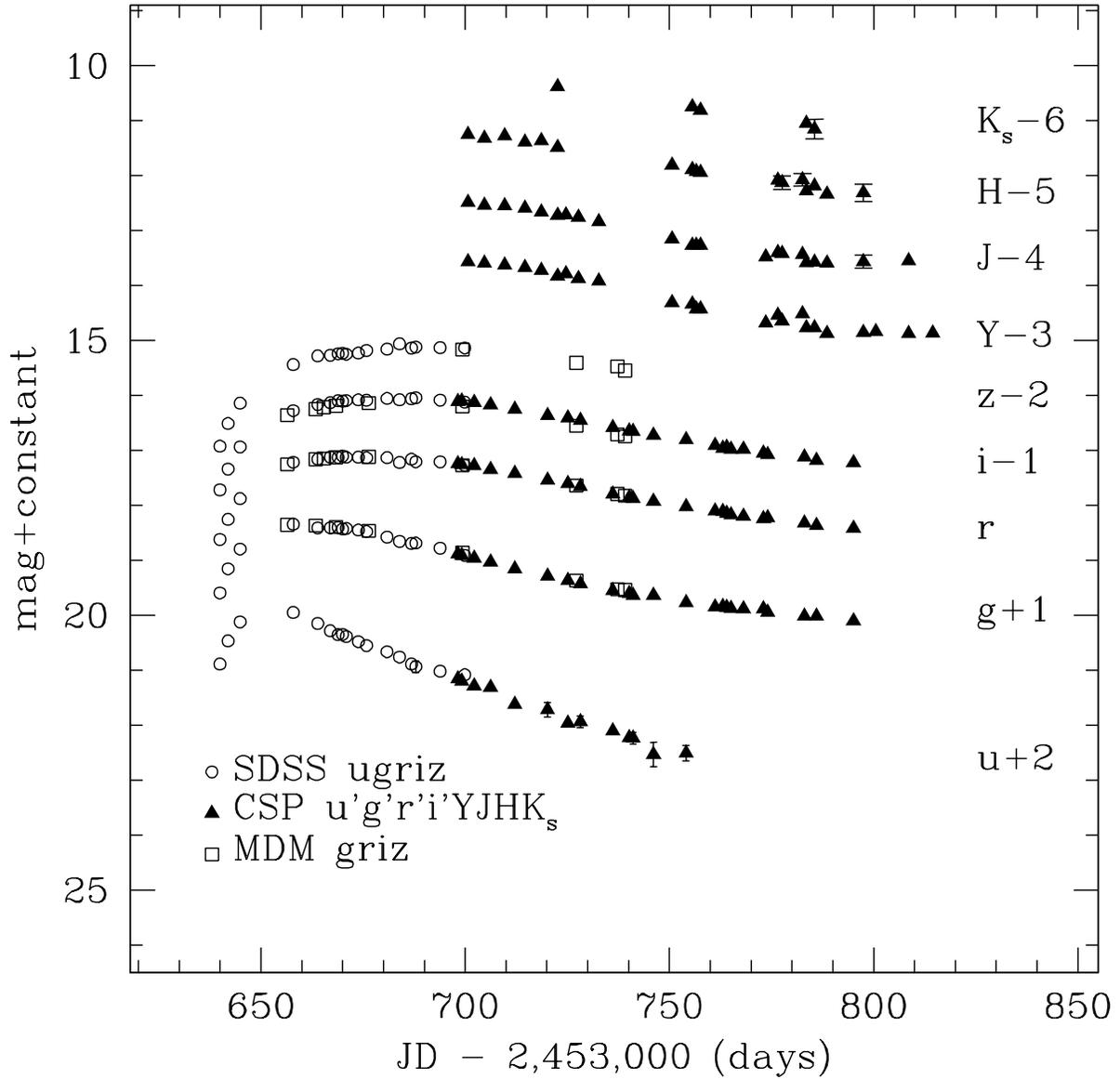}
\caption{Observed light curves of SN~2005gj from SDSS ({\it open
circles}), MDM ({\it open squares}) and CSP/Swope ({\it filled
triangles}). The error bars are smaller than the symbols. For clarity,
the light curves have been shifted by an arbitrary constant. }
\label{fig:lc}
\end{figure*}

Figure~\ref{fig:lc} shows the early SDSS and MDM $ugriz$ light curves
combined with the late time coverage given by CSP $u'g'r'i'YJHK_{s}$
photometry. They give excellent multi-wavelength optical and NIR
coverage and sampling of the first $\sim$150~days after discovery
($\sim$140 rest-frame days after the time of explosion).

In Table~\ref{tab:lpar} we give important parameters derived from the
light curves. We have a good estimate of the time of explosion of the SN
at JD~2,453,637.93$\pm$2.02 (September 24.4 UT, 2005) calculated as the
average time between the epoch when the supernova is detected at $>
5\sigma$ in all filters at JD~2,453,639.94, and the last pre-discovery
observation of the field at JD~2,453,635.91. The times and observed
magnitudes at maximum in different filters are presented in
Table~\ref{tab:lpar}. They are calculated from a high order polynomial
fit to each light curve. To estimate the errors we assume a gaussian
distribution for the magnitude uncertainty at each epoch and filter
(assuming they are not correlated from epoch to epoch), then we draw
randomly $\sim$1000 simulated light curves and fit each one with the
same high order polynomial. The $1\sigma$ uncertainties are taken as the
rms deviation of the mean values calculated from each simulated light
curve.

The risetimes, defined as the time between explosion and maximum light,
become longer at redder wavelengths. They are 13.5, 19.7, 33.7, and
46.9~days in $ugri$ filters, respectively, which correspond to 12.7,
18.5, 31.7, and 44.2~days in the rest-frame of the supernova.

We also give in Table~\ref{tab:lpar} the rest-frame magnitudes at
maximum in different filters. They have been corrected by Galactic
extinction in the line of sight, using $E(B-V)_{Gal}=0.121$
\citep[SFD:][]{sfd98} and the \citet{ccm89} (CCM) reddening law assuming
$R_{V}=3.1$, and $K$-corrections (see below), which are not negligible
in this object since it is at redshift $z\sim 0.062$. We assume that the
host galaxy extinction is negligible \citep{aldering06}.

\begin{figure*}[t]
\epsscale{1.0}
\plotone{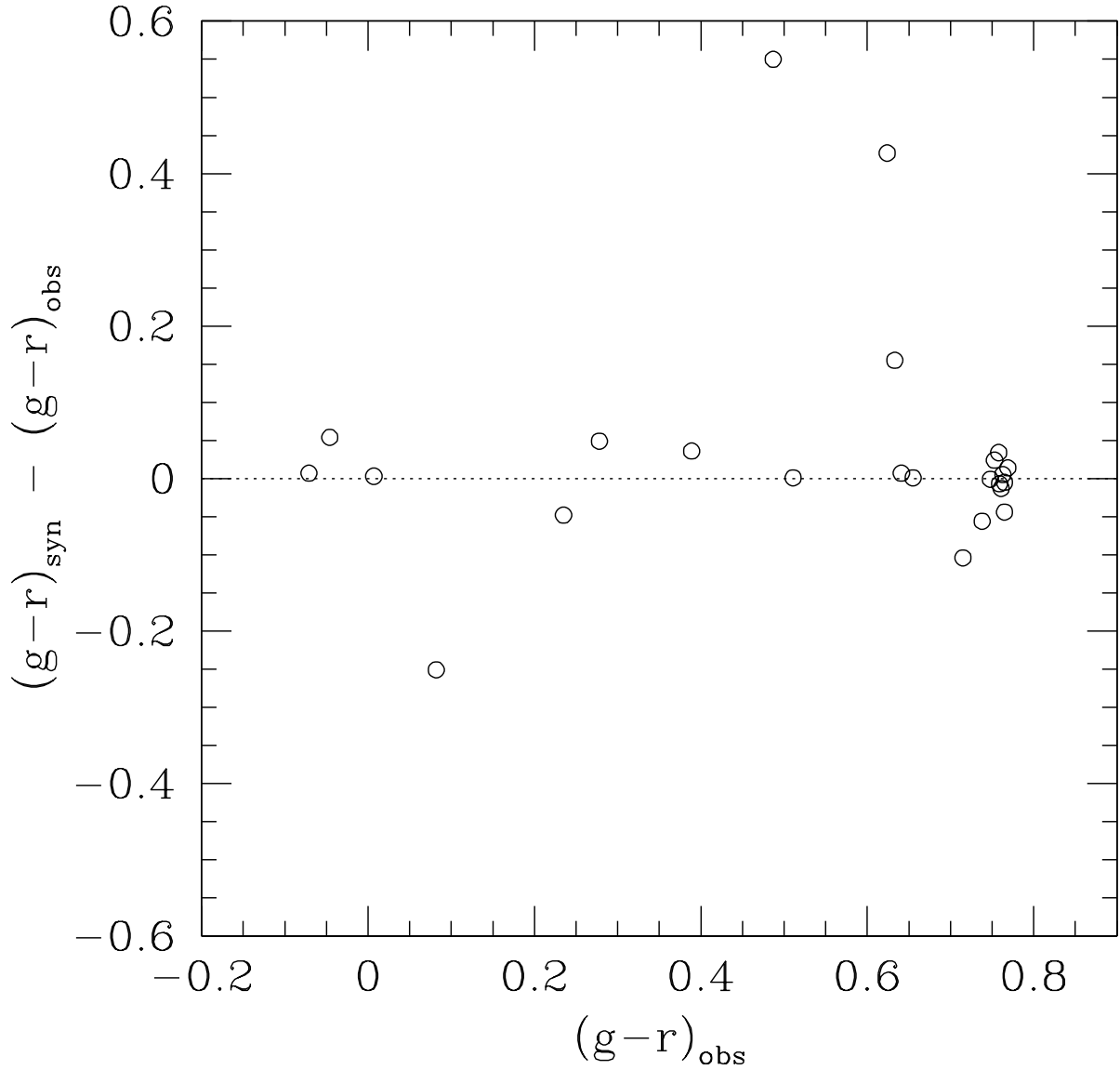}
\caption{Difference between the synthetic $g-r$ color calculated from
the spectra and the observed color from the photometry. We do not
include the latest spectra obtained on Jan 24. and Mar. 6 because there is 
no contemporaneous photometric data.}
\label{fig:specphot}
\end{figure*}

$K$-corrections have been calculated from the multi-epoch spectra
presented in \S~\ref{sec:spec}. In order to estimate accurate
$K$-corrections we need good spectrophotometric calibration in all the
wavelength range. Figure~\ref{fig:specphot} shows the differences
between the observed $g-r$ photometric colors obtained from the light
curves and the synthetic colors calculated directly from the spectra, as
a function of the observed photometric colors. We use an 8th order
polynomial fit of the light curves to obtain the observed $g-r$ color at
the epoch of a given spectrum to better than \mbox{$\sim 0.03$~mag}. We
can see that most of the spectra have good spectrophotometric
calibration in the wavelength range of $g\,,r$ filters (3800--7000~\AA),
with a residual color of $\sim$0.05 mag rms (difference between observed
and synthetic colors), but there are some obvious outliers. To correct
the spectra and produce a better spectrophotometric calibration
consistent with the observed spectral energy distribution (SED) obtained
from broadband photometry we {\it warp} the spectra multiplying by a
smooth function to match the observed colors, a technique commonly used
for calculating $K$-corrections in Type~Ia SN \citep{nugent02}.

First, we extrapolate the continuum in the blue and red sides of each
spectrum presented in \S~\ref{sec:spec} with a low order polynomial to
have complete coverage of the SDSS $ugriz$ filters
(2000--11000~\AA). Using this extended version of the spectra we apply
the CCM reddening law iteratively until the synthetic $g-r$ color
matches the observed color in each spectrum. This procedure does not
ensure that the calibration is good in the complete wavelength range;
therefore we multiply by a smooth spline with knots at the effective
wavelength of the SDSS filters until the synthetic $u-g$, $g-r$, $r-i$
and $i-z$ colors match the observed colors obtained, again by using
polynomial interpolation of the light curves.

The $K$-corrections for the same filter \citep{hamuy93} calculated from
the modified, spectrophotometrically calibrated spectra using SDSS
passbands are listed in Table~\ref{tab:kcorr}. The $K$-corrections are
probably accurate to $\pm$0.05~mag for $g$ and $r$ filters, and to
$\pm$0.1~mag for $u$, $i$ and $z$ where we had to extrapolate the
spectra. However, our estimate is not precise and we can not exclude the
possibility of even larger errors. After fitting with a low order
polynomial we use the results to transform the observed SDSS magnitudes
in Table~\ref{tab:phot1} to the rest-frame.

We corrected the CSP $u'g'r'i'$ magnitudes to rest-frame SDSS $ugri$
magnitudes using cross-filter $K$-corrections according to the
prescription of \citet{nugent02}. These take into account the difference
between the CSP and SDSS passbands convolved by the SED of SN~2005gj and
allow us to put all the rest-frame magnitudes in the same system. We
find that the differences between the same- and cross-filter
$K$-corrections are small ($\lesssim 0.03$~mag) at all epochs.

\begin{figure*}[t]
\epsscale{1.0} 
\plotone{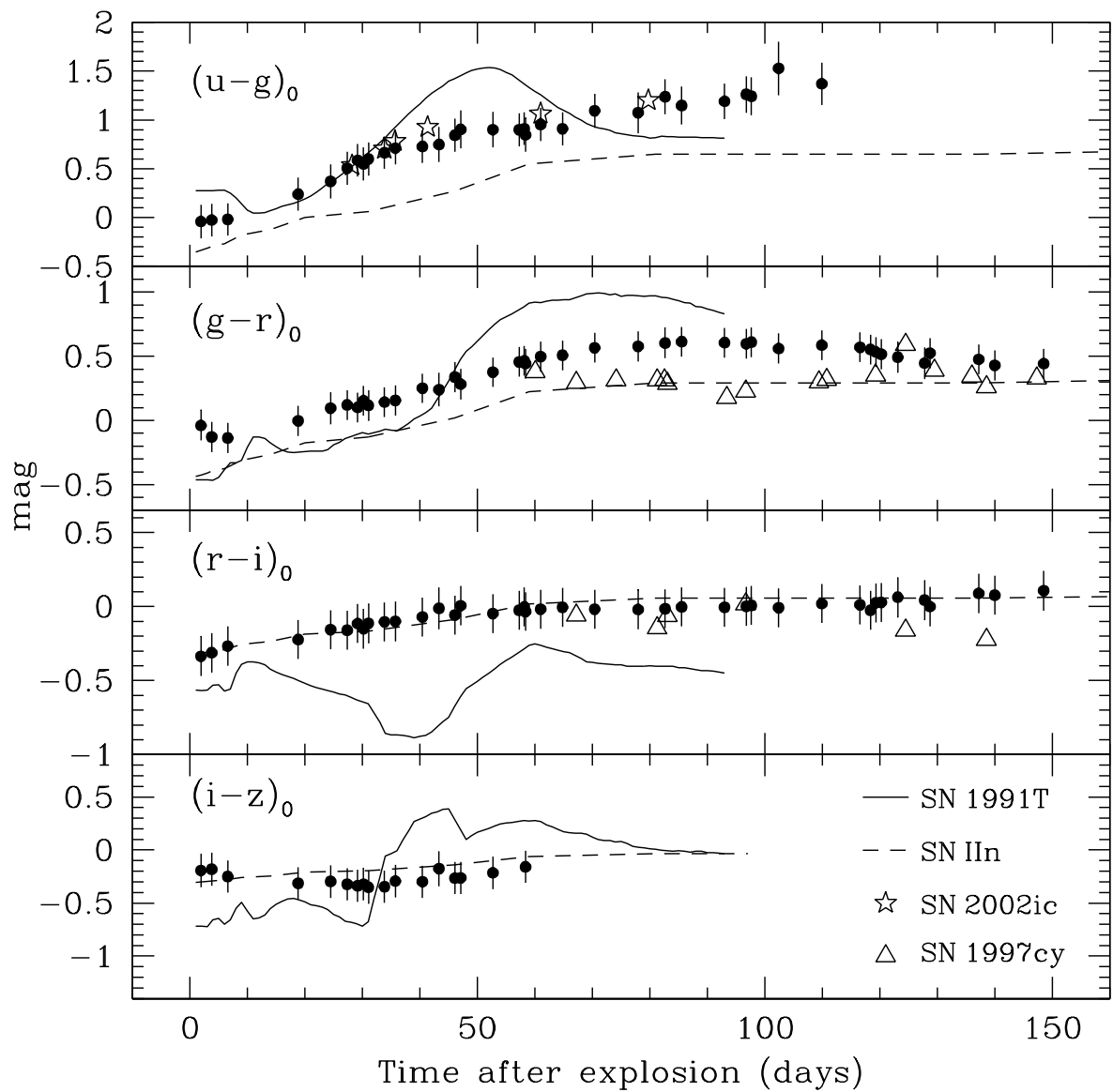}
\caption{Time evolution of the colors of SN~2005gj ({\it filled
circles}). For comparison we also show the color evolution of the
overluminous Type~Ia SN~1991T ({\it solid line}), the Type~IIn SN~1999el
({\it dashed line}), and two previous cases of Type~Ia strongly
interacting with its circumstellar medium, SN~2002ic ({\it stars}) and
SN~1997cy ({\it triangles}).}
\label{fig:colors}
\end{figure*}

Figure~\ref{fig:colors} shows the evolution of the rest-frame colors of
SN~2005gj as a function of time after explosion. We have corrected the
magnitudes by $K$-corrections and Galactic extinction in the line of
sight. As a comparison we also plot the color evolution of the
overluminous Type~Ia SN~1991T, a typical Type~IIn SN~1999el
(\citealt{dicarlo02}) (both obtained from spectral templates of P.~Nugent %
\footnote{\scriptsize{{\tt http://supernova.lbl.gov/$\sim$nugent/nugent\_templates.html}}} %
), and two previous cases that are thought to be Type~Ia explosions in
a very dense environment: SN~2002ic and SN~1997cy. For SN~2002ic we use
the published $BVI$ photometry \citep{hamuy03} corrected by
Galactic extinction ($E(B-V)=0.06$; SFD). We calculate cross-filter
$K$-corrections using the calibrated spectra of SN~2005gj to transform
observed magnitudes of the SN at $z=0.0667$ \citep{kotak04} to
rest-frame SDSS magnitudes: $B\rightarrow u$, $V\rightarrow g$,
$I\rightarrow i$. The time of explosion of SN~2002ic is assumed to at
$\approx$JD~2,452,581.5 (2002, Nov. 3 UT; \citealt{deng04}).  We obtain
rest-frame colors in the SDSS system for SN~1997cy
\citep{germany00,turatto00} by transforming the $K$-corrected $VRI$
magnitudes in \citet{germany00} to $gri$ magnitudes using
$S$-corrections calculated directly from spectra of SN~1997cy available
online in the SUSPECT database %
\footnote{\scriptsize{{\tt http://suspect.nhn.ou.edu/$\sim$suspect/}}}. %
We supplement this with synthetic colors from the spectra. We
also correct by Galactic extinction in the line of sight ($E(B-V)=0.02$;
SFD) and assume the time of explosion of SN~1997cy to be JD~2,450,582.5
(1997, May. 14 UT; \citealt{germany00}), which is very uncertain, since it is
taken as the time of detection of the gamma-ray burst GRB~970514, which may not have been 
associated with the SN. Magnitudes for SN~2002ic and SN~1997cy are not corrected 
by extinction in their host galaxies, which is unknown.

Initially the evolution in rest-frame $u-g$ and $g-r$ colors of
SN~2005gj, up to $\sim$30-40~days after explosion ($\sim$10-20~days
after maximum for a SN~Ia), is roughly consistent with the colors of
SN~1991T but $\sim 0.2$~mag redder in $g-r$, and evolves to redder
colors at later times. SN~1991T reaches its maximum colors of
($u-g$)$=$1.5~mag and ($g-r$)$=$1.0~mag at $\sim$50~days (30~days after
maximum), and after that it enters the nebular phase and becomes
bluer. At late times the $u-g$ color of SN~2005gj has a slow linear
increase and becomes systematically redder than SN~1991T at $> 70$~days,
while the $g-r$ color stays approximately flat at ($g-r$)$=$0.5~mag (and
bluer than SN~1991T) between $60-110$~days. The evolution in $u-g$ color
of SN~2002ic is very similar and consistent with SN~2005gj.  SN~1997cy
has a similar color evolution but it is $\sim$0.2 mag bluer in $g-r$
color between $60-100$~days. However, \citet{germany00} give
$K$-correction errors of $\sim$0.15~mag and we have further applied
$S$-corrections, therefore this does not imply a significant difference
between SN~1997cy and the other two Type~Ia/IIn supernovae.

The rest-frame $r-i$ and $i-z$ color evolution of SN~2005gj is very
different from SN~1991T and closely follows the evolution of a SN~IIn.
The $r-i$ colors of SN~1997cy are also consistent with SN~2005gj. From
these comparisons it is clear that the colors, a proxy for the
temperature of the photosphere, of SN~2005gj and two earlier cases of
SN~Ia strongly interacting with their circumstellar medium are dominated
by the radiation coming from the ejecta-CSM interaction.

\begin{figure*}[t]
\epsscale{1.0}
\plotone{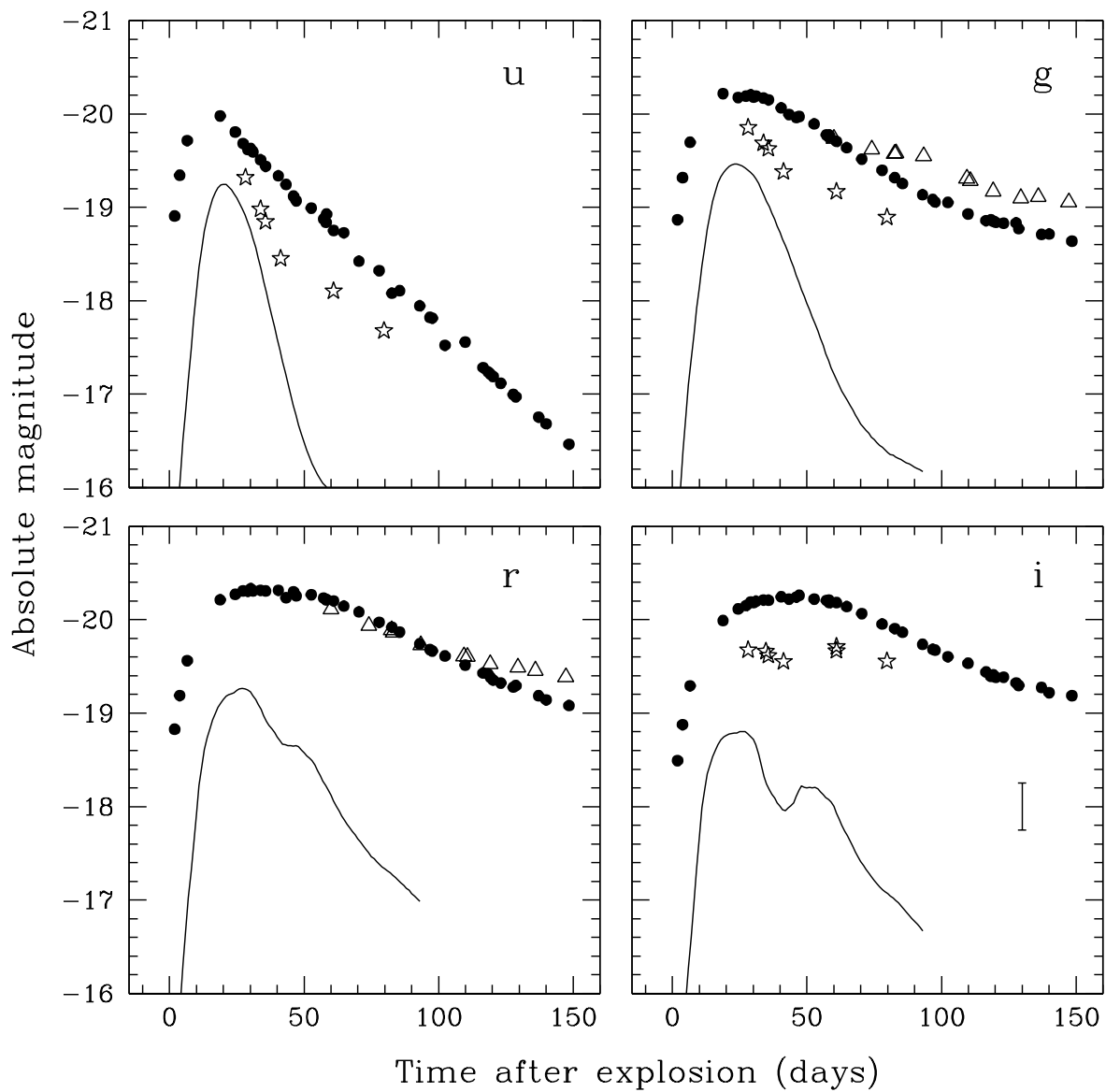}
\caption{Absolute $ugri$ light curves of SN~2005gj ({\it filled
circles}). For comparison we also show the absolute light curves of the
overluminous Type~Ia SN~1991T ({\it solid line}), SN~2002ic ({\it
stars}) and SN~1997cy ({\it triangles}). The error bar in the lower
right pannel represents the typical error in the absolute magnitudes
dominated by a $10\%$ uncertainty in the Hubble constant.}
\label{fig:abslc}
\end{figure*}

In Figure~\ref{fig:abslc} we present the $ugri$ light curves of
SN~2005gj in absolute magnitudes. We also show the light curves of
SN~1991T, SN~2002ic and SN~1997cy obtained from the literature and
corrected to SDSS rest-frame magnitudes as explained above.

SN~2005gj has peak $ugriz$ absolute magnitudes in the range $-20.0$ to
$-20.3$~mag, this is $\sim 0.7-1.4$~mag brighter than the overluminous
Type~Ia SN~1991T. The $u$ light curve is consistent with a linear
decline after peak luminosity at a constant rate of $0.027\pm 0.001\,\,
{\rm mag \,day^{-1}}$.  The $g$ light curve has a $\sim 20$~day plateau
with roughly constant luminosity after maximum ($20-40$~days after
explosion), then the light curve declines linearly between $40-100$~days
at $0.018\pm 0.001\,\, {\rm mag \,day^{-1}}$ and continues its linear
decay at later times, but with a shallower slope of $0.007\pm 0.002\,\,
{\rm mag \,day^{-1}}$. The $r$ and $i$ light curves have a similar
plateau shape between $20-60$~days and a constant linear decay at later
times of $0.013\pm 0.001\,\, {\rm mag \,day^{-1}}$. The change in slope
observed in the $g$-band at late times is less clear in $ri$, but still
present. The secondary maximum present in $ri$ light curves of SN~1991T
and other SN~Ia is completely absent in SN~2005gj.

The light curves of SN~2002ic are fainter than SN~2005gj at all times by
$0.3-0.6$~mag depending on the filter ($0.3-0.8$~mag brighter than
SN~1991T at peak). The initial decline rates from maximum of the $u$ and
$g$ light curves of SN~2002ic are intermediate between those of SN~1991T
and SN~2005gj until around 40~days after explosion; after day 50 when
the ejecta-CSM interaction had become dominant in SN~2002ic
\citep{hamuy03}, they closely resemble the decline rates of SN~2005gj in
the same bands. The $i$ band light curve of SN~2002ic showed definite
evidence for a weak secondary maximum, which is again intermediate in
morphology between the strong secondary maximum observed in SN~1991T,
and the absence of such a feature in SN~2005gj. The $gr$ light curves of
SN~1997cy are consistent with linear decay of $\sim 0.008\,\, {\rm mag
\,day^{-1}}$, being $\sim 0.4$~mag brighter than SN~2005gj at
150~days. Unfortunately, the light curves of SN~1997cy start at $\sim
60$~days after explosion and there is no information near peak to
compare with the luminosities of SN~2005gj and SN~2002ic. However, if we
extrapolate the $gr$ light curves to the time of peak luminosity we find
that the absolute magnitudes at maximum of SN~1997cy would be within
$\sim 0.2$~mag of SN~2005gj. This has to be taken with caution because
of the extrapolation and uncertain time of explosion of SN~1997cy.

\subsection{NIR light curves}
\label{sec:lcnir_res}

We used the CSP $JHK_{s}$ photometry obtained between 59-166~days after
the explosion (rest-frame) to construct the absolute magnitude light
curves of SN~2005gj.  The observed magnitudes were corrected by Galactic
reddening in the line of sight ($A_{J}=0.108$, $A_{H}=0.070$, and
$A_{Ks}=0.044$~mag) and $K$-corrections. We calculated $K$-corrections
for the same filter using the spectral templates of P.~Nugent for the
Type~IIn SN~1999el (which are derived from black-body curve fits to the
photometry), because as we showed in \S~\ref{sec:lc_res} the synthetic
colors obtained from the spectral templates approximate reasonably well
the evolution of the redder optical colors of SN~2005gj (i.e., $r-i$ and
$i-z$). The values of the $K$-corrections are consistent with being
constant in this time range: $K_{J}\approx-0.12$~mag
,$K_{H}\approx-0.14$~mag, $K_{K_{s}}\approx 0.16$~mag.

\begin{figure*}[t]
\epsscale{1.0}
\plotone{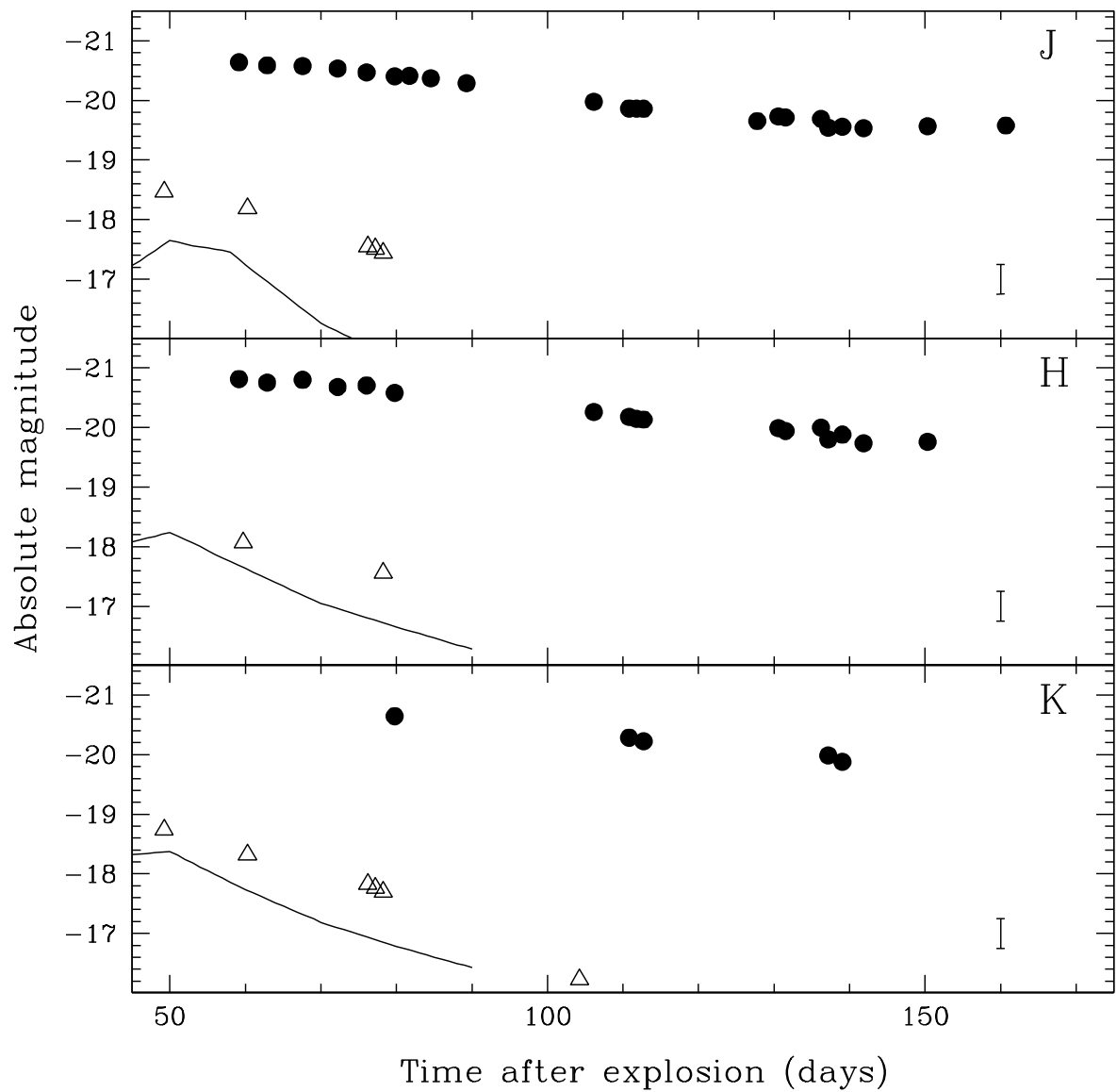}
\caption{Absolute light curves of SN~2005gj in the NIR: $J$ ({\it top
panel}), $H$ ({\it middle panel}) and $K_{s}$ ({\it bottom panel}). For
comparison we also show the absolute light curves of a normal Type~Ia
({\it solid line}) and the Type~IIn SN~1999el ({\it open
triangles}). The error bar in the lower right of each panel represents
the typical error in the absolute magnitudes dominated by a $10\%$
uncertainty in the Hubble constant.}
\label{fig:abslc_nir}
\end{figure*}

The $JHK_{s}$ absolute magnitude light curves of SN~2005gj are presented
in Figure~\ref{fig:abslc_nir}. For comparison we show the NIR light
curves of a normal Type~Ia obtained from synthetic photometry of
spectral templates and the Type~IIn SN~1999el \citep{dicarlo02}. The
Type~Ia light curves have been shifted in magnitudes to match the mean
absolute magnitude at maximum of SN~Ia \citep{krisciunas04}. SN~2005gj
is $1.7-3$~mag brighter than a normal SN~Ia and SN~IIn at 60~days after
the explosion and declines in a slower fashion at later times. A linear
decay is a good fit in all NIR bandpasses at this late times ($>
60$~days after explosion), with decline rates of $\sim 0.014\,\, {\rm
mag \,day^{-1}}$ in $J$, $\sim 0.013\,\, {\rm mag \,day^{-1}}$ in $H$,
and $\sim 0.011\,\, {\rm mag \,day^{-1}}$ in $K_{s}$. These values are
similar to the decline rates in the optical $ri$ bands.

Since there are no template images of the host galaxy obtained in the
NIR bands, the light curves are preliminary and the analysis has to be
taken with caution.

\subsection{Bolometric light curve}
\label{sec:bollc} 

The SDSS and CSP magnitudes were used to produce a quasi-bolometric
light curve of SN~2005gj covering the optical wavelength range from
$3000-10,000$~\AA ($u \to z$). We corrected the magnitudes by Galactic
extinction in the line of sight and $K$-corrections to obtain magnitudes
in the rest-frame $ugriz$ filters (see \S~\ref{sec:lc_res} for
details). We applied small corrections to transform the magnitudes based
in the SDSS photometric system into the $AB$ system obtained from the SDSS
website %
\footnote{\scriptsize{{\tt http://www.sdss.org/dr5/algorithms/fluxcal.html}}. }. %
The $AB$ magnitudes derived in this way are transformed directly to bandpass
averaged fluxes using the definition of the $AB$ system
\citep{oke_gunn83} and they are assigned to the frequencies that
correspond to the effective wavelengths of the SDSS $ugriz$ filters,
calculated from the filters using the definition in \citet{fukugita96}:
3567, 4735, 6195, 7510, and 8977~\AA. We use the trapezoidal rule to
obtain the integrated flux from $u$ to $z$, this is between
$\lambda_1=3340$~\AA\ and $\lambda_2=9596$~\AA, where the limits of the
wavelength coverage are obtained from $\lambda_1=\lambda_{eff,u} -
\Delta\lambda_{u}/2$ and $\lambda_2=\lambda_{eff,z} +
\Delta\lambda_{z}/2$. We extrapolate linearly the $u$ and $z$ light
curves at late times to fill in the lack of coverage of the SDSS in $z$,
and CSP in $u$ and $z$ bands, including the MDM $z$-band data at these
epochs. The integrated fluxes are converted to luminosity assuming a
luminosity distance to SN~2005gj of 268.5~Mpc and a spherically
symmetric distribution of the output energy. We present the integrated,
quasi-bolometric luminosities from $u$ to $i$ ($L_{(u\rightarrow i)}$)
and from $u$ to $z$ ($L_{(u\rightarrow z)}$) as a function of time in
Table~\ref{tab:lum}.

In order to estimate bolometric UVOIR luminosities we calculate time
dependent {\it bolometric corrections} to include the energy output of
the SN at wavelengths bluer than $u$-band ($\lambda < 3340$~\AA) and
redder than $z$-band ($\lambda > 9596$~\AA). We find that black-body
distributions with different temperatures are a reasonable approximation
of the spectral energy distribution of SN~2005gj. We use
$\chi^{2}$-minimization to fit the optical SED with a two parameter
black-body function: temperature and a multiplicative scaling
factor. The scaling factor is a combination of fundamental constants and
the square of the angular radius of the spherical black-body:
$(R_{bb}/d)^{2}$, where $R_{bb}$ is the radius of the black-body and $d$
is the distance to the SN. We calculate time-dependent bolometric
corrections by integrating the black-body distributions in the
ultraviolet and NIR/IR regions, and converting the integrated fluxes to
luminosities as explained above. For the CSP data we also include NIR
flux densities derived from the reddening and $K$-corrected $YJHK_{s}$
magnitudes after fitting the light curves with low order polynomials.
 
At early phases before peak the bolometric corrections account for $\sim
53-65\%$ of the total integrated luminosity, of which $85-93\%$ is from
the ultraviolet part of the spectrum and only $7-15\%$ from the
NIR/IR. As the supernova evolves, the ejecta expands and shocks the
circumstellar gas. The energy emitted in the ultraviolet/blue part of
the spectrum declines quickly after maximum light and most of the energy
is emitted in the optical region, coinciding with the appearance in the
spectrum of emission/absorption features of the intermediate and
iron-peak elements (see \S~\ref{sec:spec_res}). Between $25-150$~days
after explosion the bolometric corrections account for $\sim 32-45\%$ of
the total output luminosity, with the NIR/IR correction dominating
completely over the blue/ultraviolet at $> 60$~days.
 
The bolometric UVOIR luminosities, black-body temperatures and radii
derived from the fits are presented in Table~\ref{tab:lum}. The
uncertainties in black-body temperatures and radii are calculated using
the diagonal terms of the covariance matrix obtained from the $\chi^{2}$
minimization. We add a $10\%$ error in the distance to the SN to the
error in the black-body radii, which comes from the random and
systematic uncertainties in the value of the Hubble constant
\citep{freedman01,riess05}. The uncertainties in the bolometric
luminosities were estimated by propagating errors through the
trapezoidal integration of the SED, taking into account: uncertainties
in the photometry, light curve interpolation and fitting, Galactic
extinction, $K$-corrections, and distance to the SN. To approximately
take into account the errors introduced by the bolometric corrections we
multiplied these values by $\sqrt{\chi^{2}_{\nu}}$ when the reduced
$\chi^{2}$ is greater than 1.

\begin{figure*}[t]
\epsscale{1.1}
\plotone{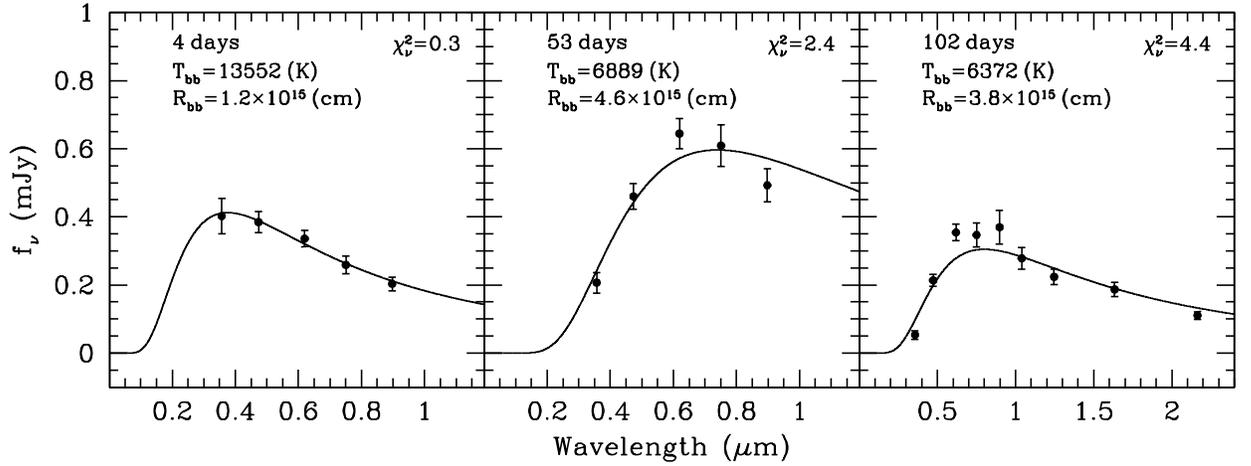}
\caption{Examples of black-body fits ({\it solid line}) to the SED of
SN~2005gj obtained by transforming the rest-frame $ugriz$ magnitudes to
monochromatic fluxes at the effective wavelength of the filters ({\it
filled circles}). These examples show the quality (i.e., goodness-of-fit)
range of the black body fits at different epochs: $\chi^{2}_{\nu}=$0.3
({\it left panel}), 2.4 ({\it middle}), 4.4 ({\it right}). The units of
flux density in the y-axis are ${\rm mJy} = 10^{-26}\,\, {\rm
erg\,s^{-1}\,cm^{-2}\,Hz^{-1}}$.}
\label{fig:bbfits}
\end{figure*}

In Figure~\ref{fig:bbfits} we show some examples of black-body fits to
the optical SED at different epochs. At early times, shortly after
explosion, the SED is very well fit by a hot $\sim$13000~K
black-body. The temperature starts to decrease steadily close to the
time of explosion to a constant value of $\sim 6500$~K at 60~days after.
A black-body is still a reasonable approximation of the SED at later
times, but the fits become poorer when emission/absorption features
start to dominate the spectrum, which is represented in the
$\chi^{2}_{\nu}$ of the fits (see Table~\ref{tab:lum}).

\begin{figure*}[t]
\epsscale{1.0}
\plotone{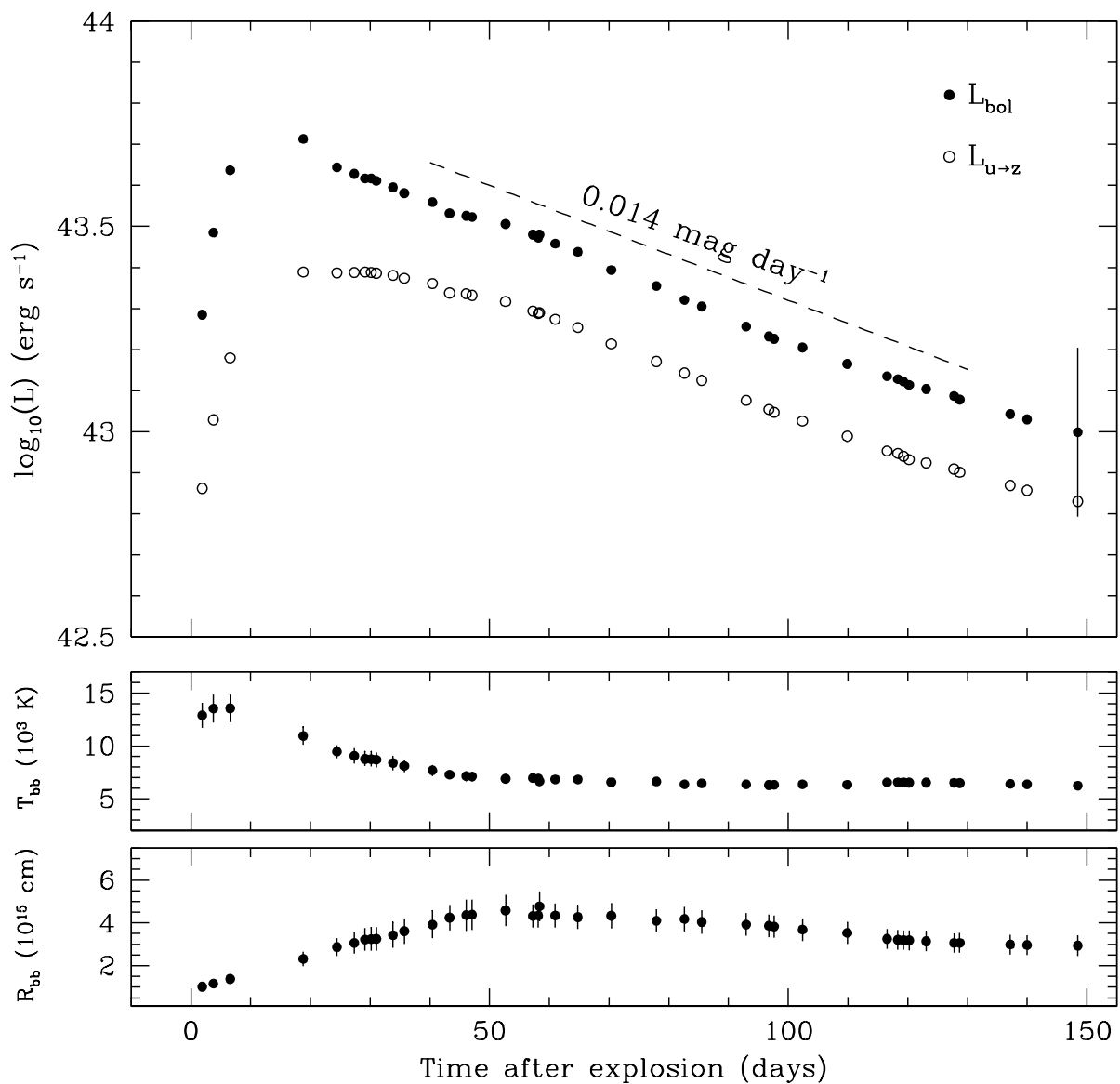}
\caption{{\it Top panel:} Quasi-bolometric ({\it open circles}) and
bolometric light curves of SN~2005gj ({\it filled circles}). The
bolometric light luminosities were obtained after applying bolometric
corrections calculated from black-body fits to the optical SED obtained
from the $ugriz$ photometry. The {\it dashed} line shows the best-fit
linear decay of $0.014\,\,{\rm mag\,day^{-1}}$. The {\it middle} and
{\it bottom panels} show the evolution of the black-body temperature and
radius, respectively.}
\label{fig:bollc}
\end{figure*}

Figure~\ref{fig:bollc} shows the bolometric light curve of SN~2005gj in
the top panel and the evolution in temperature and radius from the
black-body fits in the lower panels. The early data of the bolometric
light curve are well fit by an exponential rise in luminosity,
$L(t)\propto e^{0.17\,t}$.  The time of maximum bolometric luminosity
occurs between 6.6--18.8~days after explosion. After maximum, the
bolometric light curve is very well approximated by an exponential decay
in luminosity, linear in the logarithmic scale shown in
Figure~\ref{fig:bollc}, $L(t)\propto e^{-0.013\,t}$ ($0.014\,\, {\rm mag
\,day^{-1}}$). This is consistent with the exponential density
distribution of the ejecta of Type~Ia SN \citep{dwarkadas98}, whereas
the distribution of ejecta around core-collapse supernovae is better
approximated by a power-law \citep{chevalier03}. Extrapolating the pre-
and post-maximum fits we get a maximum bolometric luminosity of
$L_{bol}^{max}=5.6\times 10^{43}\,\,{\rm erg s^{-1}}$, which is $\sim$3
times more luminous than the Type~Ia SN~1991T at maximum light
\citep{contardo00,stritzinger06}. Assuming a bolometric correction of
50\% at maximum, we find that SN~2005gj was $\sim$1.5 times more
luminous than SN~2002ic.

\subsection{Optical spectroscopy}
\label{sec:spec_res}

\begin{figure*}[t]
\epsscale{1.0}
\plotone{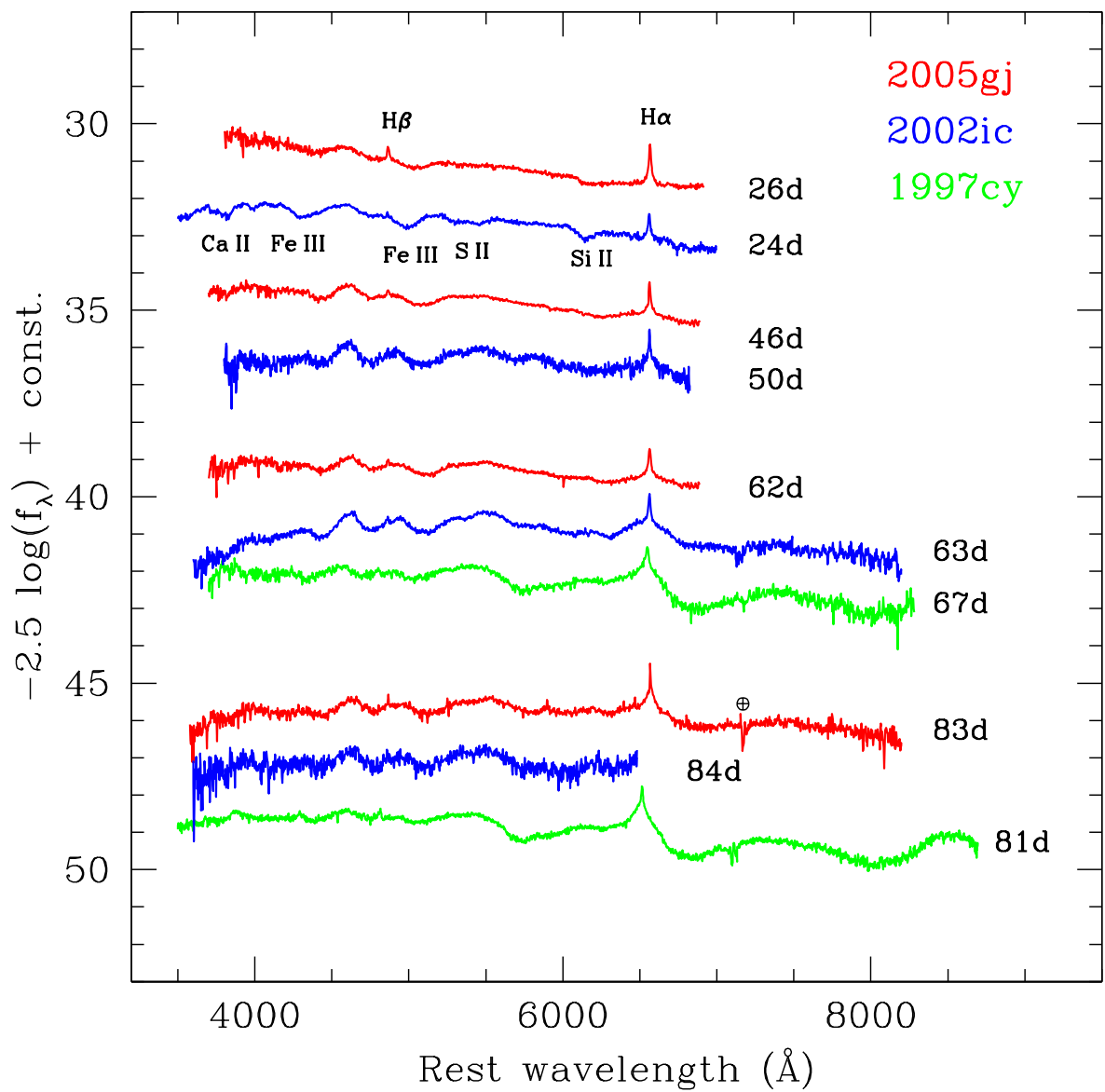}
\caption{Comparison of spectra of SN~2005gj at 26, 46, 62 and 83~days
after explosion with comparable epoch spectra of SN~2002ic (Hamuy et
al. 2003) and SN~1997cy (from SUSPECT database). The spectra are plotted
on a logarithmic flux scale and shifted by an arbitrary constant. The
wavelength was shifted to the restframe using $z=0.0616$ of the host
galaxy (Aldering et al. 2006).}
\label{fig:spec_comp}
\end{figure*}

In Figure~\ref{fig:spec_comp} we show a comparison of the spectra of
SN~2005gj with spectra of SN~2002ic and SN~1997cy obtained at similar
times after explosion. The spectra of SN~2005gj and SN~2002ic are very
similar at all times. They are characterized by strong and broad
Hydrogen-Balmer lines \Halpha\ and \Hbeta\ in emission\footnote{As shown
in Figure~\ref{fig:spec1}, \Hgamma\ is also visible in the early spectra
of SN~2005gj.} and a blue continuum at early times that becomes redder
and increasingly dominated by absorption/emission P-cygni profiles from
Fe-peak ions (e.g., \ion{Fe}{2}, \ion{Fe}{3}, \ion{Ni}{3}, \ion{Si}{2},
\ion{S}{2}).

\subsubsection{Classification}
\label{subsec:class}

\citet{benetti06} proposed that SN~2002ic-like events are well explained
by the core-collapse of a massive star in a dense medium, casting doubt
in the previous classification of SN~2002ic as a Type~Ia supernova. The
authors found relatively good agreement at all times between the spectra
of SN~2004aw \citep{tautenberger06}, a Type~Ic supernova, and SN~2002ic.

We used the SuperNova IDentification code, SNID (\citealt{matheson05};
\citealt{miknaitis07}; \citealt{blondin07}), to find the spectra that
best match SN~2005gj at different epochs. SNID cross-correlates an input
spectrum with a library of supernovae spectra. In the library we
included spectra of 5 normal SN~Ia, two 1991T-like objects, two
1991bg-like objects, 4 broad-lined SN~Ic (or hypernovae), and 3 normal
SN~Ic (including SN~2004aw), that were chosen to span a wide range of
observed properties of SN~Ia and SN~Ic. In Table~\ref{table:speclib} we
present the supernovae and the epochs of the spectra in the library. We
fixed the redshift of SN~2005gj at $z=0.0616$ and allowed for a range
around the mean redshift of $\Delta z=0.02$ to find cross-correlation
peaks. The Balmer lines in emission were clipped from the input
SN~2005gj spectra to avoid spurious cross-correlation signal with library
spectra that contain emission lines from the host galaxy.

\begin{figure*}[t]
\epsscale{1.0}
\plotone{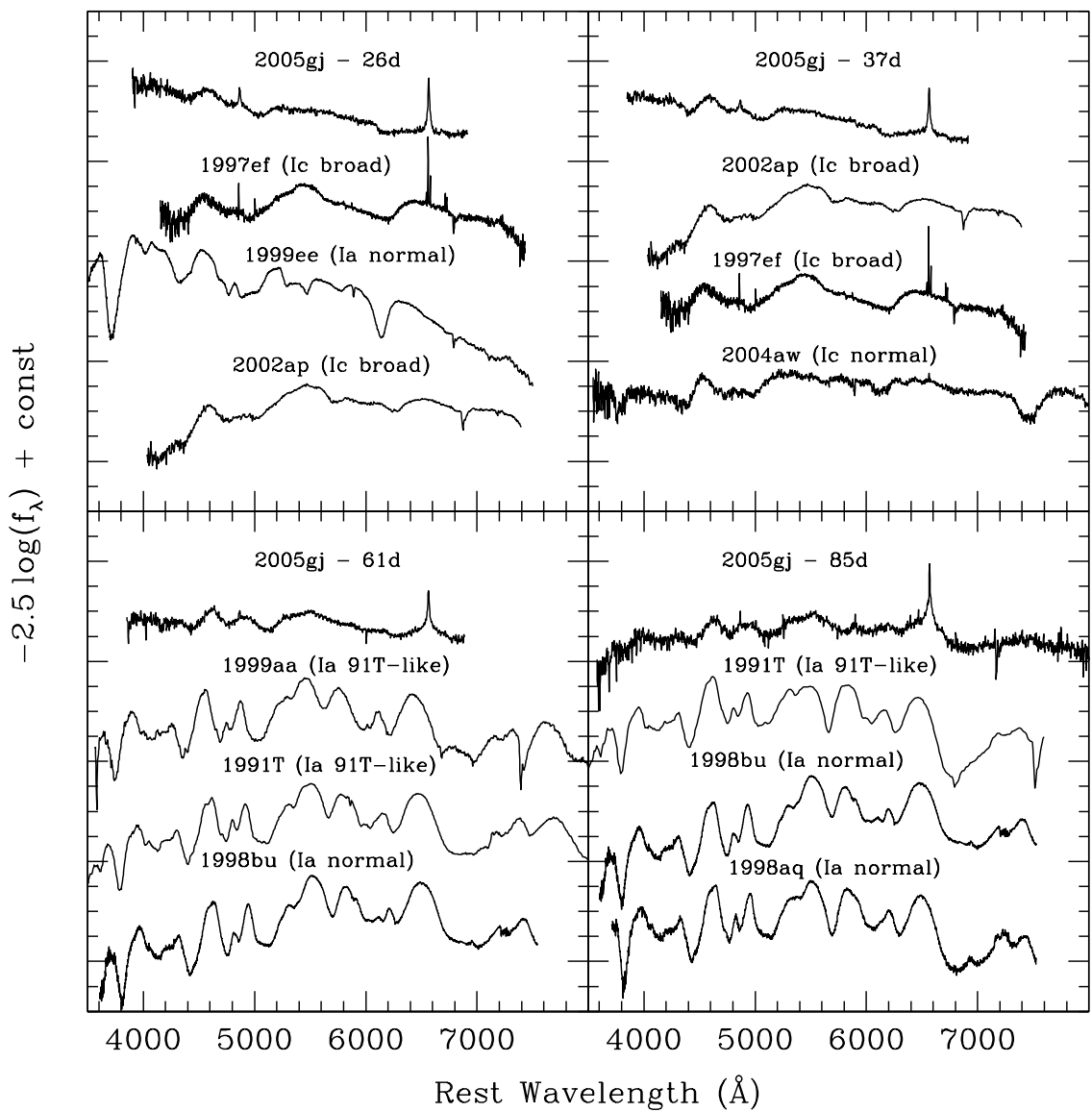}
\caption{Results of SNID. We show the spectra of SN~2005gj at four
epochs (26, 37, 61, and 85~days after explosion) and their best three
cross-correlation library spectra. The spectra are plotted on a
logarithmic flux scale and shifted by an arbitrary constant.}
\label{fig:snid_fits}
\end{figure*}

Figure~\ref{fig:snid_fits} shows the spectra of SN~2005gj at four epochs and
the library spectra with the highest cross-correlation significance from
SNID. Type~Ia supernovae spectra are a better match to SN~2005gj at
most epochs, with 20 of the 26 (77\%) epochs having best matched a
type~Ia spectrum (45\%-91T, 45\%-normal, 10\%-91bg) and a similar
fraction for the next-best matches. The broad-lined SN~Ic 1997ef
and 2002ap are the best matching spectra for 6 epochs, all of them
between 26--46~days after explosion. We repeated the same procedure
using five spectra of SN~2002ic obtained between 24--84~days after
explosion \citep{hamuy03}. All the spectra of SN~2002ic are better
matched with SN~Ia, in contradiction with the results obtained by
\citet{benetti06}. 

\begin{figure*}[t]
\epsscale{1.0}
\plotone{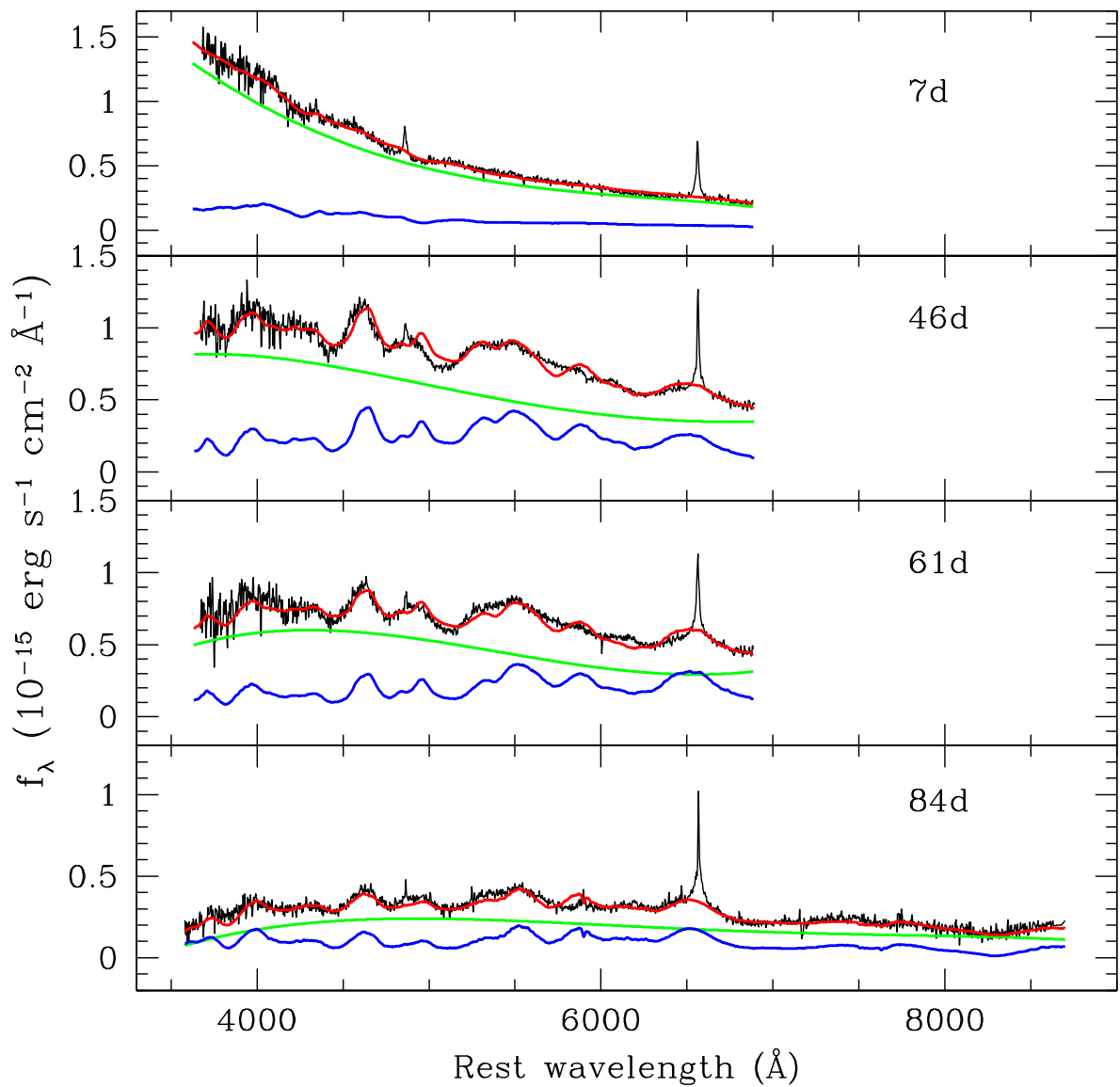}
\caption{Fits to the spectra of SN~2005gj. We model the spectra as the
sum of two components: (1) SN~1991T spectrum at the same epoch after
explosion as SN~2005gj scaled by an arbitrary constant ({\it blue
line}); (2) fourth order polynomial ({\it green line}). The results of
the fits are in {\it red} and the spectra of SN~2005gj, corrected by
Galactic extinction in the line of sight are in {\it black}. The epochs
of the spectra are shown in the upper right of each panel.}
\label{fig:cont_fits}
\end{figure*}

The continuum of SN~2005gj is well approximated at all times by the sum
of a scaled spectrum of the overluminous Type~Ia SN~1991T, at the same
epoch after explosion as SN~2005gj, and a fourth order polynomial. A
normal SN~Ia does not fit as well as SN~1991T. This procedure is very
similar to the fits to SN~2002ic \citep{hamuy03} and SN~2005gj
\citep{aldering06} presented in previous studies. In
Figure~\ref{fig:cont_fits} we show examples of the spectra decomposition
at four epochs. We excluded from the fit a region of $\pm$100~\AA\ around
the \Halpha\ and \Hbeta\ lines and obtained a good fit for the remainder
of the spectrum.

\subsubsection{Balmer lines}
\label{subsec:balmer}

We analyzed the Balmer emission features in the spectra using the sum of
two Gaussian components to model the line profiles. This decomposition
gives much better fits for \Halpha\ at all epochs than a single Gaussian
and it is physically motivated \citep{chugai97b,chugai97a}. The spectra
of Type~IIn SN show Balmer features with both a narrow and broad
component that can be explained by radiation coming from different
regions of the ejecta/CSM, whether it is direct emission from the
shock-heated CSM (broad component) or emission from un-shocked gas
photoionized by the SN radiation (narrow component). The \Hbeta\ line is
unresolved or only marginally resolved for most of the
spectra. Therefore a single Gaussian component was used to fit the line
profile. We used a third order polynomial to model the local continuum
around each line that was included in the Gaussian fits. It is important
to stress that at late times there is a broad \ion{Fe}{2} feature
intrinsic to the supernova spectrum in the region of \Halpha\ (see
spectra in Figure~\ref{fig:cont_fits}) that makes the definition of the
continuum less reliable and may affect the line measurements. 

The results of the Gaussian fits to the \Halpha\ and \Hbeta\ emission
features, integrated fluxes and FWHM, are shown in
Table~\ref{tab:balmer} as a function of epoch of the spectra. We have
excluded the two spectra with better resolution because they show
P-Cygni profiles (see below). We used the flux calibrated spectra
corrected to match the observed $g$, $r$ magnitudes (as explained in
\S~\ref{sec:lc_res}) and corrected for Galactic reddening in the line of
sight. The FWHM of the Gaussian profiles listed in
Table~\ref{tab:balmer} were corrected by the resolution of the
spectrographs (from Table~\ref{tab:spec}). We do not present the values
when the component is unresolved. The errors quoted for the integrated
fluxes are obtained by adding in quadrature an estimate of 10\% error
assigned to the absolute flux calibration and the rms deviation of the
Gaussian fits.
 
\begin{figure*}[t]
\epsscale{1.0} 
\plotone{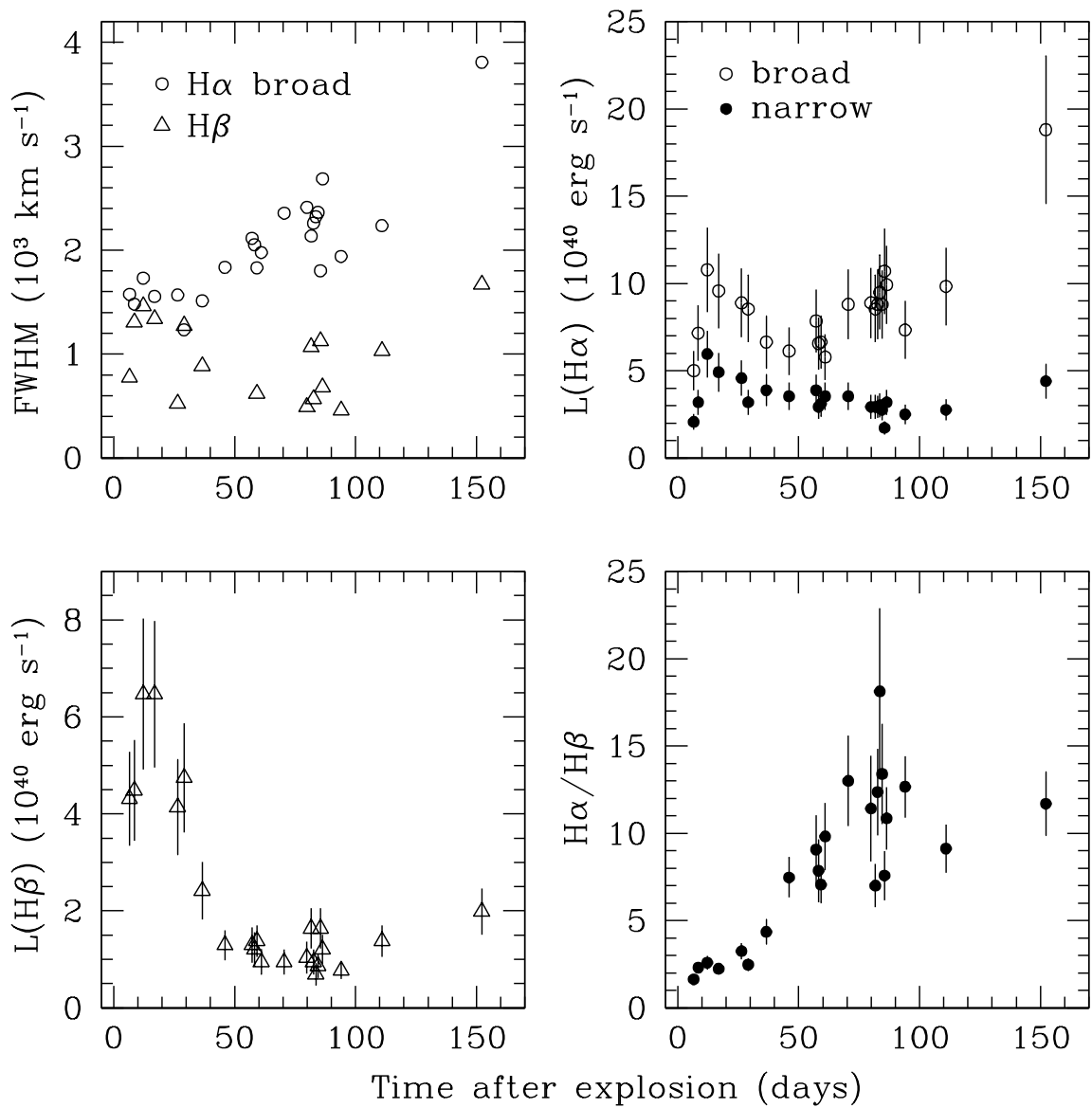}
\caption{Results from the Gaussian fits to \Halpha\ and \Hbeta\ emission
features as a function of time. {\it Top left panel:} FWHM of the
\Halpha-broad and \Hbeta. {\it Top right panel:} Luminosity of the
narrow and broad Gaussian components of \Halpha. {\it Bottom left
panel:} Luminosity of \Hbeta. {\it Bottom right panel:} Balmer
decrement, ratio of total fluxes in \Halpha\ and \Hbeta\ lines.}
\label{fig:lines_fluxes}
\end{figure*}

Figure~\ref{fig:lines_fluxes} shows the evolution in time of the FWHM
(top left panel), \Halpha\ and \Hbeta\ luminosities (top right and
bottom left panels) and the Balmer decrement (bottom right panel). The
FWHM of \Halpha\ varies between $\sim 130-500\,\, {\rm km\ s^{-1}}$
(narrow component) and $\sim 1200-3800\,\, {\rm km\ s^{-1}}$ (broad
component), with the broad component showing a slow increase in time.
The FWHM of \Hbeta\ varies between $\sim 470-1700\,\, {\rm km\ s^{-1}}$
and does not show evident evolution. The luminosities of \Halpha-narrow
and \Hbeta\ lines evolve in a similar fashion, increasing at early times
to peak at $\sim 12$~days with luminosities $6-6.5\times 10^{40}\,\,{\rm
erg s^{-1}}$, then they decay and stay roughly constant after
50~days. The evolution of \Halpha-broad is similar during the first
50~days, peaking at $1.1\times 10^{41}\,\,{\rm erg s^{-1}}$, but it
shows an increase at later times. Compared with the \Halpha\
luminosities observed in SN~2002ic, both components are $\sim 4$ times
more luminous. 

The Balmer decrement, the ratio between \Halpha\ (sum of narrow and
broad components) and \Hbeta\ fluxes, stays approximately constant
during the first $30-40$~days (${\rm mean}=2.5$ and ${\rm rms}=0.5$) and
is consistent with the theoretical value in Case B recombination of
H$\alpha$/H$\beta = 2.86$ \citep{osterbrock89}. At later times it shows
an steady increase, reaching ${\rm H} \alpha/{\rm H}\beta \sim 7-13$ at
$\sim 80$~days. In Case B recombination a Balmer decrement ${\rm H}
\alpha/{\rm H}\beta > 2.86$ is usually interpreted as evidence for the
presence of internal extinction in the host; however, the large values
observed at late times would produce an \ion{Na}{1}~D interstellar
absorption doublet easily detectable in the spectra, which is not
observed (see also \citet{aldering06}), and the evolution in time is not
expected. \citet{aldering06} proposed that the H level populations are
in Case C recombination, where the optical depth in the \Halpha\ line is
high implying high densities and greater importance of collisional
processes. In this scenario, the observed change in the Balmer decrement
could indicate that collisional excitation becomes increasingly
important at later times \citep{branch81,turatto93}. SN~2002ic
\citep{deng04} and other SN~IIn, like SN~1988Z \citep{aretxaga99} and
SN~1995G \citep{pastorello02}, have also shown large values of the
Balmer decrement and therefore may have similar physical processes
affecting the formation of the Balmer lines.
 
\begin{figure*}[t]
\epsscale{1.0} 
\plotone{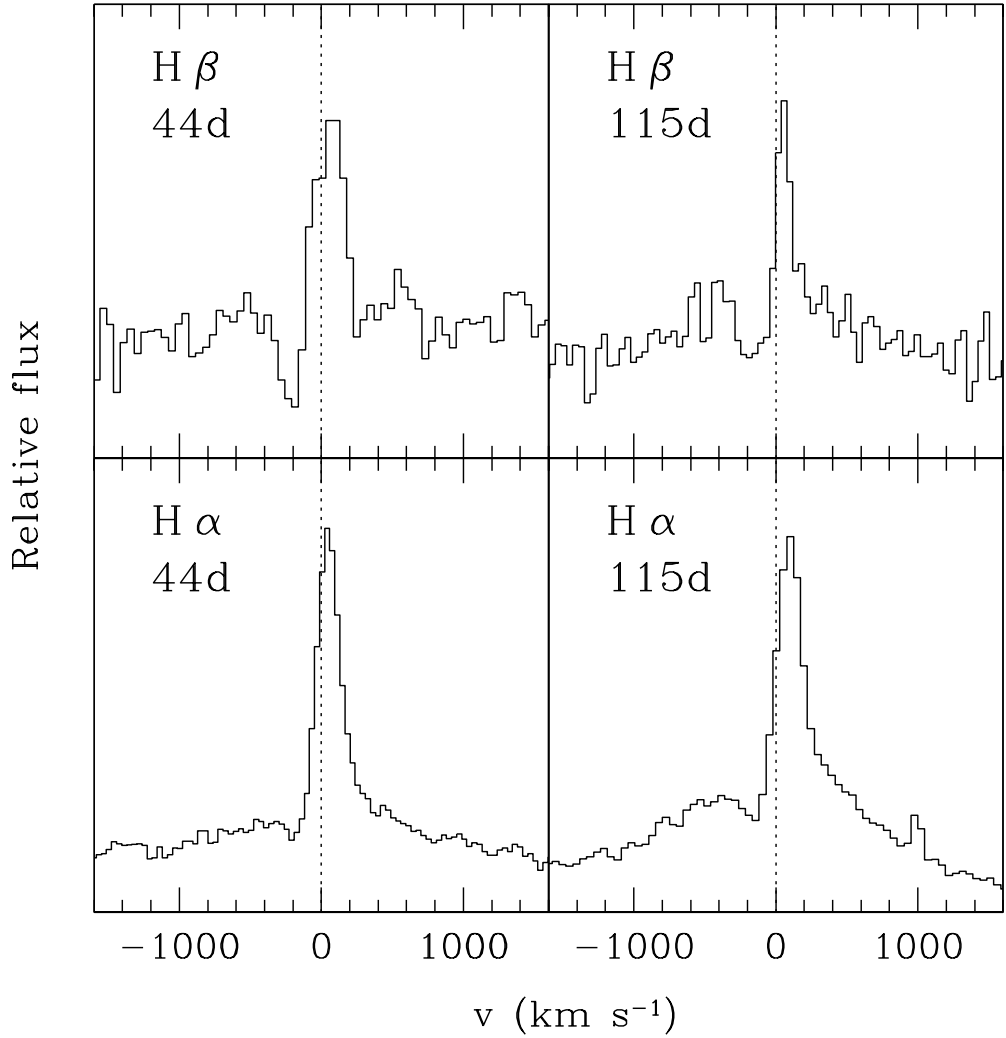}
\caption{Line profiles of \Hbeta\ ({\it top}) and \Halpha\ ({\it
bottom}) in the two highest resolution spectra of SN~2005gj obtained
with WHT+ISIS on day 44 ({\it left}) and Magellan+LDSS-3 on day 115
({\it right}). The features show clear P-cygni profiles with weak
absorption minima at $\sim -200\,\,{\rm km s^{-1}}$, demonstrating the
presence of a slowly moving outflow.}
\label{fig:pcygni}
\end{figure*}

In Figure~\ref{fig:pcygni} we show the regions around \Halpha\ and
\Hbeta\ features in the best resolution spectra from ISIS and LDSS-3,
obtained at 44 and 115~days after explosion, respectively. We clearly
detect P-Cygni profiles in all these features, which indicates the
presence of an outflow moving at $\sim$150-200~${\rm km\, s^{-1}}$;
however, these measurements are limited by the resolution of the spectra
between $\sim$130-180~${\rm km\,s^{-1}}$ (FWHM). After correcting for
the resolution we obtain an outflow velocity of 60-70~${\rm km\,
s^{-1}}$. The detection of P-Cygni-like absorption rules out an
\ion{H}{2} region in the line of sight that could be producing the
narrow emission/absorption features; the line profiles are intrinsic to
the SN. \citet{aldering06} detected P-Cygni profiles in \ion{He}{1},
\Halpha\ and \Hbeta, in a high resolution spectrum obtained with
LRIS+Keck 71~days after the explosion. They derived a wind velocity of
$v_{w}\approx 60$~${\rm km\,s^{-1}}$ consistent with our
estimate. \citet{kotak04} also detected a P-Cygni profile in the a
spectrum of SN~2002ic obtained 256~days after explosion.

\subsubsection{Line identification near maximum}

We used the parameterized resonance-scattering code SYNOW
\citep{fisher97,fisher00} to identify the lines in the spectra obtained
near maximum light of SN~2005gj. SYNOW is a fast supernova
spectrum-synthesis code used for direct (empirical) analysis of
supernova spectra, mainly to identify the lines, their formation
velocities and optical depths. The code is based on simple assumptions:
spherical symmetry, homologous expansion, a sharp photosphere that emits
a black-body continuous spectrum, and line formation by
resonance-scattering, treated in the Sobolev approximation. We have used
the latest version of the code that includes a Gaussian distribution of
optical depths.

\begin{figure*}[t]
\epsscale{1.0} 
\plotone{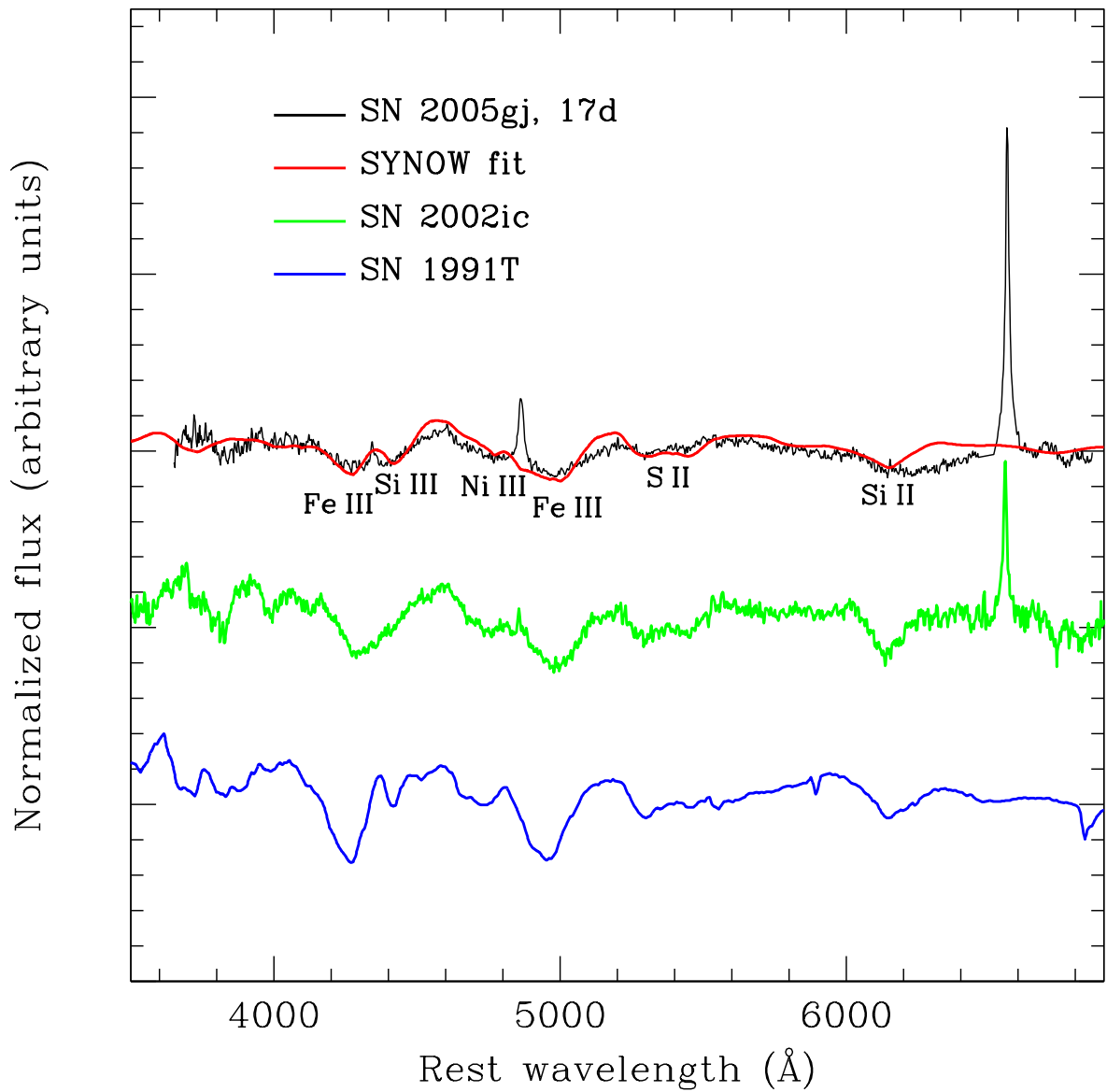}
\caption{Identification of lines in the spectrum of SN~2005gj obtained
at 17~days after explosion (2~days before $g$ maximum). The {\it red
line} shows the best fit synthetic spectrum generated with the SYNOW
code. The lines of SN~2005gj are typical of SN~1991T around maximum
light ({\it blue line}), and very similar to the spectrum of SN~2002ic
around maximum ({\it green line}). All the spectra have been locally
normalized. We have subtracted a constant value to the spectra of
SN~2002ic and SN~1991T for clarity.}
\label{fig:synow}
\end{figure*}

Figure~\ref{fig:synow} shows the spectrum of SN~2005gj at 17~days after
explosion (2~days before $g$ maximum) and the best synthetic spectrum
obtained with SYNOW. We also show for comparison the spectrum of
SN~1991T obtained at -3~days with respect to the time of $B$
maximum. The spectra have been locally normalized as in
\citet{jeffery06}. The synthetic spectrum has a black-body continuum
temperature $T_{bb} = 11000\,{\rm K}$, photospheric velocity
$v_{phot}=10000\,\,{\rm km\ s^{-1}}$, and excitation temperature
$T_{exc}=10000\,\, {\rm K}$. We find a reasonably good match with the
spectrum of SN~2005gj using the following lines/multiplets:
\ion{Fe}{3}~$\lambda$4404 and $\lambda$5129, \ion{Si}{3}~$\lambda$4561,
\ion{Ni}{3}, \ion{S}{2}~$\lambda$5468 and $\lambda$5633, and
\ion{Si}{2}~$\lambda$6355. These lines are characteristic of the
overluminous and spectroscopically peculiar, Type~Ia SN~1991T around
maximum light with strong \ion{Fe}{3} features and weak \ion{S}{2}
doublet and \ion{Si}{2} \citep{jeffery92,mazalli95,fisher99}.
  
The main discrepancy between the SYNOW modeling of SN~2005gj and
SN~1991T is in the optical depths of the lines. The fit to SN~2005gj
needs unphysically small optical depths, approximately 1/10th of the
values used for SN~1991T around maximum light. We interpret this as an
effect of the {\it extra} continuum radiation that is added by the
ejecta-CSM interaction, which is {\it veiling} \citep{branch00} the
supernova lines \citep[e.g.,][]{hamuy03,aldering06}. This interpretation
is supported by the good agreement obtained from fitting the spectra of
SN~2005gj using a simple polynomial continuum added to the spectra of
SN~1991T at the same epochs after explosion (see
Figure~\ref{fig:cont_fits}).

\section{Discussion}
\label{sec:discus}

We have presented extensive spectroscopy and optical/NIR photometry of
SN~2005gj obtained by the SDSS-II and CSP supernova groups during the
first $\sim$150~days after explosion, and also an X-ray observation at
74~days that gives an upper limit on the X-ray luminosity. We have shown
the remarkable similarity in spectroscopic and photometric properties
between SN~2005gj and SN~2002ic, which is thought to be the first clear
case of a thermonuclear supernova explosion embedded in a dense CSM. The
observational properties of SN~2005gj support this interpretation, they
are summarized as follows:

\begin{itemize}

\item Spectroscopic evidence for a shock propagating into an
Hydrogen-rich medium close to the site of the explosion inferred from the
presence of Balmer lines with narrow (FWHM$\sim 200-500\,\,{\rm
km\,s^{-1}}$) and broad (FWHM$\sim 1500-3000\,\,{\rm km\,s^{-1}}$)
components at all times. The Balmer lines show P-Cygni profiles in the
highest resolution spectra obtained at 44 and 115~days after explosion,
these detections show the presence of a slow ($\sim 100\,\,{\rm
km\,s^{-1}}$) moving outflow. Both observations support the
interpretation of the supernova ejecta interacting with a dense
circumstellar material.

\item Spectrum evolves from a very blue continuum (13000~K black-body)
similar to SN~IIn at $\sim$7~days after explosion to a redder continuum
at later times with P-Cygni absorption/emission profiles. The strongest
lines present around maximum are identified with singly and doubly
ionized iron-peak elements (especially strong \ion{Fe}{3}, weak
\ion{S}{2} and \ion{Si}{2}) and the spectra are well matched by the
overluminous Type~Ia SN~1991T {\it diluted} with a polynomial continuum
at similar times after explosion.

\item Very luminous and slowly declining bolometric light curve. The
linear decay in luminosity after peak ($\sim 0.014\,\, {\rm
mag\,day^{-1}}$) suggests an exponential density distribution of the
ejecta, which is consistent with the ejecta-density profiles obtained
from simulations of SN~Ia.

\end{itemize}

The data presented here on SN~2005gj makes the interpretation of
2002ic-like events as thermonuclear supernovae in a dense CSM, initially
proposed by \citet{hamuy03}, stronger. In contrast with
\citet{benetti06}, we find that the overall shape of the spectra of
SN~2005gj are more consistent with spectra of SN~Ia at different
epochs. Specifically, Type~Ic SNe usually do not show the \ion{S}{2}
doublet at $\sim 5500$~\AA\ around maximum light; in fact SN~2004aw
shows only two very weak notches at the wavelengths of \ion{S}{2} near
maximum \citep{tautenberger06}. This is one of the identifying features
in SN~Ia spectra, also present in the overluminous SN~1991T
\citep{phillips92}. In the spectrum of SN~2005gj obtained at 17~days
(see Figure~\ref{fig:synow}) we detect a weak double absorption that we
identify with \ion{S}{2}, that is much stronger in the spectrum of
SN~2002ic around maximum light. We can see on the top of
Figure~\ref{fig:spec_comp} that the spectrum of SN~2002ic obtained
24~days after explosion clearly shows this feature. Other conspicuous
features observed in SN~2005gj and SN~2002ic around maximum are
\ion{Fe}{3} and \ion{Si}{2}. These features are present in SN~1991T, but
\ion{Fe}{3} is not observed and \ion{Si}{2} is generally weaker in
SN~Ic.

SN~2005gj has stronger ejecta-CSM interaction than SN~2002ic. The peak
bolometric luminosity is $\sim 1.5$ times brighter and the broad and
narrow components of \Halpha\ are $\sim 4$ times more luminous in
SN~2005gj. The fact that the SN~1991T features are weaker in SN~2005gj
compared with SN~2002ic at similar epochs is consistent with this
interpretation, because the supernova features are more {\it diluted} by
the stronger continuum. The absence of evidence for a secondary maximum
in SN~2005gj, whereas the $i$ band light curve of SN~2002ic does show a
hint of such a feature, is likewise consistent with the ejecta-CSM
interaction in SN~2005gj having been stronger than in SN~2002ic.

\subsection{Structure of the CSM}
\label{sec:struc_csm}

The circumstellar interaction of core-collapse supernovae with a
circumstellar medium has been studied in detail in the literature (see
\citet{chevalier03} for a review). When the fast moving ejecta
encounters the approximately stationary CSM, a forward shock moving into
the CSM (also called circumstellar shock) and a reverse shock develops.
The fast-moving shockwave implies large post-shock temperatures,
therefore radiating energy in the X-ray regime. The density distribution
of the ejecta and the CSM can be well described by power-laws in radius,
which leads to a set of self-similar analytical solutions for the
evolution of the shock radius in time \citep{chevalier82}. The physics
of the ejecta-CSM interaction in the case of thermonuclear supernovae is
basically the same, the main difference is in the distribution of the
ejected material which follows an exponential function in velocity
\citep{dwarkadas98}. In this case the solutions are no longer
analytic. The density profile of the shocked region is different in the
case of exponential ejecta expanding into a constant density medium, but
the similarity increases for expansion into a wind profile whose density
decreases as $\propto r^{-2}$.

A simple self-similar model of a SN shock expanding into a medium with a
power-law density decline, as suggested for core-collapse SNe by
\citet{chevalier82}, is ruled out for this object by several
observations: the exponential decrease in luminosity, suggesting an
exponential ejecta density profile; the strange behavior of the broad
and narrow \Halpha\ components; and the decrease in the blackbody radius
at later times. Detailed calculations of the SN-CSM interaction would
require highly detailed hydrodynamic modeling, which are beyond the
scope of this paper. Instead herein we focus on trying to explain the
basic features of SN-CSM interaction as deduced from the observational
data.

The initial velocity of a SN shock wave as it breaks out from the
surface is at least of the order of $2\times 10^4\,\,{\rm km\,
s}^{-1}$. The broad \Halpha\ velocities that are seen in the first week
or so are of the order of $1500\,\,{\rm km \,s}^{-1}$, and increase to
more than twice this value after $\sim 50$~days. These velocities are
almost an order of magnitude smaller than expected SN blast wave
velocities in the early stages, and a factor of few smaller even after
$\sim 50$~days. Furthermore, the SN shock velocity would be expected to
gradually decrease as the shock moves outwards, whereas the \Halpha\
profile actually indicates an increasing velocity after $\sim 50$~days.

For these reasons, we suggest that the broad \Halpha\ lines do not
indicate the SN velocity. Instead, we put forward a scenario of a shock
expanding into a two-component ambient medium: a low density wind in
which are embedded high-density clumps. In this picture, there should
theoretically exist three different velocity components: a broad
velocity component, which is not easily seen in this case, and is
related to the velocity of the blast wave itself; an intermediate
velocity component, which is what we have referred to as the broad
\Halpha\, and is related to the velocity of the shock driven into the
clumped material; and a narrow velocity component, which may be related
to the narrow \Halpha\, and is representative of the velocity of the
ambient medium. This scenario is like the scenario put forward by
\citep{chugai94} to explain the origin of the broad, intermediate and
narrow line components in SN~1988Z. The large \Halpha\ luminosity of
SN~2005gj at late times is very similar to that seen in other Type~IIn
SNe, and is especially large considering that this was a
Type~Ia. However, there are significant differences. We do not see a
really broad line component representative of the SN velocity, although
there are some suggestions that this may be appearing at late times. In
particular, the \Halpha\ profile of the spectrum obtained at $\sim
150$~days is better fitted by three components, including a very broad
component with ${\rm FWHM} \approx 7000 \,\,{\rm km s}^{-1}$.

Our scenario envisions the Type~Ia SN shock wave expanding in a clumped
medium presumably formed by mass-loss from a companion star. The broad
component is not easily visible in \Halpha\ initially because the forward
shock is not radiative. The density of the clumps is much higher than
that of the interclump (ambient) medium. When the SN shock wave
interacts with a dense cloud or clump, it drives a strong shock into the
clump. A reflected shock is driven back into the expanding ejecta
\citep{klein94}. Assuming pressure equilibrium, the ratio of the
velocity of the clump shock to that of the blast wave is inversely
proportional to the square-root of the ratio of the clump density to
that of the interclump medium. The optical emission arises from behind
the clump shock, probably by reprocessing of the X-ray emission.

In this model, the intermediate component represents the velocity of the
clump shock, which is probably radiative. If we assume that the initial
velocity of the SN shock wave is $\sim 20000\,\,{\rm km \, s}^{-1}$ and
the broad \Halpha\ emission velocity is $\sim 1500 \,\,{\rm
km\,s}^{-1}$, then the ratio of velocities is $13-14$. This indicates
that the clump density is about $14^2$, or $\sim 200$ times the
interclump density. Note that the optical emission, which goes as
density squared, will then be $200^2$ times, or about 40,000 times
greater compared to that from the interclump medium. This is consistent
with the fact that no broad line emission is seen from the interclump
medium. If the initial velocity is much higher, as is conceivable, the
clump density could be up to $\sim 50$\% higher, and the ratio between
the emission from the dense clumps and interclump medium even larger.

What value of the clump density is suggested? A shock wave traveling at
$1500 \,\,{\rm km \, s}^{-1}$ would be radiative if it were expanding in
a medium whose density is greater than $\sim 10^6$~cm$^{-3}$, whereas a
$2500 \,\,{\rm km \, s}^{-1}$ shock would require minimum densities of
the order of $10^7$~cm$^{-3}$ \citep{draine93} in order to be
radiative. The CSM density, being two orders of magnitude smaller, would
then to be $\ga 10^4$~cm$^{-3}$. These are just minimum values, and it
is conceivable that the actual clump density is much higher. This result
is consistent with the conclusion of \citet{aldering06}.

The observations show that the broad \Halpha\ width increases after
50~days, suggesting an increase in the clump shock velocity at later
times, which could perhaps be due to a decrease in the clump
density. Conversely, however, the luminosity of the \Halpha\ also
increases, suggesting an increase in the electron density. At the same
time, we would expect the SN shock to be decreasing in velocity as it
continues its outwards expansion.

We suggest that the way to reconcile these observations is a scenario in
which the density within the clump medium starts out higher than
$10^8$~cm$^{-3}$, probably as high as $10^{10}$~cm$^{-3}$ in the first
few days, and decreases gradually outwards. The almost constant behavior
of the FWHM of the broad \Halpha\ suggests that the density profile of
the ambient medium is flatter than $r^{-2}$. Since we want the clump
shock to be radiative even when the shock velocity is almost $3000
\,\,{\rm km \, s}^{-1}$, this suggest that the density at $\sim
150$~days is greater than about $10^7$~cm$^{-3}$.  And since the density
is decreasing outwards, we infer that the density close in is even
larger.  Over the entire period of observations the clump density is
large enough that the shock driven into these clumps is always
radiative. The density of the ambient medium is two orders of magnitude
smaller, as discussed above. The high bolometric luminosity is
consistent with these values.

For the first $\sim 50$~days the \Halpha\ emission arises only from the
radiative shock driven into the dense clumps. However, by $\sim 50$~days
the SN forward shock, which is decreasing in velocity, enters the
radiative regime, and the cooling shell of material begins to contribute
to the \Halpha\ luminosity. The velocity of the SN shock is quite large,
and its contribution initially is not a large fraction of the total
\Halpha\ luminosity. But as it expands outwards, its velocity decreases
and the shock becomes more radiative, and the contribution to the total
\Halpha\ luminosity increases, more than compensating for the decreasing
density. If this conclusion is correct, then we would expect that a
broad velocity component would be visible in the \Halpha\ spectra, whose
intensity would gradually increase with time even as the FWHM
decreases. Although the underlying supernova contamination makes it hard
to isolate a broad component, it is suggestive that by day $\sim 150$
the spectrum is best fit by a third, much broader component of the
velocity, thus providing support for this line of reasoning.

Finally, in this model the narrow line emission arises from the
unshocked slowly expanding ambient material, presumably the outflow that
we find expanding at $\sim$60 km s$^{-1}$. We note that although the
width of the narrow line \Halpha\ emission as listed is higher, it is
still unresolved, and it is possible that within the limits of
resolution the narrow line component and outflow velocity are indeed the
same.

To summarize, in this model the Type~Ia SN expands in a clumped ambient
medium, with the clump density about $\sim 200$~times that of the
surrounding medium close in to the star, and decreasing as we go
outwards. The \Halpha\ emission initially arises mainly from the shock
driven into the dense clumps. The SN shock propagating into the
interclump medium begins to enter the radiative regime around day 50,
and its contribution to the \Halpha\ emission gradually increases beyond
that coming from the clumped medium, leading to the gradual rise in the
\Halpha\ emission. We note that several features of this model are
similar to the model presented by \citet{chugai04} for SN~2002ic, thus
further supporting the similarity between the two supernovae.

The upper limit on the X-ray luminosity obtained at 74~days after the
explosion can put a constrain on the mass loss rate from the precursor or
companion \citep[e.g.,][]{immler06}. Assuming that the X-ray luminosity is
dominated by emission from the reverse shock we obtain $\dot{M}\la
2\times 10^{-4}\,\,\Msun\,{\rm yr^{-1}}$ ($2\sigma$) using Equation 3.10
in \citet{fransson96}. This value has to be taken as an approximate
estimate because we are making several assumptions about the physical
properties of the ejecta-CSM interaction that should be calculated using
detailed hydrodynamical simulations: a constant velocity of the shock,
$V_{sh} \approx 8000\,\,{\rm km\,s^{-1}}$; a solar composition of the
CSM material; an electron temperature at the reverse shock of
$T_{e}=10^{7}$~K, which comes from the modeling of SN~2002ic
\citep{nomoto05}; a flat density profile of the CSM, $\rho \propto
r^{-2}$; and a power-law ejecta density profile with index $n=7$
\citep{nomoto84}. 

We can also estimate a mass loss rate from the companion using the
density of the ambient medium ($n\sim 10^{8}$~cm$^{-3}$), the initial
optical radius of the CSM ($R\approx 10^{15}$~cm), and the velocity of
the wind: \hbox{$\dot{M} = 4\pi\,R^{2}\,v\,\rho$}, this is assuming a
flat density profile for the CSM. We obtain: $\dot{M}\approx 2\times
10^{-4}\,\,\Msun\,{\rm yr^{-1}}$, which is in agreement with the
\hbox{2$\sigma$} upper limit calculated from the X-ray luminosity.


The presence of Balmer lines in emission in the first spectrum obtained
6.6~days after explosion shows that the ejecta started to interact with
the CSM at an earlier epoch \citep{aldering06}. Extrapolating linearly
to zero flux the early increase of \Halpha\ and \Hbeta\ fluxes we find
that the ejecta-CSM interaction started $3 \pm 1$~days after explosion,
which gives an internal radius of the CSM $R_{i}\approx 1.1 \times
10^{15}\,\,{\rm cm}$. The outer radius of the CSM can be estimated
assuming a constant velocity of the shock of $V_{sh} \approx
8000\,\,{\rm km\,s^{-1}}$ over the first year. We detect \Halpha\ in
emission in a spectrum obtained at 368~days after explosion, which will
be presented elsewhere, putting a lower limit on the outer radius of the
CSM, $R_{o}\ga 3\times 10^{16}\,\,{\rm cm}$. This is also consistent
with a Type~Ia SN with an exponential ejecta density profile expanding
outwards in a medium of average density $\ga 10^7$~cm$^{-3}$.

In the interpretation above we assume that the broad component of the
Hydrogen Balmer lines originate in the dense clumps, while the narrow
component arises from the photoionized un-shocked gas. However, Thompson
scattering of the lines has been considered as an alternative mechanism
that can explain relatively well the symmetric line profiles of
SN~2002ic \citep{wang04} and SN~2005gj \citep{aldering06}. In this
scenario, both components would arise from a single high density
region. The total mass of hydrogen in the emitting region would be
$M_{H} \approx 2 \times 10^{-2}\,(10^{10}/n_{e})\,\,\Msun$, where
$n_{e}$ is the average electron density in the emitting zone, as
calculated from the luminosity of the \Halpha\ line at maximum using the
Case B recombination coefficient. The electron density must be
sufficiently high, $n_{e}\approx 10^{10}$~cm$^{-3}$, to be consistent
with the line ratios of He lines observed in the spectra
\citep{aldering06}, and a high electron density would explain the
non-detection in X-ray and radio \citep{soderberg05}. However, it is
unlikely that this model would be able to explain the initial constancy
and then rise of the broad \Halpha\ luminosity.

\subsection{Rates, hosts galaxies and possible progenitors of SN~2002ic-like supernovae}
\label{sec:rates}

The SDSS-II Supernova Survey has a well understood discovery efficiency
of SN~Ia at low redshift ($z \la 0.1$), which allows us to obtain an
accurate supernova rate measurement controlling systematic errors
\citep{dilday07}. In the fall 2005 season there were a total of 16
spectroscopically confirmed SN~Ia (including 1991T-like and 1991bg-like
objects) at $z < 0.12$, one photometric identification, and the
spectroscopically confirmed peculiar events: SN~2005hk
\citep{phillips06} and SN~2005gj. Since the detection efficiency of
2002ic-like objects has not been carefully modeled, we can only put a
lower limit on the fraction of these events. The spectroscopic
confirmation of one object at $z < 0.12$ puts a lower limit of
5$^{+7}_{-4}\%$ (68\% confidence) in the fraction of 2002ic-like events
among SN~Ia at low redshift. From the previously known (2002ic) and
probable events (1997cy and 1999E) the estimated fraction is $\sim 1\%$
of SN~Ia discovered between 1997 and 2002, which is consistent with our
limit.

In the fall of 2006 we obtained the spectrum of a slowly declining
supernova that was discovered in 2005, but did not have a spectroscopic
classification, SN~7017\footnote{This is the internal name given by the
SDSS-II Collaboration. It was not announced in an IAU circular because
of the late spectroscopic classification.}. To our surprise, the late
spectrum of SN~7017 resembles that of SN~2005gj one year after explosion
and the early photometry also shows similarities which lead us to
classify it as the highest redshift SN~2002ic-like object observed to
date, at $z=0.27$ (Prieto et al. 2007, in preparation). Considering
SN~7017 in the spectroscopically confirmed sample of SN~Ia during the
2005 season, a total of 129 SNe at $z\la 0.42$, we have that 2/130
(1.5\%) are SN~2002ic like objects, which is consistent with the low
limit on the fraction at low redshift estimated before. However, this
fraction has to be taken with extreme care and probably does not reflect
the {\it true} fraction. This is because the discovery efficiency of
SN~Ia declines as a function of redshift and the total number of
spectroscopically confirmed SN~Ia does not include SNe with good Ia-like
light curves that were not spectroscopically classified. A more careful
study of the rates of SN~2005gj-like supernovae in the SDSS-II is
planned for a future publication.

The host galaxies of supernovae can provide important clues about their
progenitors. The host of SN~2005gj is a very blue, low-luminosity dwarf
($M_{B}\approx -17$), and has an irregular morphology with no well
defined core. \citet{aldering06} combined the SDSS photometry with UV
imaging from GALEX to construct an SED of the galaxy. They constrained
the metallicity to $Z < 0.3\,Z_{\odot}$, with a burst of star formation
$\sim 200$~Myr ago. SN~2002ic has a late type (Sbc) spiral host with a
well defined core. The host of SN~1997cy is also a blue, low-luminosity
($M_{V}\approx -18.2$), and low surface brightness dwarf irregular
galaxy \citep{germany00}. GALEX has imaged the positions of SN~1997cy,
2002ic and 2005gj, and their hosts galaxies are all detected in the
Near-UV (NUV) band. Their absolute magnitudes in the NUV (AB magnitudes)
are between -16.6 (SN~2005gj) and -17.3 (SN~1997cy and 2002ic). They are
low-luminosity late type galaxies, $\sim 1-1.7$ magnitudes fainter than
$L^{*}$ galaxies observed by GALEX at redshift $z < 0.1$
\citep{wyder05}. The host galaxy of SN~1999E is a late spiral with a
nuclear starburst first observed by the IRAS satellite
\citep{allen91}. From 2MASS photometry, its absolute magnitude is 1~mag
brighter than an $L^{*}$ galaxy in the $K$-band
\citep{kochanek01}. SN~7017 at redshift $z=0.27$, has a blue, dwarf-like
host galaxy with absolute magnitude in $B$ of $-17.9$.

The host galaxies of the five SN~2002ic-like objects known share some
common properties: they are late type galaxies, irregulars and late
spirals, most likely with recent star formation. Four of the host
galaxies have low luminosities, similar to the Magellanic clouds, which
indicates they are low metallicity systems. For example, a dwarf galaxy
with intrinsic luminosity $M_{B}= -18$ has an Oxygen abundance of ${\rm
12+log(O/H)} \approx 8.4$ \citep{vanzee06}, which corresponds to 1/3 the
solar Oxygen abundance \citep{delahaye06}. On the other hand, the host
galaxy of SN~1999E has a $K$-band luminosity, that when converted to
metallicity using the luminosity-metallicity relationship derived by
\citet{salzer05}, makes it consistent with the solar value. The host
luminosities are only an approximate indicator of their metallicities,
therefore spectra of the hosts are needed to infer the metallicities and
star formation rates (SFRs) of these galaxies. However, it is
interesting to note that the range of host galaxy properties of
SN~2002ic-like events seem to be inconsistent with the host galaxies of
GRBs associated with supernovae \citep{stanek06} and broad-lined
type~Ib/c SNe \citep{modjaz07}. 

Type~Ia supernovae are observed in all types of galaxies. There is a
well established correlation between the morphology of their host
galaxies and the peak luminosity of the SNe: brighter supernovae
(1991T-like) tend to explode in late type spirals and irregulars with
recent star formation, while intrinsically fainter events (1991bg-like)
are observed mainly in early type galaxies with an old stellar
population \citep{hamuy95,hamuy96b,branch96,hamuy00,gallagher05}. This
environmental effect and observations of the local supernovae rate as a
function of host galaxy properties \citep{cappellaro99,mannucci05},
motivated \citet{scannapieco05} to parametrize the delay time
distribution, time between star formation and the appearance of SNe, and
the rates with a two-component model having a piece proportional to the
SFR of the host galaxy (or {\em prompt}, they explode $\sim {\rm few}
\times 10^{8}$~yr after an episode of star formation), and a second
piece proportional to the total stellar mass ({\em delayed} component,
they explode on scales of a few Gyr after the onset of star
formation). The difference in age of the stellar populations of these
subclasses suggests that the progenitors may also be different: {\em
prompt} SN~Ia would come from more massive progenitors. The host
galaxies of all five SN~2002ic-like events known are broadly consistent
with the properties of the hosts of {\em prompt} SN~Ia, which suggest a
real association given that the best studied SN~IIa to date, SN~2002ic
and SN~2005gj, have spectral characteristics similar to 1991T-like
events. 

Several progenitors have been discussed in the literature for SN~2002ic
and SN~2005gj. \citet{livio_riess03} proposed that SN~2002ic is a rare
case of a double-degenerate binary system, a white dwarf (WD) and the
core of an AGB star spiraling-in through gravitational wave losses, in
which the explosion occurs during or immediately after the
common-envelope phase (a few hundred to a few thousand years of
duration). The difference in line strengths of the Balmer emission lines
observed for SN~2002ic and SN~2005gj makes this scenario unlikely. Also,
as \citet{aldering06} points out, in both SN~2002ic and SN~2005gj the
mass loss stopped only a few years before explosion, which is too short
compared with the timescale for gravitational wave radiation to produce
the merger of the core and the WD.

Another possible progenitor initially proposed by \citet{hamuy03} and
favored by the models of \citet{chugai04}, is the explosion of the
Chandrasekhar-mass Carbon-Oxygen core of a massive AGB star in a
degenerate medium, a supernova Type~1.5 \citep{iben83}, where the dense
Hydrogen-rich CSM would come from the outer layers of the AGB. In order
for the core to grow to the Chandrasekhar mass, the radiatively driven
winds from the AGB have to be weak enough, a condition that is only met
in a very low-metallicity environment like the Galactic halo
\citep{zijlstra04}. At face value, the range of host galaxy
metallicities for SN~2002ic-like events inferred from the
luminosity-metallicity relation does not support the SN~1.5 scenario,
although admittedly these are average metallicities and do not tell us
the actual range of metallicities of the progenitors.

\citet{han06} proposed that SN~2002ic could be produced through the
``super-soft channel'', the most common single-degenerate model for the
progenitors of SN~Ia. In this scenario the white dwarf is accreting
material from a main sequence, or slightly evolved, relatively massive
companion ($\sim 3\,\Msun $) and experiences a delayed dynamical
instability that leads to a large amount of mass-loss from the system in
the last ${\rm few} \times 10^{4}\,{\rm yr}$ before the explosion.
\citet{aldering06} notes that the estimated main-sequence mass of the
progenitor of SN~2005gj of $\sim 2\,\Msun$, calculated using the age of
the starburst of its host galaxy, is consistent with the \citet{han06}
model. Also, the predicted fraction of SN~Ia that would be produced
through the ``delayed dynamical'' channel is 0.1-1\%, consistent with
the limits we have obtain from the detection of SN~2005gj in the SDSS-II
Survey.

In general terms, the progenitor model proposed by \citet{han06}
successfully reproduces the observational properties of SN~2002ic and
SN~2005gj. However, it is still very early in the study of this new
sub-class of SN~Ia. It would be interesting to see in the near future
the results of theoretical modeling exploring other single degenerate
configurations (e.g., AGB donor) and detailed hydrodynamical modeling of
the ejecta-CSM interaction of SN~2005gj using the observations of the
early photometric and spectroscopic evolution presented in this work.

\vspace{0.5in}


We are grateful for the assistance of the staffs at the many
observatories (APO, MDM, LCO, ESO, La Palma) where data for this paper
were obtained.  We would like to thank K.~Z.~Stanek and R.~Pogee for
helpful discussions, and L.~Watson for carefully reading an earlier
version of this paper. We also thank S.~Taubenberger for making the
electronic form of the SN~2004aw data available, and S.~Blondin for
allowing us to use the SNID code before its public release. We wish to
thank H.~Tananbaum and the Chandra Observatory for the generous
allotment of Director's Discretionary Time. VVD research is supported by
award \# AST-0319261 from the National Science Foundation, and by NASA
through grant \# HST-AR-10649 awarded by the Space Science Telescope
Institute. MH acknowledges support from Centro de Astrof\'\i sica FONDAP
15010003, Proyecto Fondecyt 1060808, and Proyecto P06-045-F from
Iniciativa Cient\'\i fica Milenio. DP gratefully acknowledges the
support provided by NASA through Chandra Postdoctoral Fellowship grant
PF4-50035 awarded by the Chandra X-Ray Center, which is operated by the
Smithsonian Astrophysical Observatory for NASA under contract
NAS8-0306. This paper is based upon CSP observations supported by the
NSF under grant AST 0306969.  Use was also made of the CfA Supernova
Archive, which is funded in part by the NSF through grant AST 0606772.

Funding for the SDSS and SDSS-II has been provided by the Alfred P.
Sloan Foundation, the Participating Institutions, the National Science
Foundation (NSF), the U.S. Department of Energy, the National
Aeronautics and Space Administration, the Japanese Monbukagakusho, the
Max Planck Society, and the Higher Education Funding Council for
England.

The SDSS is managed by the Astrophysical Research Consortium for the  
Participating Institutions. The Participating Institutions are the  
American Museum of Natural History, Astrophysical Institute Potsdam,  
University of Basel, Cambridge University, Case Western Reserve  
University, University of Chicago, Drexel University, Fermilab, the  
Institute for Advanced Study, the Japan Participation Group, Johns  
Hopkins University, the Joint Institute for Nuclear Astrophysics, the  
Kavli Institute for Particle Astrophysics and Cosmology, the Korean  
Scientist Group, the Chinese Academy of Sciences (LAMOST), Los Alamos  
National Laboratory, the Max-Planck-Institute for Astronomy (MPIA),  
the Max-Planck-Institute for Astrophysics (MPA), New Mexico State  
University, Ohio State University, University of Pittsburgh,  
University of Portsmouth, Princeton University, the United States  
Naval Observatory, and the University of Washington.


\begin{deluxetable}{cccccccccccc}
\rotate
\tabletypesize{\scriptsize}
\tablecolumns{12}
\tablewidth{0pt}
\tablecaption{SDSS $ugriz$ and CSP $u'g'r'i'$ photometry of comparison stars in common in the field of SN~2005gj. \label{tab:stds1}}
\tablehead{
\colhead{Star} &
\colhead{} &
\colhead{} &
\multicolumn{2}{c}{$u$} &
\multicolumn{2}{c}{$g$} &
\multicolumn{2}{c}{$r$} &
\multicolumn{2}{c}{$i$} &
\colhead{$z$} \\
\colhead{ID} &
\colhead{$\alpha$ (J2000.0)} &
\colhead{$\delta$ (J2000.0)} &
\colhead{SDSS} &
\colhead{CSP}  &
\colhead{SDSS} &
\colhead{CSP}  &
\colhead{SDSS} &
\colhead{CSP}  &
\colhead{SDSS} &
\colhead{CSP}  &
\colhead{SDSS} }
\startdata
1 & 03 01 09.56 & --00 33 52.5 &  18.639(027) &  18.690(054) & 17.367(018) &  17.362(009) & 16.837(022) & 16.823(013) & 16.643(018) & 16.586(010) & 16.549(020) \\ 
2 & 03 01 06.29 & --00 32 59.0 &  $\cdots$ & $\cdots$ & 19.122(027) &  19.043(018) & 17.609(014) & 17.575(015) & 16.191(017) & 16.051(017) & 15.403(018) \\ 
3 & 03 01 14.03 & --00 31 48.7 &  19.843(039) &  19.899(097) &  17.676(016) &  17.653(023) & 16.745(012) & 16.726(015) & 16.404(015) & 16.362(010) & 16.215(017) \\                                
4 & 03 01 14.44 & --00 34 44.3 &  $\cdots$ & $\cdots$ & 20.116(026) & 20.038(045) & 19.579(018) & 19.556(062) & 19.356(025) & 19.113(067) & 19.207(040) \\ 
5 & 03 01 12.99 & --00 31 22.5 &  $\cdots$ &  20.262(436) & 18.856(021) & 18.770(015) & 17.345(009) & 17.323(009) & 16.595(014) & 16.505(010) & 16.145(064) \\ 
6 & 03 01 21.27 & --00 32 44.3 &  16.074(030) &  16.102(011) & 14.556(010) & 14.544(011) & 13.965(016) & 13.961(015) & 13.774(007) & 13.728(015) & 13.711(023) \\ 
7 & 03 01 07.49 & --00 30 51.5 &  19.915(041) & 20.347(285) & 18.385(019) & 18.384(017) & 17.741(020) & 17.735(010) & 17.502(016) & 17.451(013) & 17.372(021) \\ 
8 & 03 01 21.46 & --00 34 25.2 &  18.486(027) &  18.533(030) & 17.214(011) & 17.204(009) & 16.694(017) & 16.685(009) & 16.508(018)  & 16.473(009) & 16.418(016) \\ 
9 & 03 01 07.16 & --00 35 52.2 &  18.847(021) &  18.923(041) & 17.458(013) & 17.453(009) & 16.940(011) & 16.936(011) & 16.765(018) & 16.732(011) &  16.679(024) \\ 
10 & 03 01 12.26 & --00 36 14.5 &  18.285(025) &  18.306(026) & 16.956(018) &  16.935(011) & 16.412(011) & 16.404(009) & 16.208(017) & 16.181(009) & 16.125(021) \\ 
11 & 03 01 02.16 & --00 35 21.1 &  $\cdots$ &  $\cdots$ & 20.227(033) & 20.118(058) & 18.717(014) & 18.715(017) & 17.861(021) & 17.787(014) & 17.347(020) \\ 
12 & 03 01 24.01 & --00 34 27.7 &  $\cdots$ &  $\cdots$ & 18.927(020) & 18.860(015) & 17.433(014) & 17.404(020) & 16.669(018) & 16.611(010) & 16.243(024) \\ 
13 & 03 01 21.90 & --00 35 57.2 &  18.510(035) &  18.529(031) & 16.922(017) & 16.897(009) & 16.249(016) & 16.226(011) & 15.992(017) & 15.958(009) & 15.868(016) \\ 
14 & 03 01 01.95 & --00 36 08.2 &  20.030(065) & 20.081(196) &  17.833(019) & 17.819(010) & 16.882(007) & 16.864(013) & 16.515(017) & 16.470(009) & 16.306(015) \\ 
15 & 03 01 26.63 & --00 31 51.6 &  18.748(029) & 18.314(545) & 16.915(010) &  16.888(009) &  16.142(013) & 16.117(009) &  15.845(022) & 15.798(009) & 15.708(015) \\ 
16 & 03 01 22.53 & --00 30 11.5 &  18.767(032) &  18.834(109) & 17.458(013) &  17.454(011) & 16.896(015) & 16.882(009) & 16.692(013) & 16.652(009) & 16.599(018) \\\enddata
\tablecomments{Uncertainties given in parentheses are in thousandths of a
  magnitude. For the CSP photometry with the Swope they correspond to 
  the rms of the magnitudes obtained on four photometric nights, with 
  a minimum uncertainty of 0.015 mag for an individual measurement.}
\end{deluxetable}

\begin{deluxetable}{ccccccccc}
\tabletypesize{\scriptsize}
\tablecolumns{7}
\tablewidth{0pt}
\tablecaption{SDSS and MDM $ugriz$ photometry of SN~2005gj\label{tab:phot1}}
\tablehead{
\colhead{JD} &
\colhead{Epoch\tablenotemark{a}} &
\colhead{} &
\colhead{} &
\colhead{} &
\colhead{} &
\colhead{} &
\colhead{} \\
\colhead{$-2,453,000$} &
\colhead{(days)} &
\colhead{$u$} &
\colhead{$g$} & 
\colhead{$r$} & 
\colhead{$i$} &
\colhead{$z$} &
\colhead{Source} }
\startdata
616.94 & $\ldots$ & 24.00$\pm$1.24 & 25.19$\pm$0.80 & 23.81$\pm$0.61 & 23.36$\pm$0.54 & 22.15$\pm$0.53 & SDSS  \\
626.91 & $\ldots$ & 21.12$\pm$0.70 & 27.83$\pm$0.86 & 22.85$\pm$0.45 & 24.92$\pm$0.93 & 22.28$\pm$1.77 & SDSS  \\
628.90 & $\ldots$ & 23.88$\pm$1.45 & 25.72$\pm$1.23 & 22.92$\pm$0.41 & 25.98$\pm$0.45 & 21.72$\pm$0.54 & SDSS  \\
635.91 & $\ldots$ & 22.00$\pm$0.49 & 26.38$\pm$1.11 & 23.32$\pm$0.62 & 23.27$\pm$0.59 & 22.15$\pm$0.53 & SDSS  \\
639.95 & \phn1.9 & 18.887(045) & 18.592(025) & 18.621(010) & 18.718(016) & 18.924(043) & SDSS \\
641.95 & \phn3.8 & 18.468(034) & 18.154(011) & 18.256(020) & 18.340(016) & 18.511(041) & SDSS \\
644.89 & \phn6.6 & 18.121(023) & 17.795(011) & 17.879(022) & 17.935(010) & 18.141(022) & SDSS \\
656.94 & 17.9 & \ldots      & 17.355(026) & 17.255(029) & 17.360(064) & \ldots & MDM \\
657.90 & 18.8 & 17.949(039) & 17.343(012) & 17.214(012) & 17.275(016) & 17.438(024) &  SDSS \\
663.89 & 24.5 & 18.149(039) & 17.410(050) & 17.158(012) & 17.163(012) & 17.280(027) &  SDSS \\
663.92 & 24.5 & \ldots      & 17.368(025) & 17.160(020) & 17.248(055) & \ldots & MDM \\
665.91 & 26.4 & \ldots      & \ldots      & 17.146(028) & 17.215(054) & \ldots & MDM \\
666.96 & 27.3 & 18.282(043) & 17.408(012) & 17.121(012) & 17.133(013) & 17.268(015) &  SDSS \\
668.87 & 29.2 & \ldots      & 17.393(014) & 17.125(026) & 17.186(045) & \ldots & MDM \\
668.88 & 29.2 & 18.355(028) & 17.401(009) & 17.127(012) & 17.096(016) & 17.240(016) &  SDSS \\
669.97 & 30.2 & 18.347(045) & 17.430(011) & 17.101(014) & 17.105(031) & 17.231(017) &  SDSS \\
670.88 & 31.0 & 18.386(024) & 17.419(007) & 17.125(013) & 17.092(014) & 17.249(021) &  SDSS \\
673.85 & 33.8 & 18.479(028) & 17.450(009) & 17.123(012) & 17.081(014) & 17.229(015) &  SDSS \\
675.85 & 35.7 & 18.551(025) & 17.477(010) & 17.134(029) & 17.088(019) & 17.182(019) &  SDSS \\
676.90 & 36.7 & \ldots      & 17.464(043) & 17.116(026) & 17.138(055) & \ldots & MDM \\
680.86 & 40.4 & 18.663(027) & 17.578(009) & 17.133(011) & 17.054(009) & 17.156(016) &  SDSS \\
683.92 & 43.3 & 18.759(062) & 17.656(034) & 17.220(055) & 17.080(015) & 17.061(066) &  SDSS \\ 
686.86 & 46.1 & 18.887(047) & 17.695(014) & 17.160(011) & 17.060(010) & 17.138(021) &  SDSS \\
687.91 & 47.1 & 18.937(105) & 17.687(013) & 17.206(024) & 17.042(018) & 17.119(026) &  SDSS \\
693.86 & 52.7 & 19.018(081) & 17.780(009) & 17.207(015) & 17.084(014) & 17.130(018) &  SDSS \\
699.85 & 58.3 & \ldots      & 17.864(060) & 17.271(034) & 17.202(057) & 17.166(024) &  MDM \\
699.88 & 58.4 & 19.083(042) & 17.909(013) & 17.273(009) & 17.124(010) & 17.135(020) &  SDSS \\
727.70 & 84.6 & \ldots      & 18.368(025) & 17.639(043) & 17.551(053) & 17.407(015) &  MDM \\
737.75 & 94.0 & \ldots      & 18.525(038) & 17.791(033) & 17.709(063) & 17.470(057) &  MDM \\
739.64 & 95.8 & \ldots      & 18.540(037) & 17.822(035) & 17.744(065) & 17.548(027) &  MDM \\
\enddata
\tablecomments{Uncertainties given in parentheses in thousandths of a
  magnitude.}
\tablenotetext{a}{Rest-frame days since the time of explosion (JD~2,453,637.93).}
\end{deluxetable}

\begin{deluxetable}{ccccccc}
\tabletypesize{\scriptsize}
\tablecolumns{7}
\tablewidth{0pt}
\tablecaption{CSP $u'g'r'i'$ photometry of SN~2005gj\label{tab:phot2}}
\tablehead{
\colhead{JD} &
\colhead{Epoch} &
\colhead{} &
\colhead{} &
\colhead{} &
\colhead{} \\
\colhead{$-2,453,000$} &
\colhead{(days)} &
\colhead{$u'$} &
\colhead{$g'$} & 
\colhead{$r'$} & 
\colhead{$i'$}}
\startdata
698.69 & \phn57.2 & 19.159(042) & 17.884(017) & 17.243(017) & 17.106(017) \\
699.67 & \phn58.2 & 19.196(041) & 17.910(017) & 17.261(017) & 17.100(017) \\
702.68 & \phn61.0 & 19.286(042) & 17.959(017) & 17.281(017) & 17.130(017) \\
706.69 & \phn64.8 & 19.312(040) & 18.031(017) & 17.345(017) & 17.173(017) \\
712.64 & \phn70.4 & 19.619(060) & 18.157(017) & 17.418(017) & 17.249(017) \\
720.66 & \phn77.9 & 19.721(131) & 18.283(023) & 17.542(017) & 17.363(017) \\
725.65 & \phn82.6 & 19.964(079) & 18.365(017) & 17.602(017) & 17.412(017) \\
728.71 & \phn85.5 & 19.938(104) & 18.430(017) & 17.658(017) & 17.453(017) \\
736.62 & \phn93.0 & 20.104(084) & 18.554(017) & 17.794(017) & 17.586(017) \\
740.64 & \phn96.8 & 20.229(090) & 18.606(017) & 17.856(017) & 17.641(017) \\
741.59 & \phn97.6 & 20.238(103) & 18.635(017) & 17.872(017) & 17.651(017) \\
746.61 & 102.4 & 20.532(220) & 18.641(025) & 17.926(017) & 17.725(017) \\
754.58 & 109.9 & 20.509(143) & 18.770(017) & 18.021(017) & 17.803(017) \\
761.64 & 116.5 & $\cdots$ & 18.847(017) & 18.100(017) & 17.910(018) \\
763.56 & 118.3 & $\cdots$ & 18.841(017) & 18.107(017) & 17.960(017) \\
764.58 & 119.3 & $\cdots$ & 18.859(017) & 18.143(017) & 17.947(017) \\
768.60 & 123.1 & $\cdots$ & 18.882(019) & 18.200(017) & 17.978(021) \\
773.55 & 127.7 & $\cdots$ & 18.884(026) & 18.239(019) & 18.050(021) \\
774.56 & 128.7 & $\cdots$ & 18.946(031) & 18.221(017) & 18.079(022) \\
783.55 & 137.2 & $\cdots$ & 19.017(018) & 18.322(017) & 18.117(021) \\
786.53 & 140.0 & $\cdots$ & 19.015(017) & 18.363(017) & 18.178(020) \\
795.55 & 148.5 & $\cdots$ & 19.099(019) & 18.416(021) & 18.227(031) \\
\enddata
\tablecomments{Uncertainties given in parentheses in thousandths of a
  magnitude.}
\end{deluxetable}

\begin{deluxetable}{ccccccc}
\tabletypesize{\scriptsize}
\tablecolumns{7}
\tablewidth{0pt}
\tablecaption{CSP $YJHK_{s}$ photometry of SN~2005gj \label{tab:phot3}}
\tablehead{
\colhead{JD} &
\colhead{Epoch} &
\colhead{} &
\colhead{} &
\colhead{} &
\colhead{} &
\colhead{} \\
\colhead{$-2,453,000$} &
\colhead{(days)} &
\colhead{$Y$} &
\colhead{$J$} &
\colhead{$H$} &
\colhead{$K_{s}$} &
\colhead{Instrument}}
\startdata
700.71 & 59.1 & 16.565(015) & 16.484(034) & 16.253(030) & $\cdots$ & Retrocam \\
704.68 & 62.9 & 16.591(015) & 16.537(020) & 16.315(033) & $\cdots$ & Retrocam \\
709.66 & 67.6 & 16.628(015) & 16.550(023) & 16.271(029) & $\cdots$ & Retrocam \\
714.58 & 72.2 & 16.673(015) & 16.594(020) & 16.389(028) & $\cdots$ & Retrocam \\
718.65 & 76.0 & 16.725(016) & 16.658(022) & 16.364(037) & $\cdots$ & Retrocam \\
722.62 & 79.8 & 16.832(016) & 16.725(032) & 16.490(017) & 16.384(096) & WIRC \\
724.68 & 81.7 & 16.781(016) & 16.716(028) & $\cdots$ & $\cdots$ & Retrocam \\
727.69 & 84.6 & 16.872(024) & 16.757(036) & $\cdots$ & $\cdots$ & Retrocam \\
732.71 & 89.3 & 16.920(019) & 16.839(025) & $\cdots$ & $\cdots$ & Retrocam \\
750.61 & 106.1 & 17.307(016) & 17.152(016) & 16.812(016) & $\cdots$ & WIRC \\
755.57 & 110.8 & 17.337(024) & 17.263(024) & 16.891(025) & 16.745(039) & PANIC \\
756.59 & 111.8 & 17.428(016) & 17.269(016) & 16.928(017) & $\cdots$ & WIRC \\
757.60 & 112.7 & 17.423(024) & 17.270(024) & 16.938(025) & 16.807(042) & PANIC \\
773.57 & 127.8 & 17.677(016) & 17.476(016) & $\cdots$ & $\cdots$ & WIRC \\
776.54 & 130.6 & 17.538(026) & 17.399(070) & 17.081(068) & $\cdots$ & Retrocam \\
777.55 & 131.5 & 17.648(038) & 17.417(056) & 17.129(119) & $\cdots$ & Retrocam \\
782.55 & 136.2 & 17.513(039) & 17.436(073) & 17.072(114) & $\cdots$ & Retrocam \\
783.55 & 137.2 & 17.766(024) & 17.586(024) & 17.271(025) & 17.045(042) & PANIC \\
785.53 & 139.0 & 17.765(025) & 17.570(022) & 17.192(049) & 17.152(177) & WIRC \\
788.54 & 141.9 & 17.869(016) & 17.594(016) & 17.335(019) & $\cdots$ & WIRC \\
797.52 & 150.3 & 17.855(061) & 17.563(118) & 17.313(158) & $\cdots$ & Retrocam \\
800.50 & 153.1 & 17.840(041) & $\cdots$ & $\cdots$ & $\cdots$ & Retrocam \\
808.52 & 160.7 & 17.871(058) & 17.551(079) & $\cdots$ & $\cdots$ & Retrocam \\
814.49 & 166.3 & 17.864(084) & $\cdots$ & $\cdots$ & $\cdots$ & Retrocam \\
\enddata
\tablecomments{Uncertainties given in parentheses in thousandths of a magnitude.}
\end{deluxetable}

\begin{deluxetable}{lc}
\tabletypesize{\scriptsize}
\tablecolumns{2}
\tablewidth{0pt}
\tablecaption{Light-curve parameters for SN~2005gj\label{tab:lpar}}
\tablehead{
\colhead{Parameter} &
\colhead{Value}}
\startdata
Time of explosion\tablenotemark{a}    & 637.93 $\pm$ 2.02 \\ 
Time of $u_{max}$    & 651.77 $\pm$ 0.48 \\
Time of $g_{max}$    & 657.80 $\pm$ 1.28 \\
Time of $r_{max}$    & 672.49 $\pm$ 1.39 \\
Time of $i_{max}$    & 684.51 $\pm$ 1.00 \\
$u_{max}$            & 17.85 $\pm$ 0.05 \\
$g_{max}$            & 17.35 $\pm$ 0.01 \\
$r_{max}$            & 17.12 $\pm$ 0.01 \\
$i_{max}$            & 17.05 $\pm$ 0.01 \\
$u_{max}^0$\tablenotemark{b}  & 17.11 $\pm$ 0.12 \\
$g_{max}^0$          & 16.94 $\pm$ 0.09 \\
$r_{max}^0$          & 16.83 $\pm$ 0.07 \\
$i_{max}^0$          & 16.89 $\pm$ 0.11 \\
$M_{g,max}^0$        & --20.21 \\
$E(B-V)_{Gal}$\tablenotemark{c} & 0.121 $\pm$ 0.019 \\
$A_{g}{\rm (Gal)}$   & 0.45 $\pm$ 0.07 \\
\enddata
\tablenotetext{a}{JD-$2,453,000$}
\tablenotetext{b}{Magnitudes at maximum in the rest-frame, they have been corrected 
by Galactic extinction and \mbox{$K$-corrections}. We assume a negligible extinction in the 
host galaxy.}
\tablenotetext{c}{From Schlegel et al. (1998)}
\end{deluxetable}

\begin{deluxetable} {cccccc}
\tabletypesize{\scriptsize}
\tablecolumns{6}
\tablewidth{0pt}
\tablecaption{Spectroscopic observations of SN 2005gj\label{tab:spec}}
\tablehead{
\colhead{JD} & 
\colhead{Epoch} & 
\colhead{ } & 
\colhead{Wavelength} & 
\colhead{Resolution\tablenotemark{a}} & 
\colhead{Exposure} \\
\colhead{$-2,453,000$} & 
\colhead{(days)} & 
\colhead{Instrument} & 
\colhead{Range (\AA)} &
\colhead{(\AA)} & 
\colhead{(s)}} 
\startdata
644.92  &  \phn\phn6.6 & MDM-CCDS  &  3850 -- \phn7308   &  15  & 1200 \\
646.95  &  \phn\phn8.5 & ARC-DIS  &  3824 -- 10192   &  7  & 1800 \\
650.84  &  \phn12.2  & ARC-DIS  &  3600 -- \phn9597   &  7  & 1000 \\
655.87  &  \phn16.9  & MDM-CCDS  &  3823 -- \phn7283   &  15  & 1800 \\
665.92  &  \phn26.4  & MDM-CCDS  &  3883 -- \phn7341   &  15  & 2700 \\
668.83  &  \phn29.1  & MDM-CCDS  &  3886 -- \phn7346   &  15  & 2700 \\
676.79  &  \phn36.6  & MDM-CCDS  &  3882 -- \phn7338   &  15  & 3600 \\
684.73  &  \phn44.1  & WHT-ISIS  &  3924 -- \phn8901   &  4  & 1800 \\
686.79  &  \phn46.0  & MDM-CCDS  &  3858 -- \phn7315   &  15  & 2700 \\
698.67  &  \phn57.2  & duPont-ModSpec  &  3780 -- \phn7290   &  7  & 2700 \\
699.67  &  \phn58.2  & duPont-ModSpec  &  3780 -- \phn7290   &  7  & 2700 \\
700.76  &  \phn59.2  & MDM-CCDS  &  3933 -- \phn7391   &  15  & 2700 \\
702.73  &  \phn61.1  & MDM-CCDS  &  3856 -- \phn7310   &  15  & 2700 \\
712.73  &  \phn70.5  & MDM-CCDS  &  3831 -- \phn7286   &  15  & 2700 \\
722.71  &  \phn79.9  & NTT-EMMI  &  4000 -- 10200   &  9  & 2700 \\
724.66  &  \phn81.7  & duPont-WFCCD  &  3800 -- \phn9235   & 6  & 2700 \\
725.65  &  \phn82.6  & duPont-WFCCD  &  3800 -- \phn9235   & 6  & 2700 \\
726.66  &  \phn83.6  & duPont-WFCCD  &  3800 -- \phn9235   & 6  & 3600 \\
727.67  &  \phn84.5  & duPont-WFCCD  &  3800 -- \phn9235   & 6  & 3600 \\
728.67  &  \phn85.5  & duPont-WFCCD  &  3800 -- \phn9235   & 6  & 3600 \\
729.67  &  \phn86.4  & MDM-CCDS  &  3915 -- \phn7373   &  15  & 2700 \\
737.70  &  \phn94.0  & MDM-CCDS  &  3909 -- \phn7368   &  15  & 2700 \\
751.60  &  107.1     & NTT-EMMI  &  3200 -- 10200   &  9  & 2700 \\
755.62  &  110.9     & MDM-CCDS  &  3844 -- \phn7299   &  15  & 3600 \\
759.61  &  114.6     & Magellan-LDSS-3    &  3788 -- \phn9980   &  3  & 3600 \\
799.52  &  152.2     & duPont-WFCCD  &  3800 -- \phn9235   &  6  & 1200 \\
\enddata
\tablecomments{Most of the spectra are the combination of multiple
  observation, the total exposure is given.}
\tablenotetext{a}{Average resolution obtained from the FWHM of arc-lamp lines.}
\end{deluxetable}

\begin{deluxetable}{cccccc}
\tabletypesize{\scriptsize}
\tablecolumns{6}
\tablewidth{0pt}
\tablecaption{$K$-corrections of SN~2005gj \label{tab:kcorr}}
\tablehead{
\colhead{Epoch} &
\colhead{} &
\colhead{} &
\colhead{} &
\colhead{} &
\colhead{} \\
\colhead{(days)} &
\colhead{$K_{u}$} &
\colhead{$K_{g}$} &
\colhead{$K_{r}$} &
\colhead{$K_{i}$} &
\colhead{$K_{z}$}}
\startdata
\phn\phn6.6 & 0.039 & -0.077 & -0.053 & -0.141 & -0.133 \\
\phn12.2 & 0.162 & -0.097 & -0.005 & -0.196 & -0.047 \\
\phn16.9 & 0.226 & -0.052 & -0.055 & -0.135 & -0.223 \\
\phn26.4 & 0.276 & -0.015 & -0.058 & -0.116 & -0.228 \\
\phn29.1 & 0.255 & 0.008 & -0.056 & -0.104 & -0.239 \\
\phn36.6 & 0.296 & 0.034 & -0.030 & -0.118 & -0.096 \\
\phn46.0 & 0.251 & 0.063 & -0.025 & -0.079 & -0.201 \\
\phn57.2 & 0.195 & 0.106 & 0.049 & -0.123 & -0.152 \\
\phn58.2 & 0.186 & 0.101 & 0.013 & -0.068 & -0.216 \\
\phn59.2 & 0.313 & 0.090 & 0.006 & -0.072 & -0.215 \\
\phn61.1 & 0.241 & 0.091 & 0.011 & -0.096 & -0.136 \\
\phn70.5 & 0.505 & 0.086 & 0.055 & -0.113 & -0.054 \\
\phn79.9 & 0.439 & 0.112 & 0.046 & -0.090 & -0.115 \\
\phn81.7 & 0.328 & 0.118 & 0.063 & -0.096 & -0.057 \\
\phn82.6 & 0.340 & 0.115 & 0.063 & -0.104 & -0.102 \\
\phn83.6 & 0.378 & 0.105 & 0.079 & -0.099 & -0.101 \\
\phn84.5 & 0.326 & 0.104 & 0.071 & -0.104 & -0.026 \\
\phn85.5 & 0.391 & 0.112 & 0.078 & -0.110 & -0.036 \\
\phn86.4 & 0.330 & 0.124 & 0.066 & -0.094 & -0.071 \\
\phn94.0 & 0.249 & 0.126 & 0.094 & -0.081 & -0.092 \\
107.1 & 0.266 & 0.141 & 0.082 & -0.084 & -0.089 \\
110.9 & 0.337 & 0.138 & 0.097 & -0.091 & -0.007 \\
\enddata
\end{deluxetable}

\begin{deluxetable}{ccccccc}
\tabletypesize{\scriptsize}
\tablecolumns{7}
\tablewidth{0pt}
\tablecaption{Derived integrated luminosity and black-body fits. \label{tab:lum}}
\tablehead{
\colhead{Epoch} &
\colhead{$\log L_{(u\rightarrow i)}$} &
\colhead{$\log L_{(u\rightarrow z)}$} & 
\colhead{$\log L_{bol}$} & 
\colhead{$T_{bb}$\tablenotemark{a}}  &
\colhead{$R_{bb}$\tablenotemark{b}} &
\colhead{} \\
\colhead{(days)} &
\colhead{(erg s$^{-1}$)} &
\colhead{(erg s$^{-1}$)} &
\colhead{(erg s$^{-1}$)} &
\colhead{(K)} &
\colhead{(10$^{15}$ cm)} &
\colhead{$\chi^{2}_{\nu}$}\tablenotemark{c}}
\startdata
\phn\phn1.9 & 42.822 & 42.862 & 43.285(0.109) & 12898(1193)  &  1.02(0.16) & 0.7 \\
\phn\phn3.8 & 42.990 & 43.029 & 43.485(0.105) & 13552(1316)  &  1.16(0.18) & 0.3 \\
\phn\phn6.6 & 43.141 & 43.180 & 43.637(0.098) & 13569(1288)  &  1.38(0.21) & 0.2 \\
\phn18.8 & 43.344 & 43.389 & 43.713(0.096) & 10955(831)  &  2.32(0.34) & 1.2 \\
\phn24.5 & 43.335 & 43.387 & 43.644(0.125) & 9468(612)  &  2.87(0.42) & 1.2 \\
\phn27.3 & 43.335 & 43.388 & 43.628(0.135) & 9069(720)  &  3.06(0.50) & 2.2 \\
\phn29.2 & 43.335 & 43.389 & 43.617(0.140) & 8772(683)  &  3.22(0.53) & 2.3 \\
\phn30.2 & 43.334 & 43.388 & 43.617(0.137) & 8761(692)  &  3.23(0.54) & 2.3 \\
\phn31.0 & 43.332 & 43.386 & 43.611(0.122) & 8697(705)  &  3.25(0.55) & 2.6 \\
\phn33.8 & 43.325 & 43.381 & 43.595(0.140) & 8392(685)  &  3.43(0.59) & 2.9 \\
\phn35.7 & 43.316 & 43.374 & 43.581(0.160) & 8100(582)  &  3.61(0.59) & 2.4 \\
\phn40.4 & 43.299 & 43.361 & 43.559(0.152) & 7684(526)  &  3.92(0.64) & 2.5 \\
\phn43.3 & 43.270 & 43.338 & 43.532(0.143) & 7284(335)  &  4.24(0.59) & 0.6 \\
\phn46.1 & 43.269 & 43.336 & 43.526(0.101) & 7142(477)  &  4.36(0.73) & 2.8 \\
\phn47.1 & 43.263 & 43.332 & 43.523(0.139) & 7115(450)  &  4.38(0.70) & 2.1 \\
\phn52.7 & 43.247 & 43.317 & 43.506(0.123) & 6889(425)  &  4.58(0.74) & 2.4 \\
\phn57.2 & 43.219 & 43.294 & 43.480(0.095) & 6958(247)  &  4.32(0.54) & 2.0 \\
\phn58.2 & 43.212 & 43.288 & 43.473(0.096) & 6913(248)  &  4.34(0.55) & 2.0 \\
\phn58.4 & 43.215 & 43.290 & 43.480(0.133) & 6656(322)  &  4.78(0.70) & 1.6 \\
\phn61.0 & 43.198 & 43.274 & 43.458(0.107) & 6840(261)  &  4.34(0.56) & 2.4 \\
\phn64.8 & 43.176 & 43.254 & 43.438(0.087) & 6830(249)  &  4.27(0.54) & 2.2 \\
\phn70.4 & 43.133 & 43.214 & 43.394(0.109) & 6570(278)  &  4.33(0.58) & 3.4 \\
\phn77.9 & 43.088 & 43.171 & 43.355(0.098) & 6633(280)  &  4.10(0.54) & 2.8 \\
\phn82.6 & 43.057 & 43.143 & 43.321(0.127) & 6380(293)  &  4.18(0.58) & 4.2 \\
\phn85.5 & 43.038 & 43.125 & 43.305(0.096) & 6459(282)  &  4.04(0.54) & 3.5 \\
\phn93.0 & 42.987 & 43.076 & 43.256(0.103) & 6360(272)  &  3.92(0.53) & 3.6 \\
\phn96.8 & 42.963 & 43.054 & 43.232(0.101) & 6305(279)  &  3.87(0.53) & 3.9 \\
\phn97.6 & 42.955 & 43.047 & 43.226(0.105) & 6318(279)  &  3.83(0.52) & 3.8 \\
102.4 & 42.935 & 43.026 & 43.205(0.133) & 6372(332)  &  3.68(0.52) & 4.4 \\
109.9 & 42.896 & 42.989 & 43.165(0.117) & 6339(313)  &  3.53(0.50) & 4.4 \\
116.5 & 42.859 & 42.953 & 43.135(0.123) & 6549(350)  &  3.25(0.46) & 3.8 \\
118.3 & 42.855 & 42.947 & 43.128(0.138) & 6566(362)  &  3.21(0.46) & 4.0 \\
119.3 & 42.847 & 42.940 & 43.122(0.117) & 6548(357)  &  3.20(0.46) & 4.0 \\
120.2 & 42.839 & 42.932 & 43.114(0.120) & 6536(352)  &  3.18(0.45) & 3.9 \\
123.1 & 42.831 & 42.924 & 43.104(0.127) & 6533(370)  &  3.14(0.45) & 4.3 \\
127.7 & 42.818 & 42.909 & 43.087(0.155) & 6523(399)  &  3.06(0.46) & 5.0 \\
128.7 & 42.809 & 42.901 & 43.078(0.153) & 6486(399)  &  3.07(0.46) & 5.2 \\
137.2 & 42.776 & 42.869 & 43.043(0.169) & 6404(418)  &  2.99(0.46) & 6.3 \\
140.0 & 42.766 & 42.857 & 43.030(0.197) & 6377(421)  &  2.96(0.46) & 6.5 \\
148.5 & 42.738 & 42.830 & 42.999(0.205) & 6246(438)  &  2.94(0.48) & 8.0 \\
\enddata
\tablenotetext{a}{Black-body temperature from the fits to the broadband photometry; 1$\sigma$ uncertainty are given in parenthesis.}
\tablenotetext{b}{Black-body radius from the fits to the broadband photometry; 1$\sigma$ uncertainty are given in parenthesis and include a $10\%$ uncertainty in the distance to SN~2005gj.}
\tablenotetext{c}{$\chi^2$ per degree of freedom of the black-body fits.}
\end{deluxetable}

\begin{deluxetable}{lllc}
\tabletypesize{\scriptsize}
\tablecolumns{4}
\tablewidth{0pt}
\tablecaption{Library of spectra used in SNID \label{table:speclib}}
\tablehead{
\colhead{SN Name} &
\colhead{Class} &
\colhead{Epochs} &
\colhead{Reference}}
\startdata
1990N  & Ia normal & -14, -13, -8, -7, -6, 0, 4, 8, 15, 18, 39  &  1 \\
1991T  & Ia 91T    & -9, -8, -7, -6, -5, -4, -2, -1, 9, 10, 11, 12, 15, 16, 17, 18, 19, 20, 21, 22, 23, 27, 43, 44, 47, 48, 51, 69, 77  & 2, 3 \\
1991bg & Ia 91bg   &  1, 3, 16, 18, 25, 32, 33, 46, 54, 85  &  4 \\
1992A  & Ia normal & -5, -1, 3, 5, 6, 7, 9, 11, 16, 17, 24, 28  &  5 \\
1994I  & Ic normal & -6, -4, -3, 0, 1, 2, 21, 22, 23, 24, 26, 30, 36, 38 & 6 \\
1997ef & Ic broad  & -14, -12, -11, -10, -9, -6, -5, -4, 7, 13, 14, 16, 17, 19, 22, 24, 27, 41, 45, 47, 49, 75, 80, 81 & 7 \\
1998aq & Ia normal &  -9,  -8,  -3,   0,   1,   2,   3,   4,   5,   6,   7,  19,  21,  24,  31,  32,  36,  51,  55,  58,  60,  63,  66,  79,  82,  91, 211, 231, 241 & 8 \\
1998bu & Ia normal & -3, -2, -1, 9, 10, 11, 12, 13, 14, 28, 29, 30, 31, 32, 33, 34, 35, 36, 37, 38, 39, 40, 41, 42, 43, 44, 57 & 9 \\
1998bw & Ic broad  & -9, -7, -6, -3, -2, -1, 1, 3, 4, 6, 9, 11, 12, 13, 19, 22, 29, 45, 64, 125, 200, 337, 376   & 10 \\
1999aa & Ia 91T    & -11, -7, -3, -1, 5, 6, 14, 19, 25, 28, 33, 49, 47, 51  &  11 \\
1999ee & Ia normal & -9,  -7,  -2,   0,   3,   8,  10,  12,  17,  20,  23,  28,  33,  42 & 12 \\
1999ex & Ic normal &  -1, 4, 13 & 12 \\
1999by & Ia 91bg   &  -4, -3, -2, -1, 2, 3, 5, 6, 7, 8, 10, 11, 25, 29, 31, 33, 42   & 13 \\
2002ap & Ic broad  & -5, -4, 3, 8, 10, 17, 19 & 14 \\
2004aw & Ic normal & 1, 5, 6, 8, 15, 21, 22, 26, 28, 39, 35, 44, 49, 63, 64, 236, 260, 413 & 15 \\
2006aj & Ic broad  & -6, -5, -4, -3, -2, -1, 0, 2, 3 & 16 \\ 
\enddata
\tablerefs{
(1) \citet{leibundgut91}; (2) \citet{jeffery92}; (3) \citet{schmidt94};
(4) \citet{leibundgut93}; (5) \citet{kirshner93}; (6) \citet{millard99};
(7) \citet{iwamoto00}; (8) \citet{branch03}; (9) \citet{jha99}; (10) \citet{patat01}; 
(11) \citet{garavini04}; (12) \citet{hamuy02}; (13) \citet{garnavich04}; 
(14) \citet{gal-yam02}; (15) \citet{tautenberger06}; (16) \citet{modjaz06}.}
\end{deluxetable}

\begin{deluxetable}{cccccccccc}
\rotate
\tabletypesize{\scriptsize}
\tablecolumns{10}
\tablewidth{0pt}
\tablecaption{Results of the Gaussian fits to H$\alpha$ and H$\beta$ features \label{tab:balmer}}
\tablehead{
\colhead{} &
\colhead{} &
\multicolumn{2}{c}{H$\alpha$ (narrow)} &
\colhead{} &
\multicolumn{2}{c}{H$\alpha$ (broad)} &
\colhead{} &
\multicolumn{2}{c}{H$\beta$} \\
\cline{3-4} \cline{6-7} \cline{9-10} \\ 
\colhead{JD} &
\colhead{Epoch} &
\colhead{FWHM\tablenotemark{a}} &
\colhead{flux\tablenotemark{b}} &
\colhead{} &
\colhead{FWHM\tablenotemark{a}} &
\colhead{flux\tablenotemark{b}} &
\colhead{} &
\colhead{FWHM\tablenotemark{a}} &
\colhead{flux\tablenotemark{b}} \\
\colhead{$-2,453,000$}  &
\colhead{(days)}         &
\colhead{(km s$^{-1}$)} &
\colhead{(10$^{-14}$ erg s$^{-1}$ cm$^{-2}$)} &
\colhead{} &
\colhead{(km s$^{-1}$)} &
\colhead{(10$^{-14}$ erg s$^{-1}$ cm$^{-2}$)} &
\colhead{} &
\colhead{(km s$^{-1}$)} &
\colhead{(10$^{-14}$ erg s$^{-1}$ cm$^{-2}$)}}
\startdata 
644.92 & \phn\phn6.6   & \ldots & 0.24(0.02)  & & 1575 & 0.58(0.06)  & &  776   & 0.50(0.05) \\
646.95 & \phn\phn8.5   &  137   & 0.37(0.04)  & & 1481 & 0.83(0.08)  & & 1307   & 0.52(0.06) \\
650.84 & \phn12.2  &  314   & 0.69(0.07)  & & 1731 & 1.25(0.13)  & & 1462   & 0.75(0.10) \\
655.87 & \phn16.9  & \ldots & 0.57(0.06)  & & 1555 & 1.11(0.11)  & & 1339   & 0.75(0.09) \\
665.92 & \phn26.4  & \ldots & 0.53(0.05)  & & 1569 & 1.03(0.10)  & &  523   & 0.48(0.06) \\
668.83 & \phn29.1  & \ldots & 0.37(0.04)  & & 1234 & 0.99(0.10)  & & 1275   & 0.55(0.07) \\
676.79 & \phn36.6  & \ldots & 0.45(0.05)  & & 1513 & 0.77(0.08)  & &  884   & 0.28(0.04) \\
686.79 & \phn46.0  & \ldots & 0.41(0.04)  & & 1836 & 0.71(0.07)  & & \ldots & 0.15(0.02) \\
698.67 & \phn57.2  & \ldots & 0.45(0.04)  & & 2115 & 0.91(0.10)  & & \ldots & 0.15(0.03) \\
699.67 & \phn58.2  & \ldots & 0.34(0.04)  & & 2053 & 0.76(0.08)  & & \ldots & 0.14(0.03) \\
700.76 & \phn59.2  & \ldots & 0.36(0.04)  & & 1830 & 0.77(0.08)  & &  620   & 0.16(0.02) \\
702.73 & \phn61.0  & \ldots & 0.41(0.04)  & & 1978 & 0.67(0.07)  & & \ldots & 0.11(0.02) \\
712.73 & \phn70.5  & \ldots & 0.41(0.04)  & & 2357 & 1.02(0.11)  & & \ldots & 0.11(0.02) \\
722.71 & \phn79.9  & \ldots & 0.34(0.04)  & & 2413 & 1.03(0.11)  & &  490   & 0.12(0.03) \\ 
724.66 & \phn81.7  & \ldots & 0.34(0.04)  & & 2137 & 0.99(0.10)  & & 1067   & 0.19(0.03) \\
725.65 & \phn82.6  & \ldots & 0.34(0.03)  & & 2260 & 1.02(0.11)  & &  568   & 0.11(0.02) \\
726.66 & \phn83.6  & \ldots & 0.35(0.04)  & & 2322 & 1.10(0.11)  & & \ldots & 0.08(0.02) \\
727.67 & \phn84.5  & 160    & 0.32(0.03)  & & 2364 & 1.02(0.10)  & & \ldots & 0.10(0.02) \\
728.67 & \phn85.5  & \ldots & 0.20(0.02)  & & 1802 & 1.24(0.14)  & & 1127   & 0.19(0.03) \\
729.67 & \phn86.4  & \ldots & 0.37(0.04)  & & 2687 & 1.15(0.12)  & &  680   & 0.14(0.02) \\
737.70 & \phn94.0  & \ldots & 0.29(0.03)  & & 1941 & 0.85(0.09)  & &  459   & 0.09(0.01) \\
755.62 & 110.9 & \ldots & 0.32(0.03)  & & 2236 & 1.14(0.12)  & & 1031   & 0.16(0.02) \\
799.52 & 152.2 &  525   & 0.51(0.05)  & & 3809 & 2.18(0.22)  & & 1669   & 0.23(0.03) \\
\enddata
\tablenotetext{a}{FWHM is not presented when the spectral resolution is
bigger than the measured value.}
\tablenotetext{b}{1$\sigma$ uncertainties are given in parentheses.}
\end{deluxetable}


\clearpage

\newpage 

\begin{thebibliography}

\bibitem[Aldering et~al.(2006)]{aldering06}
Aldering, G., et~al. 2006, ArXiv Astrophysics e-prints

\bibitem[Allen et~al.(1991)]{allen91}
Allen, D.~A., et~al. 1991,  \mnras, 248, 528

\bibitem[Aretxaga et~al.(1999)]{aretxaga99}
Aretxaga, I., et al. 1999, \mnras, 309, 343

\bibitem[Arnett(1982)]{arnett82}
Arnett, W.~D. 1982, \apj, 253, 785

\bibitem[Astier et~al.(2006)]{astier06}
Astier, P., et al. 2006, \aap, 447, 31

\bibitem[Barentine et~al.(2005)]{barentine05}
Barentine, J., et~al. 2005,  Central Bureau Electronic Telegrams, 247, 1 (2005).~ Edited by Green,  D.~W.~E., 247

\bibitem[Benetti et~al.(2006)]{benetti06}
Benetti, S., et~al. 2006, \apjl, 653, L129

\bibitem[Blondin \& Tonry(2007)]{blondin07}
Blondin, S. \& Tonry, J.~L. 2007, \apj, in press

\bibitem[Branch et~al.(1981)]{branch81}
Branch, D., et~al. 1981, \apj, 244, 780

\bibitem[Branch et~al.(2003)]{branch03}
Branch, D., et al. 2003, \aj, 126,  1489

\bibitem[Branch et~al.(2000)]{branch00}
Branch, D., et~al. 2000, \pasp,  112, 217

\bibitem[Branch et~al.(1996)]{branch96}
Branch, D., Romanishin, W., \& Baron, E. 1996, \apj, 465, 73

\bibitem[Cappellaro et~al.(1999)]{cappellaro99}
Cappellaro, E., Evans, R., \& Turatto, M. 1999, \aap, 351, 459

\bibitem[Cardelli et~al.(1989)]{ccm89}
Cardelli, J.~A., Clayton, G.~C., \& Mathis, J.~S. 1989, \apj, 345, 245

\bibitem[Chevalier(1982)]{chevalier82}
Chevalier, R.~A. 1982, \apj, 258, 790

\bibitem[Chevalier \& Fransson(1994)]{chevalier94}
Chevalier, R.~A. \& Fransson, C. 1994, \apj, 420, 268

\bibitem[Chevalier \& Fransson(2003)]{chevalier03}
Chevalier, R.~A. \& Fransson, C. 2003, in LNP Vol. 598: Supernovae and  Gamma-Ray Bursters, ed. K.~Weiler, 171--194

\bibitem[Chugai(1997a)]{chugai97b}
Chugai, N.~N. 1997a, \apss, 252, 225

\bibitem[Chugai(1997b)]{chugai97a}
---. 1997b, Astronomy Reports, 41, 672

\bibitem[Chugai et~al.(2004)]{chugai04}
Chugai, N.~N., Chevalier, R.~A., \& Lundqvist, P. 2004, \mnras, 355, 627

\bibitem[Chugai \& Danziger(1994)]{chugai94}
Chugai, N.~N. \& Danziger, I.~J. 1994, \mnras, 268, 173

\bibitem[Contardo et~al.(2000)]{contardo00}
Contardo, G., Leibundgut, B., \& Vacca, W.~D. 2000, \aap, 359, 876

\bibitem[Delahaye \& Pinsonneault(2006)]{delahaye06}
Delahaye, F. \& Pinsonneault, M.~H. 2006, \apj, 649, 529

\bibitem[Deng et~al.(2004)]{deng04}
Deng, J., Kawabata, K.~S., Ohyama, Y., et al. 2004,  \apjl, 605, L37

\bibitem[Di Carlo et~al.(2002)]{dicarlo02}
Di Carlo, E., et~al. 2002, \apj, 573, 144

\bibitem[Dilday et~al.(2007)]{dilday07}
Dilday, B., et.~al. 2007, in preparation

\bibitem[Draine \& McKee(1993)]{draine93}
Draine, B.~T. \& McKee, C.~F. 1993, \araa, 31, 373

\bibitem[Dwarkadas \& Chevalier(1998)]{dwarkadas98}
Dwarkadas, V.~V. \& Chevalier, R.~A. 1998, \apj, 497, 807

\bibitem[Fisher et~al.(1999)]{fisher99}
Fisher, A., et al. 1999, \mnras, 304, 67

\bibitem[Fisher et~al.(1997)]{fisher97}
Fisher, A., et al. 1997, \apjl, 481, L89+

\bibitem[Fisher(2000)]{fisher00}
Fisher, A.~K. 2000, PhD thesis, AA(THE UNIVERSITY OF OKLAHOMA)

\bibitem[Fransson et~al.(1996)]{fransson96}
Fransson, C., Lundqvist, P., \& Chevalier, R.~A. 1996, \apj, 461, 993

\bibitem[Freedman et~al.(2001)]{freedman01}
Freedman, W.~L., et al. 2001, \apj, 553, 47

\bibitem[Frieman et~al.(2007)]{frieman07}
Frieman, J.~A., et.~al. 2007, submitted to AJ

\bibitem[Fukugita et~al.(1996)]{fukugita96}
Fukugita, M., et~al. 1996, \aj, 111, 1748

\bibitem[Gal-Yam et~al.(2006)]{gal-yam06}
Gal-Yam, A., et al. 2006, ArXiv Astrophysics e-prints

\bibitem[Gal-Yam et~al.(2002)]{gal-yam02}
Gal-Yam, A., Ofek, E.~O., \& Shemmer, O. 2002, \mnras, 332, L73

\bibitem[Gallagher et~al.(2005)]{gallagher05}
Gallagher, J.~S., et al. 2005, \apj, 634, 210

\bibitem[Garavini et~al.(2004)]{garavini04}
Garavini, G., et al. 2004, \aj, 128, 387

\bibitem[Garnavich et~al.(2004)]{garnavich04}
Garnavich, P.~M., et al. 2004, \apj, 613, 1120

\bibitem[Germany et~al.(2000)]{germany00}
Germany, L.~M., et~al. 2000, \apj, 533, 320

\bibitem[Gunn et~al.(1998)]{gunn98}
Gunn, J.~E., et al. 1998, \aj,  116, 3040

\bibitem[Gunn et~al.(2006)]{gunn06}
Gunn, J.~E., et al. 2006, \aj, 131, 2332

\bibitem[Hamuy et~al.(1993)]{hamuy93}
Hamuy, M., et~al. 1993, \pasp, 105,  787

\bibitem[Hamuy et~al.(1995)]{hamuy95}
Hamuy, M., et al. 1995, \aj, 109, 1

\bibitem[Hamuy et~al.(1996a)]{hamuy96b}
Hamuy, M., et~al. 1996a, \aj, 112, 2391

\bibitem[Hamuy et~al.(1996b)]{hamuy96}
---. 1996b, \aj, 112, 2398

\bibitem[Hamuy et~al.(2000)]{hamuy00}
Hamuy et~al. 2000, \aj, 120, 1479

\bibitem[Hamuy et~al.(2002)]{hamuy02}
Hamuy, M., et al. 2002, \aj, 124, 417

\bibitem[Hamuy et~al.(2003)]{hamuy03}
Hamuy, M., et al. 2003, \nat, 424

\bibitem[Hamuy et~al.(2006)]{hamuy06}
Hamuy, M., et~al. 2006, \pasp, 118, 2

\bibitem[Han \& Podsiadlowski(2006)]{han06} Han, Z. \& Podsiadlowski, P. 2006, \mnras, 368, 1095

\bibitem[Hogg et~al.(2001)]{hogg01}
Hogg, D.~W., et al. 2001,  \aj, 122, 2129

\bibitem[Holtzman et~al.(2007)]{holtzman07} Holtzman, J., et.~al. 2007, in preparation

\bibitem[Iben \& Renzini(1983)]{iben83}
Iben, Jr., I. \& Renzini, A. 1983, \araa, 21, 271

\bibitem[Immler et~al.(2006)]{immler06}
Immler et~al. 2006,  \apjl, 648, L119

\bibitem[Immler et~al.(2005)]{immler05}
Immler, S., Petre, R., \& Brown, P. 2005, \iaucirc, 8633, 2

\bibitem[Ivezi{\'c} et~al.(2007)]{ivezic07}
Ivezi{\'c}, {\v Z}., et~al. 2007, ArXiv Astrophysics e-prints

\bibitem[Ivezi{\'c} et~al.(2004)]{ivezic04}
Ivezi{\'c}, {\v Z}., et~al. 2004, Astronomische Nachrichten, 325, 583

\bibitem[Iwamoto et~al.(2000)]{iwamoto00}
Iwamoto, K., et al. 2000, \apj, 534, 660

\bibitem[Jeffery et~al.(2006)]{jeffery06}
Jeffery, D.~J., et al. 2006, ArXiv Astrophysics e-prints

\bibitem[Jeffery et~al.(1992)]{jeffery92}
Jeffery, D.~J., et~al. 1992, \apj, 397, 304

\bibitem[Jha et~al.(1999)]{jha99}
Jha, S., et al. 1999, \apjs, 125, 73

\bibitem[Kirshner et~al.(1993)]{kirshner93}
Kirshner, R.~P., et al. 1993, \apj, 415,  589

\bibitem[Klein et~al.(1994)]{klein94}
Klein, R.~I., McKee, C.~F., \& Colella, P. 1994, \apj, 420, 213

\bibitem[Kochanek et~al.(2001)]{kochanek01}
Kochanek, C.~S., et~al. 2001,  \apj, 560, 566

\bibitem[Kotak et~al.(2004)]{kotak04}
Kotak, R., et al. 2004,  \mnras, 354, L13

\bibitem[Kraft et~al.(1991)]{Kraft91}
Kraft, R.~P., Burrows, D.~N., \& Nousek, J.~A. 1991, \apj, 374, 344

\bibitem[Krisciunas et~al.(2004)]{krisciunas04}
Krisciunas, K., Phillips, M.~M., \& Suntzeff, N.~B. 2004, \apjl, 602, L81

\bibitem[Leibundgut et~al.(1991)]{leibundgut91}
Leibundgut, B., et~al. 1991, \apjl, 371, L23

\bibitem[Leibundgut et~al.(1993)]{leibundgut93}
Leibundgut, B., et~al. 1993, \aj, 105, 301

\bibitem[Livio \& Riess(2003)]{livio_riess03}
Livio, M. \& Riess, A.~G. 2003, \apjl, 594, L93

\bibitem[Lupton et~al.(1999)]{lupton99}
Lupton, R.~H., Gunn, J.~E., \& Szalay, A.~S. 1999, \aj, 118, 1406

\bibitem[Mannucci et~al.(2005)]{mannucci05}
Mannucci, F., et al. 2005, \aap, 433, 807

\bibitem[Martini et~al.(2004)]{martini04}
Martini, P., et al. 2004, in Ground-based Instrumentation  for Astronomy. Edited by Alan F. M. Moorwood and Iye Masanori. Proceedings of  the SPIE, Volume 5492, pp. 1653-1660 (2004)., ed. A.~F.~M. {Moorwood} \&  M.~{Iye}, 1653--1660

\bibitem[Matheson et~al.(2005)]{matheson05}
Matheson, T., et~al. 2005, \aj, 129,  2352

\bibitem[Mazzali et~al.(1995)]{mazalli95}
Mazzali, P.~A., Danziger, I.~J., \& Turatto, M. 1995, \aap, 297, 509

\bibitem[Miknaitis et~al.(2007)]{miknaitis07}
Miknaitis, G., et al.(2007),  ArXiv Astrophysics e-prints

\bibitem[Millard et~al.(1999)]{millard99}
Millard, J., et~al. 1999, \apj, 527, 746

\bibitem[Modjaz et~al.(2007)]{modjaz07}
Modjaz, M., et~al. 2007, ArXiv  Astrophysics e-prints

\bibitem[Modjaz et~al.(2006)]{modjaz06}
Modjaz, M., et~al. 2006, \apjl,  645, L21

\bibitem[Morgan et~al.(2005)]{Morgan05}
Morgan, C.~W., et~al. 2005, \aj, 129,  2504

\bibitem[Nomoto et~al.(1984)]{nomoto84}
Nomoto, K., Thielemann, F.-K., \& Yokoi, K. 1984, \apj, 286, 644

\bibitem[Nomoto et~al.(2005)]{nomoto05}
Nomoto, K., et al. 2005,  in ASP Conf. Ser. 342: 1604-2004: Supernovae as Cosmological Lighthouses, ed.  M.~{Turatto}, S.~{Benetti}, L.~{Zampieri}, \& W.~{Shea}, 105--+

\bibitem[Nugent et~al.(2002)]{nugent02}
Nugent, P., Kim, A., \& Perlmutter, S. 2002, \pasp, 114, 803

\bibitem[Oke \& Gunn(1983)]{oke_gunn83}
Oke, J.~B., \& Gunn, J.~E. 1983, \apj, 266, 713

\bibitem[Osterbrock(1989)]{osterbrock89}
Osterbrock, D.~E. 1989, Astrophysics of gaseous nebulae and active galactic  nuclei (Research supported by the University of California, John Simon  Guggenheim Memorial Foundation, University of Minnesota, et al.~Mill Valley,  CA, University Science Books, 1989, 422 p.

\bibitem[Pastorello et~al.(2002)]{pastorello02}
Pastorello, A., et~al. 2002, \mnras, 333, 27

\bibitem[Patat et~al.(2001)]{patat01}
Patat, F., et~al. 2001, \apj, 555, 900

\bibitem[Perlmutter et~al.(1999)]{perlmutter99}
Perlmutter, S., et al. 1999, \apj, 517, 565

\bibitem[Persson et~al.(2002)]{persson02}
Persson, S.~E., et al. 2002, \aj, 124, 619

\bibitem[Phillips et~al.(1992)]{phillips92}
Phillips, M.~M., et al. 1992, \aj, 103, 1632

\bibitem[Phillips(1993)]{phillips93}
Phillips, M.~M. 1993, \apjl, 413, L105

\bibitem[Phillips et~al.(2007)]{phillips06}
Phillips , M.~M., et al. 2007, \pasp, 119,  360

\bibitem[Pier et~al.(2003)]{pier03}
Pier, J.~R., et al. 2003, \aj, 125, 1559

\bibitem[Prieto et~al.(2005)]{prieto05}
Prieto, J., et al. 2005, \iaucirc, 8633

\bibitem[Riess et~al.(1998)]{riess98}
Riess, A.~G., et al. 1998, \aj, 116, 1009

\bibitem[Riess et~al.(2004)]{riess04}
Riess, A.~G., et al. 2004, \apj, 607, 665

\bibitem[Riess et~al.(2005)]{riess05}
{Riess}, A.~G., et al. 2005, \apj, 627, 579

\bibitem[Riess et~al.(2006)]{riess07}
Riess, A.~G., et al. 2006, ArXiv Astrophysics e-prints

\bibitem[Rigon et~al.(2003)]{rigon03}
Rigon, L., et~al. 2003, \mnras, 340, 191

\bibitem[Salzer et~al.(2005)]{salzer05}
Salzer, J.~J., et~al. 2005, \apj, 624, 661

\bibitem[Scannapieco \& Bildsten(2005)]{scannapieco05}
Scannapieco, E. \& Bildsten, L. 2005, \apjl, 629, L85

\bibitem[Schlegel(1990)]{schlegel90}
Schlegel, E.~M. 1990, \mnras, 244

\bibitem[Schlegel et~al.(1998)]{sfd98}
Schlegel, D.~J., Finkbeiner, D.~P., \& Davis, M. 1998, \apj, 500, 525

\bibitem[Schmidt et~al.(1994)]{schmidt94}
Schmidt, B.~P., et al. 1994, \apjl, 434, L19

\bibitem[Smith et~al.(2002)]{smith02}
Smith, J.~A., et al. 2002, \aj, 123

\bibitem[Soderberg \& Frail(2005)]{soderberg05}
Soderberg, A.~M. \& Frail, D.~A. 2005, The Astronomer's Telegram, 663, 1

\bibitem[Spergel et~al.(2006)]{spergel06}
Spergel, D.~N., et al. 2006, ArXiv Astrophysics e-prints

\bibitem[Stanek et~al.(2006)]{stanek06}
Stanek, K.~Z., et al. 2006, Acta Astronomica, 56, 333

\bibitem[Stritzinger et~al.(2006)]{stritzinger06}
Stritzinger, M., et~al. 2006,  \aap, 460, 793

\bibitem[Taubenberger et~al.(2006)]{tautenberger06}
Taubenberger, S., et~al.(2006), \mnras, 371, 1459

\bibitem[Tucker et~al.(2006)]{tucker06}
Tucker, D.~L., et al. 2006, Astronomische Nachrichten, 327, 821

\bibitem[Turatto, et~al.(1993)]{turatto93}
Turatto, M., et al. 1993, \mnras, 262, 128

\bibitem[Turatto et~al.(2000)]{turatto00}
Turatto, M., et~al. 2000, \apjl, 534, L57

\bibitem[van Zee et~al.(2006)]{vanzee06}
van Zee, L., Skillman, E.~D., \& Haynes, M.~P. 2006, \apj, 637, 269

\bibitem[Wang et~al.(2004)]{wang04}
Wang, L., et~al. 2004, \apjl, 604, L53

\bibitem[Wood-Vasey et~al.(2007)]{wood-vasey07}
Wood-Vasey, W.~M., et al. 2007, ArXiv Astrophysics e-prints

\bibitem[Wood-Vasey et~al.(2004)]{wood-vasey04}
Wood-Vasey, W.~M., Wang, L., \& Aldering, G. 2004, \apj, 616, 339

\bibitem[Wyder et~al.(2005)]{wyder05}
Wyder, T.~K., et al. 2005, \apjl, 619, L15

\bibitem[York et~al.(2000)]{york00}
York, D.~G., et al. 2000, \aj, 120, 1579

\bibitem[Zijlstra(2004)]{zijlstra04}
Zijlstra, A.~A. 2004, \mnras, 348, L23

\end{thebibliography}
\end{document}